\documentclass[11pt,a4paper]{article}
\usepackage{jcappub}
\usepackage[english]{babel}
\usepackage[utf8]{inputenc}
\usepackage{graphicx}
\usepackage{amsmath,amssymb,amsfonts,amsthm}
\usepackage{bm,bbm}
\usepackage{tikz,stackrel}
\usepackage{pstricks}
\usepackage{pifont} 
\usepackage{empheq}
\newcommand\nn{\nonumber\\}
\newcommand\non{\nonumber}
\newcommand{\bc}{\begin{center}}
\newcommand{\ec}{\end{center}}
\newcommand{\be}{\begin{equation}}
\newcommand{\ee}{\end{equation}}
\newcommand{\ba}{\begin{eqnarray}}
\newcommand{\ea}{\end{eqnarray}}
\def\bs{\begin{subequations}}
\def\es{\end{subequations}}
\newcommand{\ben}{\begin{equation*}}
\newcommand{\een}{\end{equation*}}
\newcommand{\ban}{\begin{eqnarray*}}
\newcommand{\ean}{\end{eqnarray*}}
\renewcommand{\leq}{\leqslant}
\renewcommand{\geq}{\geqslant}
\def\a{\alpha}
\def\b{\beta}
\def\de{\delta}
\def\g{\gamma}
\def\la{\lambda}
\def\k{\kappa}
\def\e{\epsilon}
\def\ve{\varepsilon}
\def\Om{\Omega}
\def\om{\omega}
\def\De{\Delta}
\def\G{\Gamma}
\def\t{\theta}  
\def\s{\sigma}

\def\vp{\varphi}
\def\N{\nabla}
\def\Hes{{\hat\cH}}
\def\cA{\mathcal{A}}
\def\cB{\mathcal{B}}

\def\cF{\mathcal{F}}
\def\cG{\mathcal{G}}
\def\cH{\mathcal{H}}

\def\cK{\mathcal{K}}
\def\cL{\mathcal{L}}

\def\cO{\mathcal{O}}

\def\cR{\mathcal{R}}

\def\ds{d_\textsc{s}}

\def\p{\partial}

\def\B{\Box}
\newcommand{\Eq}[1]{(\ref{#1})}
\newcommand{\Eqq}[1]{eq.~(\ref{#1})}
\newcommand{\Eqqs}[1]{eqs.~(\ref{#1})}

\def\cob{\color{blue}}

\newcommand{\au}[2]{#1.~#2}
\newcommand{\book}[5]{\emph{#1}, #2, #3, #4 (#5)}

\newcommand{\oarX}[1]{\href{http://arxiv.org/abs/#1}{{\ttfamily\cob arXiv:#1}}}
\newcommand{\arX}[1]{\href{http://arxiv.org/abs/#1}{{\ttfamily\cob arXiv:#1}}}
\newcommand{\doin}[6]{\href{http://dx.doi.org/#1}{{\cob {\it #2} {\bf #3 #4} (#6) #5}}}
\newcommand{\doinn}[5]{\href{http://dx.doi.org/#1}{{\cob {\it #2} {\bf #3} (#5) #4}}}
\newcommand{\doij}[5]{\href{http://dx.doi.org/#1}{{\cob {\it #2} {\bf #3} (#5) #4}}}
\newcommand{\ndoin}[6]{\href{#1}{{\cob {\it #2} {\bf #3 #4} (#6) #5}}}
\newcommand{\ndoinn}[5]{\href{#1}{{\cob {\it #2} {\bf #3} (#5) #4}}}
\newcommand{\procsinm}[5]{in \emph{#1}, #2 (eds.), #3, #4 (#5)}
\newcommand{\proc}[6]{in \emph{#1}, #2 (ed.), #3, #4, #5 (#6)}
\newcommand{\procm}[6]{in \emph{#1}, #2 (eds.), #3, #4, #5 (#6)}
\newcommand{\tia}[1]{\textit{#1},}
\newcommand{\boxd}[1]{\boxed{\phantom{\Biggl(}#1\phantom{\Biggl)}}}

\def\rme{e}
\def\rmd{d}
\def\rmi{i}
\def\Re{{\rm Re}}
\def\Im{{\rm Im}}

\newcounter{listcounter}

\allowdisplaybreaks

\begin{document}

\title{Ultraviolet-complete quantum field theories with fractional operators}

\author[a]{Gianluca Calcagni,}
\emailAdd{g.calcagni@csic.es}
\affiliation[a]{Instituto de Estructura de la Materia, CSIC, Serrano 121, 28006 Madrid, Spain}

\author[b]{Les{\l}aw Rachwa\l{} }
\emailAdd{grzerach@gmail.com}
\affiliation[b]{Departamento de Física -- Instituto de Ciências Exatas, Universidade Federal de Juiz de Fora, Campus Universitário, Rua José Lourenço Kelmer, s/n -- São Pedro 33036-900, Juiz de Fora, MG, Brazil}

\abstract{We explore quantum field theories with fractional d'Alembertian $\Box^\gamma$. Both a scalar field theory with a derivative-dependent potential and gauge theory are super-renormalizable for a fractional power $1<\gamma\leq 2$, one-loop super-renormalizable for $\gamma>2$ and finite if one introduces killer operators. Unitarity is achieved by splitting the kinetic term into the product of massive fractional operators, eventually sending the masses to zero if so desired. Fractional quantum gravity is also discussed and found to be super-renormalizable for $2<\gamma\leq 4$ and one-loop super-renormalizable for $\gamma>4$. To make it unitary, we combine the splitting procedure with a fractional generalization of the Anselmi--Piva procedure for fakeons. Among new technical results with wider applications, we highlight the Leibniz rule for arbitrary powers of the d'Alembertian and the Källén--Lehmann representation for a propagator with an arbitrary number of branch cuts.}


\maketitle

\tableofcontents


\section{Introduction}\label{intro}

The search for a quantum field theory (QFT) of gravity has produced a number of theories with extremely interesting features but, also, a plethora of difficult or very difficult challenges, both at the conceptual and at the technical level. String theory is based on QFT tools but applied to extended objects; there a spacetime emerges only at low energies and is higher-dimensional \cite{GSWi,Pol98,BBSb,Zwi09}. The notion of spacetime is non-fundamental also in loop quantum gravity and spin foams, where the basic degrees of freedom are not geometric and follow a discrete combinatorial harmony \cite{Rov07,Per13}, quantized non-perturbatively via the Hamiltonian and the path-integral formalism, respectively. The same holds in group field theory \cite{Ori09,BaO11,Fousp}, where the fundamental degrees of freedom are fields living on a group-valued manifold; notwithstanding this novel setting, a traditional QFT approach is indeed used in this case. In causal sets, neither the basic degrees of freedom nor their dynamics have the resemblance of gravitons in continuous spacetimes \cite{Dow13,Surya:2019ndm}. Causal dynamical triangulations \cite{AGJL4,Loll:2019rdj} and asymptotic safety \cite{NiR,Eichhorn:2018yfc,Bonanno:2020bil} follow non-perturbative quantization procedures, still within the realm of traditional but highly specialized field theories. In all these cases, the price to pay for complexity at the conceptual and mathematical levels is an inherent difficulty in extracting physics, ranging from something as basic as recovering Einstein's theory at the classical level to getting testable or even falsifiable predictions.

A recent trend in this research has been to move further away from the conceptual and mathematical complexity of some of the above major contenders and to renew the effort in applying perturbative QFT to gravity in four-dimensional spacetime. Classic negative results on the perturbative renormalizability of Einstein's gravity \cite{tHooft:1974toh,Deser:1974cz,Deser:1974cy,Deser:1974xq,Goroff:1985sz,Goroff:1985th,vandeVen:1991gw} (holding also for scalar-tensor theories \cite{Deser:1974cz}) and on the unitarity of its quadratic and higher-order extensions \cite{Stelle:1976gc,Stelle:1977ry,Julve:1978xn,Asorey:1996hz,Accioly:2002tz} have discouraged persistence along these lines. To get renormalizability, one must include higher-derivative curvature operators, but these introduce higher-order derivatives which, in turn, lead to an Ostrogradsky instability at the classical level and to a loss of unitarity at the quantum level. For quite some time, it seemed impossible to conciliate the two requirements of renormalizability and unitarity but results in the last two decades have paved the way to renewed hopes that gravity, after all, can be quantized just like the other fundamental forces of Nature. 

Four proposals in this direction are non-local quantum gravity \cite{Kra87,Kuz89,Tom97,Modesto:2011kw,BGKM,Modesto:2017sdr,Calcagni:2021hve,BasiBeneito:2022wux}, local gravity with fakeons \cite{Anselmi:2017yux,Anselmi:2017lia,Anselmi:2018kgz,Anselmi:2018ibi,Anselmi:2018bra,Anselmi:2021hab}, local gravity with complex ghosts \cite{Modesto:2015ozb,Modesto:2016ofr,Liu:2022gun} and quantum gravity with fractional operators \cite{revmu,Calcagni:2021ipd,mf2}. Although superficially these theories may look as technically difficult as the proposals mentioned at the beginning, in fact, they are not. The four of them use standard perturbative QFT on Minkowski spacetime with a well-defined notion of Feynman diagrams, renormalizability, perturbative unitarity, and so on. What makes them perhaps unfamiliar to the casual reader is, on one hand, their lesser popularity in the community due to their very young age and, on the other hand, the deviations from traditional QFTs in the choice of kinetic term, propagator prescription or techniques for computing scattering amplitudes in momentum space. However, such novelties aside, these four frameworks of quantum gravity are rather conservative.

In this paper, we further explore QFTs with a fractional d'Alembertian operator. In these theories, the dynamics is non-local but different from the above-mentioned non-local quantum gravity, since here the kinetic terms are made of derivative operators of integer as well as fractional order, or of integer as well as fractional powers of the d'Alembertian (or Laplace--Beltrami) operator $\B^\g$, where $\g$ is real.\footnote{Scalar field theories with fractional operators received attention already three decades ago as interesting models of mathematical physics \cite{Gia91,BG,doA92,BOR,Barci:1996ny} but it has not been until recently that they have been revived in the context of quantum gravity \cite{Calcagni:2009qw,frc1,frc2,Tri12,Belenchia:2014fda,Saravani:2015rva,Belenchia:2016sym,Tri17,Belenchia:2017yvv,Saravani:2018rwm,Tri18,ACo,mf1}, within the frameworks of Ho\v{r}ava--Lifshitz gravity \cite{Calcagni:2009qw}, multi-fractional spacetimes \cite{frc1,frc2,mf1}, non-commutative geometry \cite{Tri12,Tri17,Tri18,ACo} and causal sets \cite{Belenchia:2014fda,Saravani:2015rva,Belenchia:2016sym,Belenchia:2017yvv,Saravani:2018rwm}. Each of these paradigms sets the form of the kinetic term differently from the other proposals. Fractional operators have been applied to the fully non-linear gravitational sector only in \cite{mf2,Munkhammar:2010gq,Vacaru:2010wn} at the classical level and \cite{mf2,Calcagni:2009qw} at the quantum level.} Unitarity and renormalizability have been studied only for a scalar-field case with monomial potential $V(\phi)\propto \phi^N$ and it was found that they might be incompatible with each other \cite{mf1}. According to these results, preliminary conjectured to hold also in the case of gravity \cite{mf2}, it seemed as if fractional non-locality improving renormalizability ($\g>2$) would be unable to preserve unitarity ($\g<1$) at the same time. However, some one-loop diagrams were found to be finite even in the range $\g<1$ where the theory is unitary, thus leaving the question of the ultraviolet (UV) behaviour of this proposal open.

After further recalling the physical motivation and relevance of fractional theories in section \ref{sec1mot}, we consider a scalar field with derivative interactions as a toy model for QFTs invariant under a local symmetry group, such as gauge theories and gravity, which are also studied explicitly here. The scalar and gauge-theory actions are presented, respectively, in section \ref{sec1a} and \ref{sec1b}. Once secured that the usual BPHZ renormalization scheme and power-counting argument apply to fractional theories (appendix \ref{secbphz}), we find that the scalar theory is super-renormalizable (finite number of superficial divergences \cite{Kaku:1993ym,PeSc}, when written as diffeomorphism-invariant operators) for $1<\g<2$ and one-loop super-renormalizable (i.e., with divergences only at one loop) for $\g>2$ (section \ref{sec2}). Finiteness (no divergent diagrams at any loop order, all beta functions vanishing, all Green functions convergent off-shell) can be achieved introducing killer operators of sixth order in the field, as shown in section \ref{killers}. Next, we discuss the allowed range of $\g$ for the consistency of the tree-level propagator using the Källén--Lehmann representation (section \ref{sec3a}), both for the Feynman prescription (section \ref{sec3b}) and for the Anselmi--Piva prescription (section \ref{sec3c}). Surprisingly, the Anselmi--Piva procedure works also for non-integer powers of the $\B$, in the sense that the degrees of freedom associated with $\B^\g$ amount to a gas of quasi-particles (not particles, as in local gravity with fakeons) that remain virtual at all loop levels. However, both the Feynman and the Anselmi--Piva prescription are consistent only for $\g<1$, thus making the theory either non-renormalizable or ill-defined. This problem is fixed if we split the fractional operator into operators with different masses and fractional exponents distributed according to the partition of the unity (section \ref{sec4}). We do so via a very general recipe based on generic branch cuts at any angle in the complex plane. The resulting scalar or gauge theory, for which the power-counting argument of renormalization does not change, is well-defined (no global bound $\g<1$), tree-level unitary and (super-)renormalizable. Sending the masses to zero at the end of the calculation leads to a massless theory. Quantum gravity is discussed in section \ref{secgrav}, where we show that a systematic mismatch of the fractional derivative powers in the interaction terms requires technical modifications of the results obtained for the scalar model. In particular, renormalizability and unitarity can be achieved simultaneously after some work. Conclusions are in section \ref{sec5}. 

Two novel and very general technical results reported in the appendices, that can have applications in physics and mathematics beyond the present context, are the Leibniz rule for integer as well as arbitrary non-integer powers of the d'Alembertian in Minkowski spacetime (appendix \ref{appA}) and the theory of branch cuts at arbitrary angles in the complex plane, which allows one to write the Källén--Lehmann representation of a propagator with an arbitrary number of cuts (appendix \ref{appC}). The other appendices include technical results to be used in the course of the paper.


\section{Fractional field theories}\label{sec1}

After spelling out the physical motivation of fractional field theories, in this section we introduce the scalar and the gauge models as propaedeutic for the case of gravity discussed later.


\subsection{Motivation}\label{sec1mot}

The program of multi-fractional spacetimes \cite{revmu} aims at reproducing and controlling a feature emerging in all approaches to quantum gravity ---the change of spacetime dimension with the probed scale \cite{tH93,Car09,fra1,trtls,Car17,MiTr}. In some cases, this so-called dimensional flow corresponds to a spacetime with a multi-fractal geometry. Multi-fractional spacetimes pose the following question: If we assume dimensional flow to be a fundamental rather than an emergent characteristic of spacetime when the gravitational interaction is quantized, can we obtain a consistent QFT of gravity? Early attempts to define fractal spacetimes showed the difficulty of a direct combination of totally disconnected fractal geometry and fields \cite{Sti77,Svo87,Ey89a,Ey89b,BA07}. The use of fractional calculus made the task much easier because it places the fields on a continuum \cite{fra1}. Multi-fractional spacetimes are characterized by a multi-fractional measure and exotic kinetic operators, which may contain derivatives of both integer and fractional order.

Field theories with fractional operators are not introduced \emph{ad hoc} but are super-selected among other realizations of the multi-fractional paradigm according to a logical four-step reduction of all available possibilities. {\bf (1)} First, dimensional flow with a non-standard spacetime measure and second-order kinetic terms is possible but, contrary to early expectations, it does not improve renormalizability of any sector \cite{frc9} including gravity, which is the foremost goal of the program. This reduces the choice to higher- or fractional-order kinetic terms. {\bf (2)} A kinetic term with a higher-order or non-integer power has a clear physical interpretation: the spectral dimension of spacetime changes with the probed scale, to a value determined by the higher-order or fractional exponent. In particular, field theories with fractional derivatives describe matter and gravity interacting in spacetimes with ordinary measure but an intrinsic multi-fractal structure \cite{revmu,Calcagni:2021ipd,mf1}. This can be seen by computing the spectral dimension $\ds$ of spacetime from the modified dispersion relation of the theory, {as done with standard techniques in quantum gravity \cite{mf1,AJL4,LaR5,Ben08,Mod08,fra6,frc4}}. The fractional order $\g$ of the derivatives is related to the spectral dimension $\ds(\g)$, so that understanding the behaviour of the theory as a function of $\g$ gives information on the geometry of spacetime. At different scales, the value of $\ds$ changes, {ranging from \cite{mf1}
\be
\ds\simeq \frac{4}{\g}
\ee}
in the UV to $\ds\simeq 4$ in the infrared (IR), the large-scale limit coinciding with the topological dimension $D=4$. This variation of the spectral dimension is due to the presence of one or more fundamental scales defining the fabric of spacetime. {Indeed, the multi-fractional paradigm is a sub-case of the larger framework of multi-scale spacetimes, where the presence of a non-trivial UV scaling in dispersion relations or in the spacetime measure requires length scales which are part of the definition of the geometry on which the fields live.} In the theories studied in this paper, there is only one length scale $\ell_*$, hidden in dimensionful coupling constants $a_{\g}$ and $\lambda_{\g-1}$ in the case of the scalar field. This scale $\ell_*$ defines the division between the UV and the IR regimes. {Since we are interested in a theory of quantum gravity without spoiling the known phenomenology of the other fundamental interactions, the size of $\ell_*$ is expected to be above the LHC scale, possibly of order of the Planck scale. Constraints on this scale in the other multi-fractional theories have spanned a number of experiments and observations and a summary of the results can be found \cite{revmu,Calcagni:2021ipd}.

{\bf (3)} The presence of a kinetic term with anomalous scaling makes a multi-fractional spacetime measure unnecessary, thus avoiding the difficulties of a QFT with non-trivial spacetime measure \cite{frc9}. {\bf (4)} Last, analyticity of the propagator requires these fractional operators to be Lorentz-invariant, thereby selecting a fractional d'Alembertian instead of fractional derivatives \cite{mf0,mf1}. 

Therefore, the analysis of QFTs with varying Hausdorff and spectral dimension carried in the literature showed that a theory with fractional d'Alembertian is a somewhat mandatory choice within the multi-fractional paradigm. At this point, one can legitimately ask (and rigorously answer) whether this setting with a natural dimensional flow enjoys unitarity and improved renormalizability.}

We work in Minkowski spacetime in $D=4$ topological dimensions with a mostly-plus Lorentzian metric $\eta_{\mu\nu}={\rm diag}(-,+,+,+)_{\mu\nu}$. Theories with fractional operators can be defined with different symmetries but imposing diffeomorphism invariance simplifies the problem greatly \cite{mf1}. Therefore, here we will consider fractional powers of the d'Alembertian. On Minkowski spacetime, the ordinary d'Alembertian is defined as usual as $\B\coloneqq\N_\mu\N^\mu=\eta_{\mu\nu}\p^\mu\p^\nu=-\p_t^2+\p_1^2+\p_2^2+\p_3^2$ when acting on a rank-0 tensor. In Lorentzian momentum space, $-\B\to k^2=-k_0^2+|\bm{k}|^2$.


\subsection{Fractional scalar theory}\label{sec1a}

Although our primary goal is gravity, we introduce the basic properties of the theory considering scalar and gauge fields first. This is done both for completeness, to show how fractional physics works in all particle sectors, and because gauge theory is a sort of laboratory where to sketch the main features of gravity, since diffeomorphism invariance can be regarded as a special but complicated case of gauge invariance. In turn, fractional gauge theory can be studied through a scalar toy model with a very specific choice of interaction. Following the order of progression of the tensorial structure, we start with a scalar (rank 0), continuing with a gauge field (rank 1) and ending with gravity (rank 2).

The action of the scalar model we consider for renormalization is
\be
S=\!\int\!\rmd^{4}x\left[\frac{1}{2}\phi\left(\B+a_{\g}\B^{\g}\right)\phi-\phi^{2}\left(\la_0+\lambda_{\g-1}\B^{\g-1}\right)\phi^{2}\right],\label{eq:model}
\ee
where $\phi$ is a real scalar with energy dimensionality
\be
[\phi]=1\,,
\ee
so that the coefficients in the action have dimensionality
\be
[\lambda_{0}]=0
\ee
 and (note here a hidden dependence on the fundamental scale $\ell_*^{2(\g-1)}$)
\be
[a_{\g}]=[\lambda_{\g-1}]=2(1-\g)\,.
\ee
The fractional exponent $\g$ is positive and non-integer, $0<\g\notin\mathbb{N}$. In $a_\g$ and $\lambda_{\g-1}$, we have absorbed a complex phase $\exp(\rmi\pi\g)$ coming from $(-\B)^\g=\exp(\rmi\pi\g)\B^\g$. In this way, in momentum space the action is real, and so is the ratio $\lambda_{\g-1}/a_\g$, which will appear later in beta functions. Note that there is no mass term in \Eq{eq:model} and one neglects odd-leg-number interaction vertices, because of the $\mathbb{Z}_{2}$ symmetry imposed on the scalar field.

The particular choice in \Eq{eq:model} of a quartic interaction endowed with a fractional power of the d'Alembertian differing by exactly 1 with respect to the power in the kinetic term is motivated by the fact that it reproduces the structure of the action of the perturbative longitudinal mode in fractional gauge theory, as we will show in \Eqq{gauguv}. Therefore, the quantum properties of the scalar model \Eq{eq:model} will be the same as the gauge model \Eq{gauge}. In particular, the constraint on the relative difference of derivative powers comes from gauge invariance. Also, a quartic interaction term not only contains the smallest possible number of scalar fields consistent with the $\mathbb{Z}_2$ symmetry $\phi\to-\phi$, but is also compatible with the interaction order of vector gauge bosons. Later, to identify the scalar with the graviton polarization modes, we will have to generalize the scalar model. In section \ref{genpowco}, we will perform a power-counting analysis for a generic interaction and independent fractional coefficients in the kinetic and interaction terms. Although this will result in a different range of $\gamma$ for super-renormalizability, the technical tools and the conceptual framework with which to study perturbative renormalizability and unitarity will essentially be the same. That is why we will study the simple model \Eq{eq:model} in detail.

For $\g<1$, the fractional part of the action \Eq{eq:model} would be dominant in the IR. The theory would become a plain model in the UV and, obviously, its renormalization properties would be the same as for a standard scalar $\lambda_0\phi^{4}$ theory. This case will not be further analyzed here. Therefore, for $\g>1$ the IR limit of the action \Eq{eq:model} is
\be\label{SUVaction}
S_{\textsc{ir}}=\int\rmd^{4}x\left[\frac{1}{2}\phi\B\phi-\lambda_{0}\phi^{4}\right],
\ee
while the leading part of the action in the UV regime (important for divergences) is
\be\label{suv}
S_{\textsc{uv}}=\int\rmd^{4}x\left[\frac{1}{2}a_{\g}\phi\B^{\g}\phi-\lambda_{\g-1}\phi^{2}\B^{\g-1}\phi^{2}\right].
\ee
{Note that the kinetic term cannot be transformed into a canonical one with two derivatives without changing the physics. In fact, while local field theories connected by a non-linear local field redefinition are physically equivalent since their S-matrix is the same \cite{Chisholm:1961tha,Kamefuchi:1961sb}, the same is true only for certain non-local transformations \cite{Chisholm:1961tha,Bergere:1975tr} but not in general. Only linear non-local equations can be made fully local by a field redefinition, while for non-linear ones non-locality is transferred to interaction terms, the problem of initial conditions is changed and, ultimately, so is the physical spectrum. At the quantum level, the measure of the path integral is transformed and the quantization of the two theories is inequivalent. For instance, non-local field redefinitions made of entire form factors do not alter the spectrum, as is well-known in non-local quantum gravity with exponential or asymptotically polynomial operators \cite{BasiBeneito:2022wux}, while a redefinition in a non-local, purely phenomenological model with non-entire form factors given by integer inverse powers of the d'Alembertian leads to a different theory \cite{Zhang:2016ykx}. Fractional operators are not entire and, if one applied a field redefinition to make the kinetic term canonical, all the complications we will find in the unitarity analysis of the free action would be moved to the interacting theory and, hence, to the issue of perturbative unitarity, with no guarantee that the new theory be physically equivalent to the original. Another problem with such field redefinitions is that they would be inapplicable in highly non-linear equations of motion such as those stemming in the gauge and gravitational theories studied in sections \ref{sec1b} and \ref{secgrav}. In particular, they would obscure the parallelism between the model \Eq{suv} and the gauge-field UV action \Eq{gauguv} derived below.}


\subsection{Fractional gauge theory}\label{sec1b}

{Studying gauge theories to illustrate certain physical or technical features of quantum gravity theories is a common procedure we will follow also here. This is not the only reason why to consider gauge fields. Supposing the gravitational sector was indeed endowed with fractional operators, a unified theory including the Standard Model should have the same type of operators or, at least, there is no natural reason why it should not. Since the fundamental scale $\ell_*$ attached to the fractional operator is expected to be of order of the Planck scale, any such UV modification of the gauge sector would be irrelevant for particle-physics phenomenology, for instance at the LHC. Thus, the direct purpose of fractional gauge theories is not to improve the Standard Model (although there may be a number of reasons to do so \cite{Modesto:2015lna,Modesto:2015foa}) but, on one hand, to illustrate the behaviour of fractional operators in sectors with non-trivial gauge symmetries and, on the other hand, to pave the way for a unified formulation of a QFT of all fundamental interactions.}

In \cite{mf1}, we considered non-derivative scalar potentials $V(\phi)$, in particular $V\propto \phi^3$ to compute some basic Feynman diagrams. From the form of the kinetic term, this theory was dubbed ``$T[\B+\B^\g]$.'' The model \Eq{eq:model} still belongs to the same class and arises as an illustration for theories with gauge symmetry and fractional powers of the gauge-covariant d'Alembertian, such as
\be\label{gauge}
S=\int\!\rmd^{4}x\left[F_{\mu\nu}\left(b_{0}+b_{\g-1}\B^{\g-1}\right)F^{\mu\nu}\right].
\ee
The scalar field above is analogous to the gauge fluctuation $\delta A_{\mu}\longleftrightarrow\phi$. To show this, consider the UV limit of the action \Eq{gauge} (setting $b_{\g-1}=1$):
\be\label{gaugeuv}
S_\textsc{uv}=\int \rmd^{4}x\,F_{\mu\nu}\B^{\g-1}F^{\mu\nu}\,.
\ee
Writing the gauge field around the flat gauge connection background,
\be
A_{\mu}=\overline{A}_{\mu}+a_{\mu}\,,\qquad \overline{A}_{\mu}=0\,,
\ee
the gauge field strength is schematically 
\be
F_{\mu\nu}\sim\p_{\mu}a_{\nu}+ g_A a_{\mu}a_{\nu}\,,
\ee
where $g_A$ is the non-Abelian gauge coupling, so that the action \Eq{gaugeuv} can be expanded to the quadratic order:
\be\label{suvF1}
S_\textsc{uv}^{(2)}=\int \rmd^{4}x\,a_{\mu}\cO^{\mu\nu}a_{\nu}\,,
\ee
where the operatorial kernel ${\cal O}^{\mu\nu}$ is of the general form
\be
\cO^{\mu\nu}=b_{\g,1}\eta^{\mu\nu}\B^{\g}+b_{\g,2}\N^{\mu}\N^{\nu}\B^{\g-1}\,,
\ee
and $b_{\g,1}$ and $b_{\g,2}$ are numerical coefficients. Assuming that there exists a constant spacetime vector $L^{\mu}$ normalized to one $(L_{\mu}L^{\mu}=1)$, we can project the gauge vector on its direction according to the decomposition
\be\label{aparal1}
a_{\mu}=a_{\mu\parallel}+a_{\mu\perp}\coloneqq L_{\mu}L^{\nu}a_{\nu}+\left(a_{\mu}-L_{\mu}L^{\nu}a_{\nu}\right)\,,\qquad L^{\mu}a_{\mu\perp}=0\,,\qquad a_{\mu\perp}a_{\parallel}^{\mu}=0\,.
\ee
From this, we can define the scalar
\be\label{aparal2}
a_{\parallel}=L^{\nu} a_{\nu}\,,
\ee
so that the quadratized action \Eq{suvF1} becomes
\be
S_\textsc{uv}^{(2)}=\int \rmd^{4}x\,a_{\parallel}\left(b_{\g,1}\B^{\g}+b_{\g,2}L_{\mu}L_{\nu}\nabla^{\mu}\nabla^{\nu}\B^{\g-1}\right)a_{\parallel}+\ldots\,,\label{suvF2}
\ee
where we only look at the parallel components and drop terms in $a_{\mu\perp}$. Assuming that
\be
L_{\mu}\sim\frac{\N_{\mu}}{\sqrt{\B}}\qquad \Longrightarrow\qquad L_{\mu}L_{\nu}\sim\frac{\N_{\mu}\N_{\nu}}{\B}\,,
\ee
we get
\be
S_\textsc{uv}^{(2)}=\left(b_{\g,1}+b_{\g,2}\right)\int \rmd^{4}x\,a_{\parallel}\B^{\g}a_{\parallel}\eqqcolon b_\g\int \rmd^{4}x\,a_{\parallel}\B^{\g}a_{\parallel}\,.
\ee
This reproduces the kinetic term in \Eqq{suv}. Expanding \Eq{gaugeuv} to fourth order in $a_{\mu}$ and using $\mathbb{Z}_{2}$ symmetry, we get
\be\label{gauguv}
S_\textsc{uv}^{(4)}=\int \rmd^{4}x\left(b_\g\, a_{\parallel}\B^{\g}a_{\parallel}+g_A^2a_{\parallel}^{2}\B^{\g-1}a_{\parallel}^{2}\right),
\ee
where we dropped an $O(g_A)$ cubic term, because it is sub-dominant both in the IR and in the UV and we used the assumptions of $\mathbb{Z}_{2}$ symmetry. Here we have $[a_{\mu}]=[a_{\parallel}]=[\phi]=1$, coinciding with the dimension of the canonically normalized scalar field in $D=4$ dimensions. Identifying $a_{\parallel}=\phi$, we get \Eqq{suv}. Note that we obtain the correct interaction term only in the case of a non-Abelian theory ($g_A\neq 0$), otherwise the scalar theory is free.

In section \ref{gaukil}, we will add $O(F^4)$ operators to the action \Eq{gauge} to make the theory UV-finite.



\section{Renormalizability and super-renormalizability}\label{sec2}

The renormalization of standard, local QFTs can be studied either directly via the explicit calculation of Feynman diagrams or through the shortcut of power counting, the classification of all possible divergences of diagrams at all loop orders solely based on the superficial degree of divergence of momentum integrals in the cut-off regularization scheme. If the objective is to check whether a theory is renormalizable or not, regardless of the values of the beta functions, power counting is very convenient. However, it cannot be taken for granted in non-local theories, since its validity is based upon renormalization schemes usually proven only for local theories. One such scheme is the regularization-independent Bogoliubov--Parasiuk--Hepp--Zimmermann (BPHZ) renormalization \cite{Bogoliubov:1957gp,Hepp:1966eg,Zimmermann:1969jj}, a special case of algebraic renormalization (see \cite{Lowenstein:1975ug,Piguet:1995er} for reviews)
equivalent to augmenting the bare action with local regularization-dependent counter-terms.

Violations of the BPHZ scheme could lead to several problems such as the following. Consider a non-local theory where all one-loop divergences have been absorbed by a finite number of counter-terms. A theory is said to be renormalizable at higher loops once all one-loop divergences are taken care of in all possible places of a big diagram, together with possible overall genuine higher-loop divergences. At the two-loop level, for instance, there may be some overlapping divergences or sub-divergences (i.e., divergent parts of a sub-diagram) related to the divergences already met at the one-loop level. If the assumptions for the standard BPHZ procedure hold, then these divergences of one-loop origin can be absorbed by a standard one-loop counter-term Lagrangian and the theory is renormalizable at two loops. However, if the BPHZ scheme fails, then these divergences are a problem for the renormalization of the theory at the two-loop and higher levels. What could happen is that they may not be absorbable in one-loop-looking local counter-terms, or that such counter-terms may be non-local, or that they do not exist at all. In the second case, this problem is strictly related also to an issue of non-analyticity of the loop integrals. In other words, the condition for integrability of counter-terms (in the sense of originating from some local action) would be violated in generic non-local theories and the power-counting argument would not faithfully account for all possible divergences of Feynman diagrams.

In appendix \ref{secbphz}, we show that the BPHZ scheme holds unmodified in fractional theories. In turn, this guarantees the validity of the power counting of divergences, which we apply in section \ref{powcou}. We find that the theory \Eq{suv} is renormalizable for any $\g>1$ with the superficial degree of divergences $\om$ of any loop diagram bounded by $\om\leq4$. Logarithmic divergences ($\om=0$) are the only ones universal with respect to gauge-fixing parameters, regularization- and renormalization-scheme dependencies. In four dimensions and in dimensional regularization, we will argue that the divergent part of the quantum effective action reads
\be
\Gamma_{{\rm div}}=\frac{1}{\varepsilon}\!\int\!\rmd^{4}x\left[\frac{1}{2}\beta_{1}\phi\B\phi-\beta_{\lambda_{0}}\phi^{4}\right]\!,\label{eq:divergences}
\ee
with generally non-zero coefficients $\beta_{1}$ and $\beta_{\lambda_{0}}$. We will consider only the highest (superficial) level of divergence and ignore divergences proportional to the cosmological constant (absent at the classical level) and to the mass parameter of the theory (as noted above, the absence of a mass is reminiscent of gauge symmetries underlying fractional gauge theories). 


\subsection{Power counting}\label{powcou}

Given the UV limit \Eq{suv} valid for $\g>1$, the superficial degree of divergence of any Feynman graph $\G$ is
\be\label{sudedi}
\om_\G\leq4L+N_V(2\g-2)-2N_I\g\,,
\ee
where $L$ is the number of loops, $N_V$ is the number of four-leg vertices and $N_I$ is the number of internal lines. The topological relation
\be\label{pcrel1}
N_V-N_I=1-L
\ee
implies $4L=-4N_V+4N_I+4$, while the relation 
\be\label{pcrel2}
4N_V=2N_I+N_E
\ee
implies $-4N_V+2N_I=-N_E$, where $N_E$ is the number of external legs with scalar fields. Then, we get
\ba
4L+N_V(2\g-2)-2N_I\g
&=&4L+2(\g-1)\left(N_V-N_I\right)-2N_I\nn
&=&-2(\g-1)(L-1)+4L-2N_I\nn
&=&4-2(\g-1)(L-1)-4N_V+2N_I\nn
&=&4-2(\g-1)(L-1)-N_E\,,
\ea
and the superficial degree of divergence becomes
\be\label{eq:pc}
\om_\G\leq 4-2(\g-1)(L-1)-N_E\,.
\ee
Power-law divergences in the UV cut-off of loop integrals ($\om_\G> 0$) are not universal and their coefficients depend on the details of the regularization scheme; in some schemes such as dimensional regularization, they are completely absent \cite{Leibbrandt:1975dj,Barvinsky:1985an}. Only logarithmic UV divergences ($\om_\G=0$) are universal in a sense of their independence on the details of the regularization and renormalization procedures. Therefore, in this paper we will be interested only in logarithmic UV divergences.

Now we are in a position to show that the form of the divergent part of the effective action is as in \Eqq{eq:divergences}. We start with the observation that the divergent part of the effective action in $D=4$ dimensions must be the sum of operators of the form $\phi\cO_i\phi$ with $[\cO_i]=2$. For local QFTs (possibly with higher derivatives in their defining classical tree-level action), it is well known that the UV divergences can be completely expressed as a local functional of the quantum fields. Therefore, to build the operators $\cO_i$ in the divergent part of the effective action, we can only use integer positive powers of the fields (here the scalar field $\phi$) and non-negative powers of various differential operators acting on these fields (for instance, the partial derivative $\p_\mu$, or the d'Alembertian $\B$), while we cannot use negative powers of such differential operators. This theorem \cite{Weinberg:1959nj} holds mathematically for local higher-derivative models but, by an explicit computation, we can extend its range of validity also to the weakly non-local theories considered here. The flag example is given by fractional theories with the operator $\B^\g$ with $\g>0$, where in this way we do not introduce any strong non-locality, i.e., operators with $\g<0$.\footnote{Models with strong non-locality \cite{Frolov:1979tu} are used phenomenologically to describe the late-time evolution of the universe \cite{Belgacem:2020pdz} but they are notoriously difficult to embed in a QFT, even at the level of effective field theory, due to their peculiar IR regime \cite{Barvinsky:2003kg,Maggiore:2016fbn,Belgacem:2017cqo}. In section \ref{arcsinfty}, we will see a basic example of the problems one can face when quantizing strongly non-local theories.}

The situation with weakly non-local theories (of which theories with fractional derivatives are a sub-case) is under control since we know that the form of the UV Lagrangian is still local, and it develops neither weak nor strong non-locality. This can be understood intuitively from the fact that the UV regime of the theory is when the energies are very high, corresponding to very small distances between points. In other words, UV divergences express the behaviour of the theory when, for example, the two-point Green's functions are computed at two spacetime points $x$ and $x'$ and when the mathematical limit $\lim_{x'\to x}$ is taken. To write the result of such limit with $x'\to x$, we can use only local functionals of the fields containing two quantum fields $\phi=\phi(x)$ (evaluated at the same point $x$) and only with differential operators acting on them in such a way that also their action is local. This last condition means that we can use only positive and finite powers of differential operators like $\B$ and $\p_{\mu}$ acting on the field $\phi=\phi(x)$  at the same spacetime point $x^\mu$. The claim that in weakly non-local (including fractional) theories the form of the UV divergences is still local is proven in section \ref{feydia} by an explicit computation using Feynman diagrams. 

As explained above, in fractional theories the power-counting analysis gives a bound on the superficial degree of divergence of any divergent graph of the theory, according to formula \Eq{eq:pc}. There, we see that the dimension of the divergent terms in the Lagrangian (and also the corresponding counter-terms) is bounded by $4$ in $D=4$ spacetime dimensions. From the local character of divergences, we know that it must also be non-negative. Therefore, the list of all possible local divergent terms is
\be
\Lambda\,, \quad m^2\phi^2\,, \quad \phi\B\phi\,, \quad \phi^4\,.
\ee
All these terms are local and their dimension is 4. The first term corresponds to the contribution to the cosmological constant, which we neglect here since we can always shift the vacuum energy of the QFT when taking Minkowski spacetime as the background; this operation is not observable in non-gravitational particle physics. The absence of a cosmological-constant counter-term can also be explained due to the fact that, to source such logarithmic divergences, one would need the presence of another mass or energy scale in the theory to construct a ratio of dimensionality $[\Lambda]=4$. Similarly, the second term is a UV-divergent correction to the mass of $\phi$ but, when one considers only the monomial behaviour in the UV theory, there is no reason to create a dimensionful quantity $m^2$ as written above, since in the action \Eq{suv} all the couplings have the same dimensionality. In general, when the theory contains various mass scales, various exponents $\g$, or even when only $a_\g\neq0$ and also $\lambda_0,\lambda_{\g-1}\neq0$, it is combinatorially possible to construct ratios whose dimensions agree with the dimension of the cosmological constant $\Lambda$ or of the constant mass term $m^2$. We omit such contributions motivated by the fact that, when the exponent $\g>1$ is closer to one, then the combinatorics of such terms become more and more complicated. We avoid it here for the sake of simplicity of our presentation. Therefore, we concentrate only on two types of logarithmic UV divergences, $\phi\B\phi$ or $\phi^4$. This leads to \Eqq{eq:divergences}.

To proceed further, we neglect vacuum diagrams ($N_E=0$), since their contribution to vacuum energy can be shifted away and one can just redefine the vacuum state without any observational consequences. In gravitational theories, this shift is reabsorbed into a cosmological constant. We also note that one-legged (tadpole) diagrams ($N_E=1$) vanish for a quartic interaction, due to $\mathbb{Z}_2$ symmetry. Then, the worst behaviour in \Eq{eq:pc} happens when $N_E=2$, so that
\be\label{omfin}
\om_\G\leq2-2(\g-1)(L-1)\,.
\ee
At one loop ($L=1$), the power-counting argument cannot exclude the presence of divergences. For $L>1$ and $\g>1$, there are no divergences provided $2-2(\g-1)(L-1)<0$, i.e., $(\g-1)(L-1)>1$. Therefore, the condition for power-counting super-renormalizability is 
\be\label{gmin}
\g>\frac{L}{L-1}\,.
\ee
When this inequality is saturated, the model has logarithmic divergences ($\om_\G=0$). In general, for
\be\label{lmax}
\frac{L_{\rm max}+1}{L_{\rm max}}<\g\leq\frac{L_{\rm max}}{L_{\rm max}-1}\,,
\ee
the theory is $L_{\rm max}$-loop super-renormalizable: It has divergences at $L=1,\ldots,L_{\rm max}$ loops while all higher-loop diagrams at levels $>L_{\rm max}$ are superficially convergent. In particular:
\begin{itemize}
\item For $\g>2$, the model has divergences only for $L=1$ and is one-loop super-renormalizable.
\item For $\frac{3}{2}<\g\leq2$, the model has divergences at $L=1,2$ loops.
\item For $\frac{4}{3}<\g\leq\frac{3}{2}$, the model has divergences at $L=1,2,3$ loops.
\item For $\frac{5}{4}<\g\leq\frac{4}{3}$, the model has divergences at $L=1,2,3,4$ loops.
\end{itemize}
And so on. In contrast, in the limit $\g\to 1^+$ divergences persist at all orders in perturbation theory but the model is still renormalizable. In fact, from the bound \Eq{omfin} we have $\om_\G\leq 2$, which is smaller than 4 at any loop order $L$ and does not grow with $L$. Here we only consider logarithmic divergences for which $\om=0$ and this means that the number of derivatives on external legs (that is, what we have in counter-terms) is still bounded by $4$, the energy dimension of possible counter-terms in \Eq{eq:divergences}. Hence the number of covariant structures needed to absorb UV-divergences does not grow with $L$. In all these cases, the form of the UV-divergent part of the quantum effective action is as in \Eqq{eq:divergences}.

Therefore, for $\g<1$ the theory is non-renormalizable (i.e., it has divergences at all loop orders and their number grows with $L$), for $\g=1$ it is strictly renormalizable (divergences at all loop orders and their number does do not grow with $L$), for $1<\g\leq 2$ it is super-renormalizable (there are only a finite number of divergences up to some loop order $L_{\rm max}$, excluding divergent sub-diagrams), while for $\g>2$ it is one-loop super-renormalizable (only divergences at the one-loop level are present, excluding divergent sub-diagrams):
\bs\label{sfreno}\ba
&&\text{strictly renormalizable:\hspace{1.9cm} $1=\g$}\,,\\
&&\text{super-renormalizable:\hspace{2.15cm} $1<\g\leq 2$}\,,\\
&&\text{one-loop super-renormalizable:\hspace{.6cm} $2<\g$}\,.
\ea\es

In non-local quantum gravity with asymptotically polynomial operators, the presence in the action of certain higher-order local operators called killers can make the theory finite \cite{MoRa1}. Finiteness, the absence of divergences at all perturbative orders, can lead to Weyl conformal invariance in the UV \cite{MoRa3,Rachwal:2022pfe}, which in turn can have momentous consequences for phenomenology \cite{Englert:1975wj,narlikar:1977nf,tHooft:2009wdx,Mannheim:2011ds,tHooft:2011aa,Bars:2011mh,Penrose:2010zz,Bars:2013yba,Prester:2013fia,Mannheim:2016lnx,Bambi:2016wdn,Bambi:2016yne,Rachwal:2018gwu,ConCos1,ConCos2}, but it is not necessary in order to have a predictive, fully controllable QFT in the UV. The model \Eq{eq:model} is finite for any $\g$ only when it is free, that is, when $\lambda_{\g-1}=0=\lambda_{0}$. In this degenerate case, no beta functions are generated. In section \ref{killers}, we will introduce killer operators to make the interacting theory finite.


\subsection{Beta functions}\label{betas}

Let us give a bird's eye view of the beta functions of the theory before embarking into actual calculations. We concentrate on the beta coefficient $\beta_{1}$ for the interaction terms $\phi^{4}$ in \Eqq{eq:divergences}. A similar structure can be found for the divergent coefficient $\beta_{\lambda_{0}}$. The conclusion will be that, when $\lambda_{\g-1}\neq0$ and for any $\lambda_{0}$ (regarded here as a sub-leading term), the beta function $\b_1$ at the one-loop level is directly proportional to $\lambda_{\g-1}$, so that 
\be
\beta_{1}^{(L=1)}=c_{1}\frac{\lambda_{\g-1}}{a_{\g}}\,.\label{eq:beta1L}
\ee
The linearity is due to the structure of the self-energy diagram $\vcenter{\hbox{\includegraphics[height=.4cm]{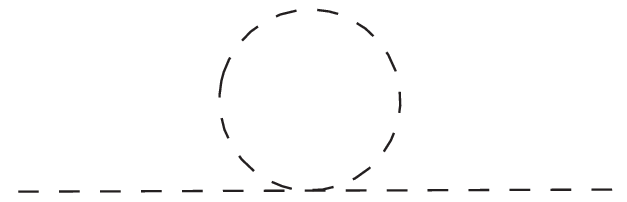}}}$ (for $\beta_{1}$) with one 4-leg interaction vertex and one propagator in the loop. Moreover, the numerical coefficient $c_{1}$ in \Eq{eq:beta1L} is proportional to $\g-1$ and vanishes continuously in the limit $\g\to1$.

At higher loops, the balance between the fractional powers in the kinetic and the interaction term breaks up except in some special cases where diagrams still display a logarithmic divergence. For all the other diagrams, the associated integral either scales as a non-integer positive power or as a negative power of integration momentum. The first case is a non-universal artifact due to cut-off regularization and the actual result in dimensional regularization is zero, as recalled above. This case appears only for $\gamma<2$. In the second case, the integral converges and it contributes to the finite part of the quantum effective action, which is not relevant for the renormalizability of the theory. Similar considerations hold also for the more general scalar model discussed in section \ref{genpowco}. Therefore, again, we can focus the discussion on logarithmically divergent graphs.

For higher-loop diagrams of this kind, if any, one expects a higher polynomial dependence of the perturbative beta function on the ratio $\lambda_{\g-1}/a_{\g}$:
\be
\beta_{1}^{(L)}=\sum_{i=1}^{L}c_{i}\left(\frac{\lambda_{\g-1}}{a_{\g}}\right)^{i}=c_{1}\frac{\lambda_{\g-1}}{a_{\g}}+c_{2}\left(\frac{\lambda_{\g-1}}{a_{\g}}\right)^{2}+\ldots+c_{L}\left(\frac{\lambda_{\g-1}}{a_{\g}}\right)^{L},
\ee
where the coefficients $c_{i}$ are all $\g$-dependent.

For the beta function $\beta_{\lambda_{0}}$, we expect a quadratic dependence on the ratio $\lambda_{\g-1}/a_{\g}$.
 At the $L$-loop level,
\be
\beta_{\lambda_{0}}^{(L)}=\sum_{i=1}^{L}d_{i}\left(\frac{\lambda_{\g-1}}{a_{\g}}\right)^{2i}=d_{1}\left(\frac{\lambda_{\g-1}}{a_{\g}}\right)^{2}+d_{2}\left(\frac{\lambda_{\g-1}}{a_{\g}}\right)^{4}+\ldots+d_{L}\left(\frac{\lambda_{\g-1}}{a_{\g}}\right)^{2L}.
\ee
where the coefficients $d_{i}$ are all $\g$-independent. 

The computation of these beta functions can be done using either the Barvinsky--Vilkovisky (BV; also called Schwinger--DeWitt or heat-kernel) technique \cite{Barvinsky:1985an,Avramidi:2000bm} for $\g\geq1$ or Feynman diagrams for any $\g$. The two methods are fully equivalent since the BV technique is nothing but the resummed version of the Feynman expansion. On one side, one works with covariant operators, while on the other side one handles non-covariant Feynman diagrams. We present both versions as a cross-check of the final result, confining the BV calculation to appendix \ref{appB}.


\subsection{One-loop effective action}\label{feydia}

In the Feynman-diagrams method, one does not make any assumption about the form of the UV divergences and computes them in a most direct and straightforward way, even in $D$ topological dimensions. When one uses other techniques such as the BV trace technology, some assumptions on the final form of the infinities are being made to ease the calculation.

For $\g>1$, the leading part of \Eq{eq:model} in the UV is \Eqq{suv}. In this regime, the propagator of the theory in Fourier space is
\be
\tilde G(-k^2)=\frac{1}{a_{\g}(-k^2)^{\g}}\,,
\ee
while the vertex of interactions is roughly (before symmetrization)
\be
\tilde{\rm V}=-24\rmi\la_{\g-1}(-k^2)^{\g-1}.
\ee
Note that, according to our $(-,+,\dots,+)$ signature convention, the coefficients $a_\g$ and $\la_{\g-1}$ carry a phase, while in the opposite convention they are real.

As said in section \ref{powcou}, we ignore vacuum and one-legged diagrams. The flat-spacetime one-loop vacuum diagram $\vcenter{\hbox{\includegraphics[height=.4cm]{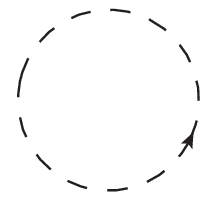}}}$ is always quartically divergent and counts the number and character of the degrees of freedom (fermionic or bosonic). The tadpole diagram vanishes due to the assumed $\mathbb{Z}_{2}$ symmetry.

\subsubsection{Two-point function}

The two-point function correction (giving rise to wave-function renormalization and to a mass) consists of the self-energy $\vcenter{\hbox{\includegraphics[height=.4cm]{self}}}$ diagram, made of one vertex and one propagator. The vertex is given by the expression
\ba
\tilde{\rm V} &=&-4\rmi\la_{\g-1}\left\{[-(p+k)^2]^{\g-1}+[-(p+r)^2]^{\g-1}+[-(p+s)^2]^{\g-1}\right.\nn
&&\left.+[-(k+r)^2]^{\g-1}+[-(k+s)^2]^{\g-1}+[-(r+s)^2]^{\g-1}\right\}.\label{vertgen}
\ea
For the topology $\vcenter{\hbox{\includegraphics[height=.4cm]{self}}}$, we can assign momenta
\ben
r=-k\,,\qquad s=-p\,,
\een
so that the vertex is
\be
\tilde{\rm V} =-8\rmi\la_{\g-1}\left\{[-(p+k)^2]^{\g-1}+[-(p-k)^2]^{\g-1}\right\},
\ee
and the whole diagram in $D$ dimensions is
\be
\parbox{4.1cm}{\includegraphics[width=4cm]{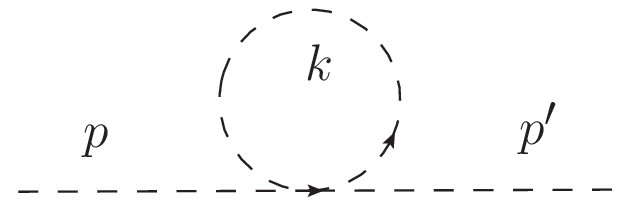}} = \rmi \Pi(p,p')=\rmi\de^D(p-p')\tilde\Pi(p),\label{selfd}
\ee
where
\be\label{tildeP}
\tilde\Pi(p)=8\lambda_{\g-1}\int\frac{\rmd^Dk}{(2\pi)^D}\frac{[-(p+k)^2]^{\g-1}+[-(p-k)^2]^{\g-1}}{a_{\g}(-k^2)^\g}\,.
\ee
Expanding the integrand for $k\gg p$, we have
\ban
&&\frac{[-(p+k)^2]^{\g-1}+[-(p-k)^2]^{\g-1}}{a_{\g}(-k^2)^\g}\\
&&\qquad\qquad=\frac{\left[-k^{2}\left(1+\frac{2p\cdot k+p^{2}}{k^{2}}\right)\right]^{\g-1}+\left[-k^{2}\left(1+\frac{-2p\cdot k+p^{2}}{k^{2}}\right)\right]^{\g-1}}{a_{\g}(-k^{2})^{\g}}\\
&&\qquad\qquad=(-k^{2})^{\g-1}\frac{\left(1+\frac{2p\cdot k+p^{2}}{k^{2}}\right)^{\g-1}+\left(1+\frac{-2p\cdot k+p^{2}}{k^{2}}\right)^{\g-1}}{a_{\g}(-k^{2})^{\g}}\\
&&\qquad\qquad=\frac{2+(\g-1)\frac{2p\cdot k+p^{2}}{k^{2}}+(\g-1)\frac{-2p\cdot k+p^{2}}{k^{2}}+(\g-1)(\g-2)\left(\frac{2p\cdot k}{k^{2}}\right)^{2}}{a_{\g}(-k^{2})}+O(p^{3})\\
&&\qquad\qquad=-\frac{2+2(\g-1)\frac{p^{2}}{k^{2}}+4(\g-1)(\g-2)\left(\frac{p\cdot k}{k^{2}}\right)^{2}}{a_{\g}k^{2}}+O(p^{3})\,.
\ean
After integrating over the angular variable, up to a solid-angle $2\pi^2$ factor this expression becomes
\be
-\frac{2+2(\g-1)\frac{p^{2}}{k^{2}}+(\g-1)(\g-2)\frac{p^{2}}{k^{2}}}{a_{\g}k^{2}}+O(p^{3})=-\frac{2+\g(\g-1)\frac{p^{2}}{k^{2}}}{a_{\g}k^{2}}+O(p^{3})\,,
\ee
so that the mass-term correction in $D=4$ dimensions is
\be
\int\!\frac{\rmd^{4}k}{(2\pi)^4}\left[-8\la_{\g-1}\frac{2}{a_{\g}k^{2}}\right]=-\frac{2\pi^2}{(2\pi)^4}16\frac{\la_{\g-1}}{a_{\g}}\!\int^{k_\textsc{uv}}\!\rmd k\,k=-\frac{1}{\pi^2}\frac{\la_{\g-1}}{a_{\g}}k_\textsc{uv}^{2},
\ee
which is quadratically divergent in the UV cut-off $k_\textsc{uv}$. Since this is a power-law divergence, we can neglect it when working in dimensional regularization. This result is exact in a monomial theory with no IR operators and $a_{\g},\lambda_{\g-1}\neq0$, while in the full theory \Eq{eq:model} there are sub-leading terms.

Let us turn our attention to the wave-function renormalization correction in four dimensions:
\ba
\int\!\frac{\rmd^{4}k}{(2\pi)^4}\left[-8\frac{\la_{\g-1}}{a_{\g}}\g(\g-1)\frac{p^2}{k^4}\right]&=&-\frac{16\pi^2}{(2\pi)^4}\frac{\la_{\g-1}}{a_{\g}}\g(\g-1)\!\int^{k_\textsc{uv}}\!\rmd k\,\frac{p^{2}}{k^{4}}k^{3}\nn
&=&-\frac{1}{\pi^2}\frac{\la_{\g-1}}{a_{\g}}\g(\g-1)p^{2}\!\int^{k_\textsc{uv}}\!\frac{\rmd k}{k}\nn
&=&-\frac{1}{2\pi^2}\frac{\la_{\g-1}}{a_{\g}}\g(\g-1)p^{2}\ln k_\textsc{uv}^2\nn
&=&\beta_1\,p^{2}\ln k_\textsc{uv}^2\,,
\ea
where
\be\label{beta1}
\b_1=-\frac{1}{2\pi^2}\frac{\la_{\g-1}}{a_{\g}}\g(\g-1)\,,
\ee
which is the first of the beta functions announced in section \ref{betas}.

\subsubsection{Self-interaction}

Consider now the corrections to the quartic self-interaction, i.e., to the four-point functions, when the external momenta can all be set zero for simplicity because the final result is independent of them. As discussed in section \ref{powcou}, only the $\phi^{4}$ counter-term is possible here. Therefore, we take the usual three bon-bon diagrams in the $s$, $t$ and $u$ channels:
\be
\parbox{4.1cm}{\includegraphics[width=4cm]{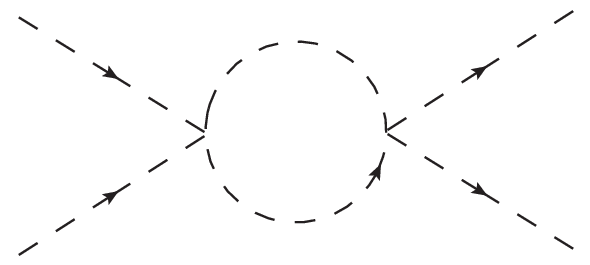}} + \parbox{4.1cm}{\includegraphics[height=3.2cm]{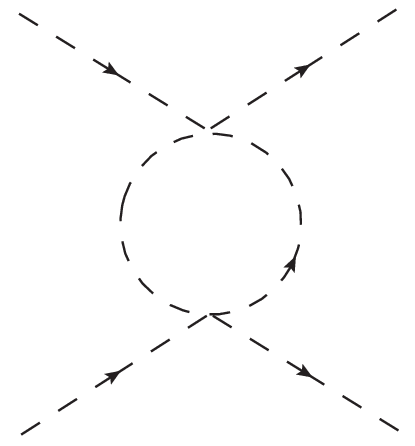}} + \parbox{4.1cm}{\includegraphics[height=3.2cm]{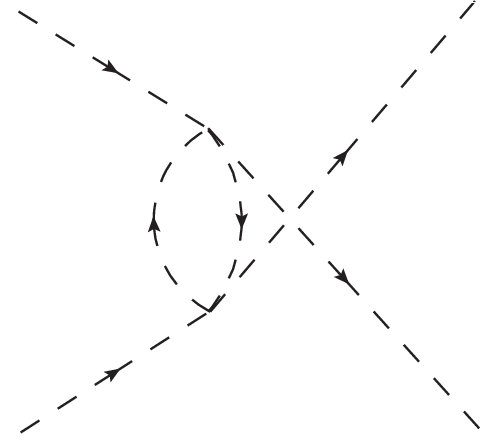}}.
\ee
Only the highest monomial term in the action matters, so that, again, we can take the asymptotic form \Eq{suv}.

The propagator part of the bon-bon diagrams for zero momenta is 
\ben
\frac{\rmi}{a_{\g}\left(-k^{2}\right)^{\g}}\frac{\rmi}{a_{\g}\left(-k^{2}\right)^{\g}}=-\frac{1}{a_{\g}^{2}\left(-k^{2}\right)^{2\g}}\,.
\een
To the vertex \Eq{vertgen}, we assign momenta
\be
p=0\,,\qquad r=-k\,,\qquad s=0\,,
\ee
so that
\be\label{bonbon}
\tilde{\rm V} =-16\rmi\la_{\g-1}(-k^2)^{\g-1}\,,
\ee
and we have three diagrams with the same contribution:
\ba
-\frac{3}{a_{\g}^{2}(-k^2)^{2\g}}\left[-16\rmi\la_{\g-1}(-k^{2})^{\g-1}\right]^2&=&\frac{3\times256\la_{\g-1}^{2}}{a_{\g}^{2}}\frac{(-k^2)^{2\g-2}}{(-k^2)^{2\g}}\nn
&=&\frac{768\la_{\g-1}^2}{a_{\g}^2}\frac{1}{(-k^2)^2}=\frac{768\la_{\g-1}^{2}}{a_{\g}^{2}}\frac{1}{k^4}\,.
\ea
Integrated over $\rmd^4k= k^{3}\rmd k\,\rmd\Om$ ($\rmd\Om$ being the solid angle part of the measure), this term gives the logarithmic divergence
\ben
\frac{384\la_{\g-1}^{2}}{a_{\g}^{2}}\ln k_\textsc{uv}^2\,,
\een
so that, multiplying by $2\pi^2/(2\pi)^4=1/(8\pi^2)$ the beta function of the $\la_{0}$ coupling is
\be\label{beta2}
\beta_{\la_{0}}=\frac{1}{8\pi^2}\frac{384\la_{\g-1}^{2}}{a_{\g}^{2}}=\frac{48}{\pi^2}\frac{\la_{\g-1}^{2}}{a_{\g}^{2}}\,.
\ee
This expression does not receive any corrections from sub-leading terms.


\section{Finiteness}\label{killers}

We saw that the theory described by the UV action \Eq{suv} can be strictly renormalizable or super-renormalizable depending on the value of $\g$. We also saw that, in $D=4$ spacetime dimensions, to achieve strict or super-renormalizability we need to include only two types of local counter-terms, namely, $\phi\B\phi$ and $\phi^4$. At this point, we ask ourselves whether a situation where all these divergences are completely absent at any perturbative order is possible.


Barring the simplistic case $\la_{\g-1}=0$ of no interactions, there is a way to kill both beta functions and make them vanish at any loop order, $\beta_{1}=0=\beta_{\la_0}$, starting from the case of one loop. The main ingredient is a new class of operators specifically added to the bare action with adjusted front coefficients such that linear relations ensure the complete cancellation of UV divergences. These operators have been dubbed killers and have been employed successfully in asymptotically polynomial non-local quantum gravity \cite{MoRa1}. A difference with respect to the present case is that the killers of asymptotically polynomial non-local quantum gravity are local operators in the UV, while ours will be non-local. The reason is that asymptotically polynomial non-local quantum gravity is local in the UV limit, while fractional gravity is not.

Let us recall what killers are and how the mechanism of divergences removal works. In a generic gravitational theory in four dimensions, operators which are cubic and quartic in curvature tensors contribute to the beta functions at the one-loop level. Using this observation, one can exploit these operators and arrive at a situation where all the one-loop beta functions are made to vanish. For this procedure to work, one needs to add these $O(\cR^3)$ and $O(\cR^4)$ operators to the bare Lagrangian and assign some coefficients $s_i$ to them. The coefficients $s_i$ are not free parameters of the theory and should be adjusted \emph{a posteriori}. 

From the power-counting analysis of $D=4$ gravitational theories, it is well known that there are quartic and quadratic power-law divergences in the UV cut-off. They are the leading terms for divergences in the UV regime but, nevertheless, all power-law divergences can be fully reabsorbed by a finite renormalization of couplings. One is left only with logarithmic divergences, for which $\om_\G=0$. These logarithmic divergences are universal, scheme-independent and give rise indirectly to physically relevant effects such as the renormalization-group flow of couplings. In other words, killer operators are not used to kill power-law divergences; they are introduced only to eradicate logarithmic divergences. Furthermore, it is obvious from the BV trace technology for covariant operators that, for example, quartic operators contribute linearly in their coefficients $s_i$ to the one-loop beta functions of counter-terms quadratic in curvatures. One can use this linearity property and adjust the killer coefficients to some special values $s_i^*$, which make the system of one-loop beta functions vanish. The clearest situation to use these killers occurs at one-loop level in super-renormalizable theories where only one-loop divergences survive \cite{MoRa1}. Then, after adding killers with specially adjusted coefficients $s_i^*$, the theory is completely UV-finite. If the theory possesses divergences also at higher loop orders, then the killers used at the one-loop level tame one-loop divergences completely, but at higher loops one needs to use more complicated operators, as we will discuss in section \ref{killmany}.

Schematically, before adding these killers, the complete expressions for the beta functions in the model \Eq{suv} are
\ba
\beta_{1}&=&\beta_{1}^{(1)}+\beta_{1}^{(2)}+\beta_{1}^{(3)}+\ldots\,,\\
\beta_{\la_{0}}&=&\beta_{\la_{0}}^{(1)}+\beta_{\la_{0}}^{(2)}+\beta_{\la_{0}}^{(3)}+\ldots,
\ea
where superscripts denote the number of loop levels at which each term is computed. At the one-loop level, the beta functions $\beta_{1}=\beta_{1}^{(1)}\neq0$ and $\beta_{\la_{0}}=\beta_{\la_{0}}^{(1)}\neq0$ can be effectively set to zero when we add the following piece of killer (``K'') Lagrangian:
\be\label{killLuv}
\cL_{\rm K}^\textsc{uv}=\cL_{{\rm K}1}^\textsc{uv}+\cL_{{\rm K}2}^\textsc{uv}=s_{1}\phi^2(\B\phi)\B^{\g-2}\phi+s_{\la_{0}}\phi^{4}\B^{\g-2}\phi^{2},
\ee
where $s_1$ and $s_{\la_0}$ are constants. These terms are at least quartic in the scalar field $\phi$, hence they do not contribute to the propagator or to the zero-vacuum expectation value. This implies that the propagator is the same and that killer operators do not change anything about the question of tree-level unitarity of the theory (sections \ref{sec3} and \ref{sec4}). 

The operators \Eq{killLuv} are strongly non-local for $\g<2$, since the power exponent on the d'Alembertian is negative. However, this is not a problem because of two reasons. First, the form of these killers should be attained only in the UV regime of the theory, when we concentrate only on the UV-divergent part of the effective action. Moreover, at lower energies (characterized by the eigenvalues of $\B$) the form of these operators may be different in such a way that they interpolate between the UV and the IR regime and the real issue is whether they create any problem in the IR end, as is typically known to happen with strong non-locality. This issue can be solved by a choice such as the following example of interpolating rational function of the box operator:
\be\label{killL}
\cL_{\rm K}=s_{1}\phi^2(\B\phi)\frac{\B^{\g}}{\B^2+\Lambda_\textsc{ir}^4}\phi+s_{\la_{0}}\phi^{4}\frac{\B^{\g}}{\B^2+\Lambda_\textsc{ir}^4}\phi^{2},
\ee
where $\Lambda_\textsc{ir}$ is some IR energy scale. One has two limits, the UV regime \Eq{killLuv},
\be
\cL_{\rm K}\,\stackrel{|\B|\gg\Lambda_\textsc{ir}^{2}}{\longrightarrow}\,\cL_{\rm K}^\textsc{uv}
\ee
and the IR regime
\be
\cL_{\rm K}\,\stackrel{|\B|\ll\Lambda_\textsc{ir}^{2}}{\longrightarrow}\,s_{1}\phi^2(\B\phi)\frac{\B^{\g}}{\Lambda_\textsc{ir}^4}\phi+s_{\la_{0}}\phi^{4}\frac{\B^{\g}}{\Lambda_\textsc{ir}^{4}}\phi^{2}.
\ee
This signifies that, for $\g>0$, these terms are completely harmless in the IR since they are suppressed by the IR energy scale $\Lambda_\textsc{ir}$ as in any standard low-energy effective field theory, so that they do not create any issue related to strong non-locality. Other interpolating functions between the IR and the UV are possible but what matters for the issue of renormalizability is only the UV form \Eq{killLuv} of the killer Lagrangian.


\subsection{First killer at one loop}

At the one-loop level, one can perform the whole computation of the contributions of the killers to the beta functions. Here we only
quote some intermediate steps and the main results.

The first and second variation of the killer Lagrangian $\cL_{{\rm K}1}^\textsc{uv}$ for the first operator in \Eq{killLuv} are, respectively, 
\be
\delta\cL_{{\rm K}1}=s_1\left[2\phi\de\phi(\B\phi)\B^{\g-2}\phi+\phi^{2}(\B\de\phi)\B^{\g-2}\phi+\phi^{2}(\B\phi)\B^{\g-2}\de\phi\right],
\ee
and
\ba
\delta^{2}\cL_{{\rm K}1}&=&s_1\left[2\de\phi\de\phi(\B\phi)\B^{\g-2}\phi+2\phi\de\phi(\B\de\phi)\B^{\g-2}\phi+2\phi\de\phi(\B\phi)\B^{\g-2}\de\phi\right.\nn
&&\qquad+2\phi\de\phi(\B\de\phi)\B^{\g-2}\phi+\phi^{2}(\B\de\phi)\B^{\g-2}\de\phi+2\phi\de\phi(\B\phi)\B^{\g-2}\de\phi\nn
&&\qquad\left.+\phi^{2}(\B\de\phi)\B^{\g-2}\de\phi\right]\nn
&=&s_1\left[2(\B\phi)(\B^{\g-2}\phi)\de\phi\de\phi+4\phi(\B^{\g-2}\phi)\de\phi\B\de\phi+4\phi(\B\phi)\de\phi\B^{\g-2}\de\phi\right.\nn
&&\qquad\left.+2(\B\phi^{2})\de\phi\B^{\g-2}\de\phi+4(\N^{\mu}\phi^{2})\de\phi\N_{\mu}\B^{\g-2}\de\phi+2\phi^{2}\de\phi\B^{\g-1}\de\phi\right]\nn
&=&s_1\left[2(\B\phi)(\B^{\g-2}\phi)\de\phi\de\phi+4\phi(\B^{\g-2}\phi)\de\phi\B\de\phi+4\phi(\B\phi)\de\phi\B^{\g-2}\de\phi\right.\nn
&&\qquad\left.+2\phi^{2}\de\phi\B^{\g-1}\de\phi\right]+\dots\,,
\ea
where we neglected total derivatives in the background quantities. As usual, the special cases of $\g=1$ and $\g=2$ should be treated separately. When $\g\neq 1,2$, we have just two relevant terms contributing to, respectively, the $\phi\B\phi$ divergence (linear in $s_1$)
and the $\phi^{4}$ divergence (quadratic in $s_1$):
\be
\delta^{2}\cL_{{\rm K}1}= 2s_1\left[2(\phi\B\phi)\de\phi\B^{\g-2}\de\phi+\phi^{2}\de\phi\B^{\g-1}\de\phi\right]+\dots\,.
\ee
The contribution of these terms to the Hessian and to the corresponding reduced Hessian read, respectively,
\be
\hat\cH_{{\rm int}}\supset 2s_1\left[2(\phi\B\phi)\B^{\g-2}+\phi^{2}\B^{\g-1}\right]\,,\qquad \hat\cH_{{\rm red}}\supset\frac{2s_1}{a_\g}\left[2(\phi\B\phi)\B^{-2}+\phi^{2}\B^{-1}\right]\,,
\ee
so that the corresponding beta functions are
\ba
\b_{1,{\rm K}1}&=&-\frac{1}{2\pi^{2}}\frac{s_1}{a_{\g}}\,,\\
\b_{\la_{0},{\rm K}1}&=&\frac{3}{\pi^{2}}\left(\frac{s_1}{a_{\g}}\right)^2.\label{betaK12}
\ea
The first condition to make the theory finite at the one-loop level is 
\be\label{linequ}
0=\beta_{1,{\rm tot}}^{(1)}=\beta_{1}^{(1)}+\beta_{1,{\rm K}1}\,,
\ee
where $\beta_{1}^{(1)}$ contains all non-vanishing contributions to the beta function of the coupling in front of the wave-function renormalization term $\phi\B\phi$, except for the contribution of the first killer operator. In the case of the minimal UV theory \Eq{suv}, $\beta_{1}^{(1)}=-\g(\g-1)\la_{\g-1}/(2\pi^2a_\g)$ is given by \Eqq{beta1}. In general, the linear equation \Eq{linequ} can always be solved for a real and unique value of $s_1$ given by
\be
s_1=s_1^*=2\pi^2a_{\g}\b_{1}^{(1)}\,.
\ee


\subsection{Second killer at one loop}

The second-order variation of the killer Lagrangian $\cL_{{\rm K}2}^\textsc{uv}$ for the second killer in \Eq{killLuv} is
\be
\delta^{2}\cL_{{\rm K}2}=s_{\la_{0}}\!\left[12\phi^{2}\delta\phi\delta\phi\B^{\g-2}\phi^{2}+8\phi^{3}\delta\phi\B^{\g-2}(\phi\delta\phi)+8\phi^{3}\delta\phi\B^{\g-2}(\delta\phi\phi)+2\phi^{4}\B^{\g-2}(\delta\phi\delta\phi)\right]\!,
\ee
where we can effectively forget about the corresponding interpolating function in \Eqq{killLuv}, which only plays the role of a spectator. Remembering that we are interested only in terms $\propto\phi^4$ in background quantities, we select only the relevant terms:
\be\label{de2Lrel}
\delta^{2}\cL_{{\rm K}2}\supset 16s_{\la_{0}}\phi^{4}\delta\phi\B^{\g-2}\delta\phi\,,
\ee
with the exception of the special case $\g=2$ where some more terms must be added. The contribution of \Eq{de2Lrel} to the Hessian and the corresponding reduced Hessian read, respectively,
\be
\hat\cH_{{\rm int}}\supset 16s_{\la_{0}}\phi^{4}\B^{\g-2}\,,\qquad \hat\cH_{{\rm red}}\supset\frac{16s_{\la_{0}}}{a_{\g}}\phi^{4}\B^{-2}\,,
\ee
so that the functional trace is
\be
{\rm Tr}\ln\hat\cH_{{\rm K}2}=\frac{16s_{\la_{0}}}{a_{\g}}\phi^{4}+\dots\,.
\ee
From this, we obtain the beta function
\be
\beta_{\la_{0},{\rm K}}=\frac{24}{\pi^{2}}\frac{16s_{\la_{0}}}{a_{\g}}=\frac{384}{\pi^{2}}\frac{s_{\la_{0}}}{a_{\g}}\,,
\ee
which is linear in the coefficient $s_{\la_{0}}$. Including the other contributions to the beta function at the one-loop level (such as \Eq{betaK12}) into a collective term $\beta_{\la_{0}}^{(1)}$, the total one-loop expression is
\be
\beta_{\la_{0},{\rm tot}}^{(1)}=\beta_{\la_{0}}^{(1)}+\beta_{\la_{0},{\rm K}2}\,.
\ee
For example, from previous considerations we know that in the minimal model with bare action \Eq{suv} we have $\beta_{\la_{0}}^{(1)}=(48/\pi^2)(\la_{\g-1}/a_\g)^2$ according to \Eqq{beta2}. Then, one can easily solve the linear equation $\beta_{\la_{0},{\rm tot}}^{(1)}=0$  for a unique and real value of the coefficient $s_{\la_{0}}$, thus making the $\phi^4$ sector of the theory finite:
\be
s_{\la_{0}}=s_{\la_{0}}^*=-\frac{1}{8}\frac{\la_{\g-1}^{2}}{a_{\g}}\,,
\ee
where the values of $\la_{\g-1}$ and $a_{\g}$ are arbitrary.

This concludes the proof that, at the one-loop level and using fractional killer operators, it is always possible to find an example of the theory which is completely finite, irrespectively of the values of the exponent $\g$, provided that $\g>0$.


\subsection{Higher loops}\label{killmany}

If $\g>2$, then the minimal UV action \Eq{suv} is one-loop super-renormalizable and it is sufficient to augment it with the killer operators \Eq{killLuv} to make it finite. However, if $1<\g<2$, divergences appear also at higher loop orders and the above killers are insufficient to achieve finiteness. In this section, we tackle the issue of whether one can extend the above construction to this case.

The situation at higher loop levels is very similar and here we describe it qualitatively without entering into details. One technical reason is that, as is well known, one does not have a simple higher-loop trace logarithmic formula for the effective action using the functional trace of the second variational derivative operator (understood as the Hessian differential operator on the vector
space of fluctuations). Therefore, a background-covariant mathematical apparatus to perform such computation is unavailable. Feynman diagrams can bypass this problem but we will not attempt such tedious calculations here. We just give arguments why the same killing procedure
works also in this case.

An important ingredient is that at the $L$-loop level the effective action is linearly proportional to the $2L$th variational derivative
of the bare action with respect to the field fluctuations, with other fields set to their background values. Therefore, on one hand we
know that 
\be
\Gamma_{{\rm div}}^{(L)}\sim\frac{\delta^{2L}S}{\delta\phi^{2L}}\,.
\ee
On the other hand, from the power-counting analysis we also know that, at any loop order, we have divergences only of the form \Eq{eq:divergences}. Therefore, suitable killers in the UV at the general $L$ loop level should have the form\footnote{A simpler form of the first killer can be $\phi^{2}\B^{\g-L}\phi^{2L}$, which however works only for $L\geq 2$.}
\be\label{killLL}
\cL_{\rm K}^{\textsc{uv}}=\sum_{L\in\{L_{\rm div}\}}\cL_{\rm K}^{\textsc{uv}(L)},\qquad \cL_{\rm K}^{\textsc{uv}(L)}=s_{1}^{(L)}\phi^2(\B\phi)\B^{\g-L-1}\phi^{2L-1}+s_{\la_{0}}^{(L)}\phi^{4}\B^{\g-L-1}\phi^{2L}\,,
\ee
where $\{L_{\rm div}\}$ is the set of loop levels where divergent diagrams appear. For instance, taking the list of examples below \Eq{lmax}, when ${3}/{2}<\g\leq2$ we have $L_{\rm div}=1,2$ and a total of four killers; when ${4}/{3}<\g\leq{3}/{2}$, we have $L_{\rm div}=1,2,3$ and a total of six killers; and so on. 

One can prove the validity of the killing system since the equations
\ba
0&=&\beta_{1,{\rm tot}}^{(L)}=\beta_{1}^{(L)}+\beta_{1,{\rm K}}^{(L)}\,,\label{kill1}\\
0&=&\beta_{\la_{0},{\rm tot}}^{(L)}=\beta_{\la_{0}}^{(L)}+\beta_{\la_{0},{\rm K}}^{(L)}\,,\label{kill2}
\ea
are realized for any loop order $L$ when the values of the coefficients are adjusted according to $s_{1}^{(L)}=s_{1}^{(L)*}$ and $s_{\la_{0}}^{(L)}=s_{\la_{0}}^{(L)*}$. As one can see, these coefficients are real and always exist as solutions of the linear independent system of killing equations \Eq{kill1}--\Eq{kill2}. Explicit numerical relations between $s_{1}^{(L)}$ and $\beta_{1}^{(L)}$ or between $s_{\la_{0}}^{(L)}$ and $\beta_{\la_{0}}^{(L)}$ can be found but they go beyond the scope of the present paper. Suffice it to recognize that the killers \Eq{killLL} contribute in a linear way to the sought divergences after taking the $2L$th variational derivatives. One can avoid the strong non-locality problem in the IR in the same way as presented above with the interpolating form factors.

In conclusion, UV-finiteness in the considered models of scalar fractional theories can be easily and always achieved at any loop
order by adding certain types of killer operators.


\subsection{Killers in gauge theory}\label{gaukil}

Adding to the action \Eq{gauge} the killer operator
\be
\cL_{\rm K}^{\textsc{uv}}=F^{\mu\nu}F_{\mu\nu}F^{\rho\s}\B^{\g-3}F_{\rho\s}\,,
\ee
we get the term
\be
a_\parallel(\B a_\parallel)a_\parallel\B^{\g-2}a_\parallel\,,
\ee
equivalent to the first one-loop killer in \Eqq{killLuv}, where $a_\parallel$ was defined in \Eqqs{aparal1} and \Eq{aparal2}. Therefore, for $\g>2$, the one-loop super-renormalizable fractional theory of gauge fields can be made finite. For $1<\g<2$, there is a finite number of divergences at higher loops and one can devise a construction similar to the one in section \ref{killmany}.


\section{Tree-level propagator}\label{sec3}

In \cite{mf1}, it was shown that, with the fractional equivalent of the standard Feynman prescription for the propagator, the scalar theory $T[\B+\B^\g]$ is well-defined and unitary only for $\g<1$ at the free level (i.e., independently of the type of self-interactions), a range more precisely defined by $0<\g<1$ if we require a positive-definite spectral dimension. At the one-loop level, some $\phi^3$ diagrams were found to be finite in this interval for $\g\neq1/2$ but we have seen above that this result was peculiar to the $V(\phi)\propto\phi^N$ model and is not reproduced in models with derivative interactions $V(\B,\phi)$. 
 Given that $\g<1$ leads to a model where fractional modifications are important in the IR rather than the UV, the conclusion was that the simplest version of the model was non-renormalizable. A possible resolution involving a non-trivial integration over a continuum of $\g$ values was proposed without elaborating the resulting theory $T[\B^{\g(\ell)}]$ too much.

In this section, our goal is to individuate ways out of the problem detected in \cite{mf1} without attempting to modify the classical theory $T[\B+\B^\g]$. The solution should be apparent already at the free level, since any trouble (including non-unitarity) spotted at the free level would almost surely propagate to all orders in perturbation theory, as found in \cite{mf1}. Vice versa, a positive result at the free level would require further checks at higher perturbative orders but it would be a solid starting point for this task. Therefore, from now on we concentrate on the tree-level propagator and ignore interactions and loop corrections. This also allows us to mass export all the above results on renormalizability to the quantum theories described in the following, since the kinetic term in the UV will still be of the type $\sim \B^\g$ and the power counting will be unaffected. We allow also for a mass $m$, which can be set to zero at any point in the calculation.


\subsection{Källén--Lehmann representation}\label{sec3a}

Consider the Fourier transform $\tilde G(-k^2)$ of the Green's function $G$ for a Lorentz-invariant theory, where $-k^2=k_0^2-|\bm{k}|^2\in\mathbb{R}$. Here we consider the free Green's function but what follows also applies to the exact Green's function with interactions. Promote $\tilde G$ to a function $\tilde G(z)$ on the complex plane and assume that $\tilde G(z^*)=\tilde G^*(z)$. Given a counter-clockwise contour (closed path) $\tilde\G$ encircling the point $z=-k^2$ and such that $\tilde G(z)$ is analytic inside and on $\tilde\G$, Cauchy's integral formula yields the Green's function as an integral on the contour:
\be\label{opt}
\tilde G(-k^2)=\frac{1}{2\pi\rmi}\oint_{\tilde\G}\rmd z\,\frac{\tilde G(z)}{z+k^2}\,.
\ee
Assume now that $\tilde G$ has poles and branch cuts on the real axis $s\coloneqq \Re\,z$. (Branch cuts on the imaginary axis, or even anywhere in the complex plane, are also possible but they do not change much this generic discussion. They will be considered in sections \ref{sec4} and \ref{douspl}.) We can deform the contour $\tilde\G$ in such a way that, if the contributions of arcs at infinity and around branch points vanish, then, for a massive theory with one branch cut, we get the Källén--Lehmann representation of the Green's function \cite{Kal52,Leh54,tV74b,Sre07,Zwi16} (see appendix \ref{appKL}):
\be\label{kalegen2}
\tilde G(-k^2)=\int_{m^2}^{+\infty}\rmd s\,\frac{\rho(s)}{s+k^2}\,,
\ee
where $\rho$ is called spectral function or spectral density.


\subsection{Feynman prescription}\label{sec3b}

\begin{figure}
\bc
\includegraphics[width=12cm]{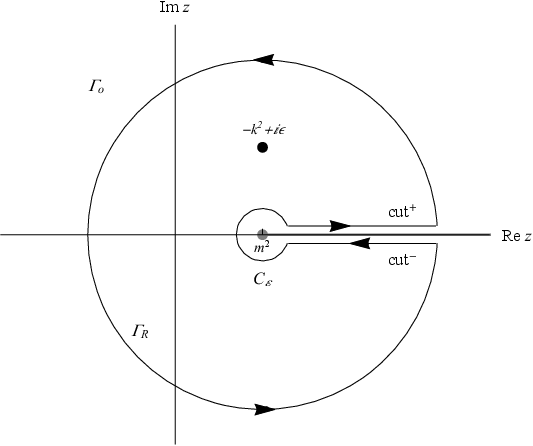}
\ec
\caption{\label{fig1} Contour $\Gamma_{\rm o}$ in the $(s=\Re\,z,\Im\,z)$ plane for a propagator $\tilde G_{\rm F}$ with a branch cut at $z\geq m^2$ (gray thick line) and the Feynman (causal) prescription $-k^2+\rmi\e$.}
\end{figure}

Let us review the problematic case of the Feynman propagator in fractional theories \cite{mf1}. Full details are given in appendix \ref{appFEY}, where we correct some typos and imprecisions in \cite{mf1}. Assume that $\tilde G$ has a branch cut at $s\coloneqq \Re\,z \geq m^2$ and no poles:
\be\label{Gz}
\tilde G(z) =\frac{1}{(m^2-z)^\g}\,,
\ee
where $\g$ is non-integer. In the Feynman prescription giving rise to what we would usually call the causal propagator, $-k^2$ is displaced in the upper half plane: $-k^2\to -k^2+\rmi\e$, where $\e>0$. The small parameter $\e$ is sent to zero at the very end of the calculation. Then, \Eqq{opt} becomes
\be\label{optF}
\tilde G_{\rm F}(-k^2)=\lim_{\e\to 0^+}\frac{1}{2\pi\rmi}\oint_{\G_{\rm o}}\rmd z\,\frac{\tilde G(z)}{z+k^2-\rmi\e}\,,
\ee
where the contour $\tilde\G=\G_{\rm o}$ is shown in Fig.~\ref{fig1}.

As shown in appendix \ref{appFEY}, the contribution $\int_{\G_R}$ of the arcs at infinity vanishes in the limit $R\to+\infty$ if
\be\label{g0}
\g>0\,.
\ee
The contribution of the branch point is proportional to the radius $\ve$ of the mini-arc around it,
\be
\tilde G_{\rm b.p.}(-k^2) \propto \ve^{1-\g}\,.
\ee
When $\g$ is non-integer, this expression diverges for $\g>1$ and vanishes for $\g<1$. Since there is no other term that can compensate this, we must make it vanish and enforce the condition
\be\label{g1}
\g<1\,.
\ee
The only non-zero contribution left is the one from the branch cut, which reads
\be
\tilde G_{\rm F}(-k^2) =\tilde G_{\rm cut}(-k^2)= \int_{m^2}^{+\infty}\rmd s\,\frac{\rho_{\rm F}(s)}{s+k^2-\rmi\e},\label{GFey}
\ee
where
\be
\rho_{\rm F}(s) = \frac{1}{|s-m^2|^\g}\frac{\sin(\pi\g)}{\pi}\,.\label{rhoFey}
\ee
Unitarity is preserved if the spectral density is positive definite, $\rho_{\rm F}(s)>0$, which is true for $2n<\g<2n+1$, $n\in\mathbb{Z}$. However, combined with the constraint \Eq{g1} from the contribution of the branch points and with the constraint \Eq{g0} from the positivity of the spectral dimension, we end up with the interval
\be\label{g01}
0<\g<1\,.
\ee
Note that $\rho_{\rm F}=\Im\,\tilde G/\pi$, so that the condition $\rho_{\rm F}(s)>0$ is equivalent to the optical theorem applied to the propagator. For completeness, we recall the higher-derivative case $\g=n$ in appendix \ref{appLOC}.


\subsection{Dyson prescription}

The same calculation can be applied to the Dyson prescription $-k^2\to -k^2-\rmi\e$ giving rise to what we would usually call the anti-causal propagator, where $\e>0$. In this case, $-k^2$ is displaced in the lower half plane and \Eqq{opt} becomes
\be\label{optD}
\tilde G_{\rm D}(-k^2)=\lim_{\e\to 0^+}\frac{1}{2\pi\rmi}\oint_{\G_{\rm o}}\rmd z\,\frac{\tilde G(z)}{z+k^2+\rmi\e}=\int_{m^2}^{+\infty}\rmd s\,\frac{\rho_{\rm F}(s)}{s+k^2+\rmi\e}\,,
\ee
where the contour $\tilde\G=\G_{\rm o}$ is exactly the same as before, except for the position of $z=-k^2$ (now in the fourth quadrant). The spectral function $\rho_{\rm F}(s)$ is \Eq{rhoFey} and one gets the same condition $\g<1$ to make the contribution of the branch point vanish.


\subsection{Anselmi--Piva prescription}\label{sec3c}

Unless we change the propagator \Eq{Gz}, the condition \Eq{g01} is basically unavoidable because, otherwise, the arc around the branch point will always give a divergent contribution for non-integer $\g$, regardless of the contour prescription adopted. However, before fixing this issue we move our attention to the Anselmi--Piva procedure \cite{Anselmi:2017yux,Anselmi:2017lia,Anselmi:2021hab,Anselmi:2022qor}, which has added value with respect to others because it can remove unwanted modes such as ghosts from the physical spectrum of higher-derivative theories. We will extend such procedure to the case of fractional QFTs.

The Anselmi--Piva procedure is non-analytic and consists in taking the average of the causal (Feynman) and anti-causal (Dyson) propagators:
\be\label{Gfake}
\tilde G_{\rm AP}(-k^2)\coloneqq \frac12\left[\tilde G_{\rm F}(-k^2)+\tilde G_{\rm D}(-k^2)\right].
\ee
Imposing this definition to the propagator of a particle field makes the latter a fakeon, i.e., a purely virtual particle which never appears in the physical spectrum of the theory. At the diagrammatic and S-matrix level, this means that the fakeon never appears in, respectively, external legs and asymptotic states. In fact, according to the Sokhotski--Plemelj formula, 
\ba
\frac{A}{k^2+m^2-\rmi\e}+\frac{B}{k^2+m^2+\rmi\e}&=&\frac{k^2+m^2}{(k^2+m^2)^2+\e^2}\,(A+B)+\rmi\,\frac{\e}{(k^2+m^2)^2+\e^2}\,(A-B)\nn
&\stackrel{\e\to 0^+}{=}& {\rm PV}\!\left(\frac{1}{k^2+m^2}\right)\,(A+B)+\rmi\pi\de(k^2+m^2)\,(A-B)\,,\nn\label{plemsok}
\ea
for any $A$ and $B$, where PV is the Cauchy principal value. This formula is valid in the sense of distributions, i.e., both sides should be applied to a test function and integrated in $k^2$. When $A=B$, \Eqq{plemsok} corresponds to \Eq{Gfake} for a massive particle with residue $A$, the latter being $A=1$ for a standard particle and $A=-1$ for a ghost. The delta contribution vanishes and one is left with the principal value of the propagator: the particle never goes on shell ($k^2\neq-m^2$). One can also view \cite{Anselmi:2018kgz} the fakeon propagator as the analytic limit of the propagator of a Lee--Wick particle \cite{Lee:1969fy,Lee:1970iw}, when the mass $M$ of the latter is sent to zero.

Thus, the prescription \Eq{Gfake} on the free propagator, which affects the physics off-shell, is accompanied by a projection taking care of the physics on-shell. Only the combination of the prescription together with the projection is consistent with unitarity. By itself, the replacement of Feynman propagators by averaged Feynman--Dyson ones is insufficient unless one makes sure not to add any on-shell contribution. It is crucial that the prescription be accompanied by a mathematical projection similar to the one done for Faddeev--Popov ghosts: unitarity holds in a reduced space of Lehmann--Symanzik--Zimmermann amplitudes, those with no external ghosts. In practice, one must specify that all the amplitudes with external unwanted particles be thrown away, just like in the case of the Faddeev--Popov ghosts. This is the reason why these virtual unwanted particles are called fakeons: they are ghosts (with negative signs of the kinetic terms) or even normal particles that can never go on shell; they do not appear in the physical spectrum as asymptotic states. In other words, a field theory with ghosts has a metric $(1,1,\dots,1,-1,-1,\dots,-1)$ between $T^\dagger$ and $T$ matrices in the optical theorem.\footnote{In a field theory with only normal particles, this metric is $(1,1,\dots,1,1,1,\dots,1)$.} The fakeon prescription changes diagrams so that to get the metric $(1,1,\dots,1,0,0,\dots,0)$. But then the states corresponding to the zeros in the above metric should not appear between external states, unless one wants to have external fakeons and  violation of unitarity due to ghosts. This second step of the procedure is the projection, where one ignores amplitudes with external fakeons and takes a sub-space with reduced metric $(1,1,\dots,1)$ \cite{Anselmi:2022qor}.

In higher-derivative theories, one can apply this procedure precisely to ghost modes (appendix \ref{appLOC}). In fractional theories, we can do the following. The field redefinition \Eq{Lneven2} is useful when we have a continuum of ghost quasi-particles that cannot be easily counted like the finite number of ghosts in higher-derivative models. Generalizing \Eqqs{Lneven1} and \Eq{Lneven2} to a non-integer $n\to \g$, we get non-local transformations of the fields and the Lagrangian
\ba
\cL &=& \frac12\,\phi\left(\B+\ell_*^{2\g-2}\B^\g\right)\phi\label{lagra}\\
&=&\frac12\,\phi_1\B\phi_1-\frac12\,\phi_2\left(\ell_*^{-2}+\ell_*^{2\g-4}\B^{\g-1}\right)\phi_2\,,\label{Lgeven1}\\
\phi_1 &=& \left(1+\ell_*^{2\g-2}\B^{\g-1}\right)\phi\,,\qquad \phi_2=-\ell_*^{\g}\B^{\frac{\g}{2}}\phi\,.\label{Lgeven2}
\ea
Note that this transformation is well defined only if $\g>1$. In momentum space, the causal propagators for $\phi_1$ and $\phi_2$ are
\be
\tilde G_{\rm F}^{\phi_1}(-k^2)=\frac{1}{k^2-\rmi\e}\,,\qquad\tilde G_{\rm F}^{\phi_2}(-k^2)=-\frac{1}{\ell_*^{-2}+\ell_*^{2\g-4}(k^2-\rmi\e)^{\g-1}}\,.
\ee
We choose to apply the Anselmi--Piva prescription \Eq{Gfake} only to the second propagator:
\be
\tilde G^{\phi_1}(-k^2)=\tilde G_{\rm F}^{\phi_1}(-k^2)\,,\qquad\tilde G^{\phi_2}(-k^2)=\tilde G_{\rm AP}^{\phi_2}(-k^2)=\frac12\left[\tilde G_{\rm F}^{\phi_2}(-k^2)+\tilde G_{\rm D}^{\phi_2}(-k^2)\right]\!.
\ee
However, we cannot apply this prescription directly to our model because, modulo a shift in $\g$, the form of the propagator of $\phi_2$ is $\propto 1/(m^{2\g}+k^{2\g})$ with $m\neq 0$, while in the UV we have a propagator $\propto 1/(m^2+k^2)^\g$ or, in the massless case, $\propto 1/k^{2\g}$. To overcome this difficulty, there are two slightly inequivalent ways. 

The first method is to introduce an auxiliary scalar field and recast \Eq{lagra} as
\be\label{lagra2}
\cL=\frac12\phi\B\phi+\frac12\ell_*^{2\g-2}\left(\phi-\frac14 \chi\right)\B^\g\chi\,.
\ee
Varying the action with respect to $\chi$ and plugging the solution $\chi$ back into \Eq{lagra2}, one reproduces \Eq{lagra}.\footnote{There is some ambiguity in the relation between $\phi$ and $\chi$. The equation of motion $\de S/\de\chi=0$ is $\B^\g\chi=2\B^\g\phi$. This implies $\chi=2\phi+\la$, where the function $\la(x)$ is the general solution of $\B^\g\la=0$. Plugging $\chi=2\phi+\la$ into \Eq{lagra2}, one gets \Eq{lagra} up to a term proportional to $\B^\g\la$, which vanishes identically.} The Lagrangian is not diagonalized in this way but, upon quantizing the system, one writes the perturbative tree-level propagator of $\chi$ and singles out the latter as the fakeon to be projected out of the physical spectrum. This procedure is followed also in quadratic quantum gravity \cite{Anselmi:2018bra,Anselmi:2018tmf}.

As an alternative, we propose a modification of the Anselmi--Piva prescription which acts only on the UV part of the total propagator:
\ba
\tilde G_{\rm AP'}(-k^2)&\coloneqq& \lim_{\e\to 0^+}\frac12\left[\frac{1}{m^2+k^2-\rmi\e-a_\g (m^2+k^2-\rmi\e)^\g}\right.\nn
&&\qquad\qquad\left.+\frac{1}{m^2+k^2-\rmi\e-a_\g (m^2+k^2+\rmi\e)^\g}\right].\label{APfull}
\ea
We make a modification confined to the UV because only there ghost-like dangerous modes are risky for unitarity, if they were present hypothetically. In principle, this has nothing to do with a decomposition of, or field redefinitions from, the scalar $\phi$. In this way, in the IR there are non-trivial, stable degrees of freedom with a standard Feynman propagator, while in the UV there is no propagating physical degree of freedom. In particular, all instabilities are removed regardless of whether all UV modes are ghost or not. Note that both the original prescription singling out ghost modes and the modified prescription \Eq{APfull} are applied to \Eqq{opt} after separating the relevant term $\tilde G(z)$ from the total propagator. That is to say, $\tilde G(z)$ in \Eqq{opt} is the propagator of the unstable mode in the case of the original Anselmi--Piva prescription, and the UV propagator if we choose the modified prescription. In the latter case, we have
\ba
\tilde G_{\rm AP'}(-k^2)&=&\lim_{\e\to 0^+}\frac{1}{4\pi\rmi}\oint_{\tilde\G}\rmd z\,\tilde G_\textsc{uv}(z)\left(\frac{1}{z+k^2-\rmi\e}+\frac{1}{z+k^2+\rmi\e}\right)\nn
&=&\frac{1}{2\pi\rmi}\oint_{\tilde\G}\rmd z\,\tilde G_\textsc{uv}(z)\,{\rm PV}\!\left(\frac{1}{z+k^2}\right).\label{optAP}
\ea

To better understand the structure of $\tilde G_{\rm AP'}$ in the UV, we note that the contour leading to this propagator looks degenerate in the two-dimensional complex plane, i.e., on a single branch of the complex integrand. However, one can consider taking the union $\tilde\G=\G_{\rm AP}\coloneqq \G_{\rm F}^+\cup \G_{\rm D}^-$, where $\G_{\rm F}^+$ is the Feynman contour in Fig.~\ref{fig1} in the principal branch of $\tilde G(z)/2$ (with $0\leq\t\leq 2\pi$) and $\G_{\rm D}^-$ is the Dyson contour in a different branch. Eventually, the branch we choose does not matter and it can even be the main one for both terms in \Eq{Gfake}. The Källén--Lehmann representation of \Eq{optAP} is obtained from \Eqqs{GFey}, \Eq{optD} and \Eq{plemsok} with $A=B=\rho_{\rm F}(s)$:
\ba
\tilde G_{\rm AP'}(-k^2) &\stackrel{\textsc{uv}}{\simeq}& \frac12\int_{m^2}^{+\infty}\rmd s\,\rho_{\rm F}(s)\left[\frac{1}{s+k^2-\rmi\e}+\frac{1}{s+k^2+\rmi\e}\right]\nn
&\stackrel{\text{\tiny\Eq{plemsok}}}{=}& \int_{m^2}^{+\infty}\rmd s\,\rho_{\rm F}(s)\,{\rm PV}\!\left(\frac{1}{s+k^2}\right),
\ea
where $\rho_{\rm F}(s)$ is given by \Eqq{rhoFey}. Since only the principal value remains in the propagator, the gas of quasi-particles associated with $\B^\g$ can never go on-shell and never appear as asymptotic states in scattering amplitudes. Therefore, unitarity is preserved regardless of the sign of $\rho_{\rm F}(s)$. 

Unfortunately, this does not ameliorate the compatibility between unitarity and renormalizability, since the condition \Eq{g1} must still hold for the theory to be consistent. In fact, for the term $1/(m^2+k^2-\rmi\e)^\g$ the contribution of the branch point is \Eq{ceps3}, while for the term $1/(m^2+k^2+\rmi\e)^\g$ it is the same but possibly with a different overall phase.
 The sum of the two contributions from the branch point is then of the form \Eq{plemsok}, which vanishes only when $A=B=0$. Therefore, once again, the total contribution of the contours around the branch point scales as $\ve^{1-\g}$ and, for non-integer $\g$, it vanishes if, and only if, inequality \Eq{g1} holds. Failing to do so leads to a divergent coefficient in front of the principal value.

To solve this problem in the case of scalar and gauge theories, we will give up the Anselmi--Piva procedure but will redefine the kinetic operator with a mass regulator (section \ref{sec4}), while in the case of gravity we will employ a two-step procedure where the Anselmi--Piva procedure will be applied after regularizing the kinetic term in a certain way (section \ref{douspl}).


\section{Fractional QFT with splitting}\label{sec4}

In this section, we propose a different representation of the kinetic operator $(-\B)^\g$ which will allow us to overcome the problem of the bound \Eq{g1} and obtain theories which are both unitary and (super-)renormalizable. We consider only the UV (fractional) part of the kinetic term, since in the IR unitarity is obviously achieved. In the massive case, we define the kinetic operator
\bs\label{m2B}\ba
\hspace{-1cm}\cK(\B) &=&-(m^2-\B)^\g\coloneqq \lim_{\vp_i\to 0}\cK_N(\B)\,,\qquad m^2> 0\,,\\
\hspace{-1cm}\cK_N(\B)&=&-\frac{1}{c}\prod_{j=1}^N\left(m^2-\rme^{-\rmi\vp_j}\B\right)^{\g_j},\qquad \sum_{j=1}^N\g_j=\g\,,\qquad 0\leq\vp_j< 2\pi\,,
\ea\es
where $c$ is a numerical constant fixed so that the eigenvalues of the kinetic term in Euclidean signature ($\B\to -k_{\rm E}^2=k_1^2+\dots+k_{D}^2$) are negative definite and a path-integral quantization is made possible:\footnote{The sign of the phase in $c$ depends on how we define the propagator (as $(\dots)^{-\g}$ or as $1/(\dots)^\g$) but this is just a matter of convention and, if followed consistently, does not affect the final result, which will be that $c=1$.}
\be\label{cphase}
c=\rme^{-\rmi\sum_j\g_j\vp_j}.
\ee
The representation \Eq{m2B} splits the fractional exponent into several components, each corresponding to a branch cut. The limits $\vp_i\to 0$ and $m\to 0$ give a split representation of the massless fractional d'Alembertian:
\be
(-\B)^\g\coloneqq -\lim_{\substack{\vp_i\to 0\\m^2\to 0}}\cK_N(\B)=\lim_{m^2\to 0}(m^2-\B)^\g\,,
\ee
where all the branch cuts are rotated back into one common position on the positive real axis at the end of the calculation, in a way that avoids the point $z=-k^2+\rmi\e$ (in Feynman prescription). More generally, we can send to zero only a subset of the masses, if we work in a massive theory. 


\subsection{Branch cuts at any angle}\label{sec4a}

Call $\tilde G_N$ the Green's function in momentum space associated with \Eq{m2B}:
\be\label{GN}
\tilde G_N(z) = c\prod_{j=1}^N(m^2-\rme^{-\rmi\vp_j}z)^{-\g_j}\,,\qquad \sum_{j=1}^N\g_j=\g\,.
\ee
Each factor in $\tilde G_N$ corresponds to a radial branch cut at an angle $\vp_j$ starting at the branch point $z=z_j\coloneqq m^2\exp(\rmi\vp_j)$. We will avoid configurations with non-radial and intersecting cuts. The calculation of the propagator \Eq{GN} is presented in appendix \ref{appGN}:
\be
\boxd{\tilde G_N(-k^2) = \sum_{i=1}^N\int_{m^2}^{+\infty}\rmd s\,\frac{\rho_i(s)}{s+\rme^{-\rmi\vp_i}k^2}=\int_{m^2}^{+\infty}\rmd s\,\cG_N(s,k^2)\,,}\label{optNfinal}
\ee
where, to have unitarity at the free level, we must have a real and positive semi-definite total spectral function:
\be\label{rhotot2}
\boxd{\cG_N(s,k^2)\coloneqq \frac{\rho_1(s)}{s+k^2}+2\sum_{(i,i^*)} \frac{|\rho_i|\left[s\cos\psi_i+k^2\cos(\psi_i-\vp_i)\right]}{s^2+k^4+2sk^2\cos\vp_i}\geq 0\,,}
\ee
where the sum is over pairs of complex conjugate branch cuts $i$ and $i^*$, i.e., such that $\vp_{i^*}=-\vp_i$ and $\g_{i^*}=\g_i$. The individual spectral functions are
\be
\rho_i(s) = \frac{\rme^{-\rmi\psi_i}}{|s-m^2|^{\g_i}\prod_{j\neq i}\big|s^2+m^4-2m^2s\cos\De_{ij}\big|^{\frac{\g_j}{2}}}\frac{\sin(\pi\g_i)}{\pi}\,,\label{master}
\ee
where the phases are
\ba
\psi_1 &=& 0\,,\qquad \text{if $\vp_1=0$ present}\,,\\
\psi_{i\neq 1} &=& \g_1\,{\rm Arg}_{1i}+\g_i\,{\rm Arg}_{i^*i}+\sum_{(j,j^*)\neq (i,i^*)}\g_j\cA_{ji}\,.\label{psi2}
\ea
The first term in the right-hand side of \Eq{psi2} is present only if the $\vp_1=0$ cut is present, in which case it is given by \Eqq{arg1i}. The second term is given by \Eqq{argeq2} with $j=i^*$, while the third term is the sum over pairs of
\be\label{Aji}
\cA_{ji}\coloneqq {\rm Arg}_{ji}+{\rm Arg}_{j^*i}
={\rm Arg}\left[m^2-s\,\rme^{\rmi(\vp_i-\vp_j)}\right]+{\rm Arg}\left[m^2-s\,\rme^{\rmi(\vp_i+\vp_j)}\right]\,.
\ee
Note that the total spectral function \Eq{rhotot2} is indeed real and semi-positive, since $\psi_{i^*} = -\psi_i$ and $|\rho_i|>0$, as shown in appendix \ref{appGN}.

The derivation of \Eq{optNfinal} also implies conditions on the fractional exponent $\g$. The vanishing of contribution of the arcs at infinity yields $\g>0$, while the vanishing of the branch-point contributions is guaranteed if, and only if,
\be\label{gN}
\g_i<1\qquad\Longrightarrow\qquad \g=\sum_{i=1}^N\g_i<N\,,
\ee
a remarkable improvement with respect to the bound \Eq{g1}. For instance, it is sufficient to take $1/N<\g_i<2/N$ for all $i$ to get a higher-loop super-renormalizable theory ($1<\g<2$), and $2/N<\g_i<1$ for $N\geq 3$ to get a one-loop super-renormalizable theory ($\g>2$). The condition \Eq{gN} is insufficient to guarantee also unitarity, which should be checked looking at the spectral function in the Källén--Lehmann representation of the propagator. 


\subsection{Unitarity}\label{secun}

All the viable ways to split the original branch cut are catalogued in appendix \ref{appC2}. It turns out that, imposing the theory to be unitary, the best we can achieve in terms of renormalizability is with the $N=2$ splitting and we get at most higher-loop super-renormalizability. This is not a limitation, since, on one hand, renormalizability as the minimal goal is achieved and, on the other hand, we can add the killer operators discussed in section \ref{killmany} to render the theory finite, without altering the present discussion.

Let us then study the $N=2$ case in detail. For any even $N$, the phase for the $\vp=0$ cut does not change because it is self-conjugate, so that its contribution to the spectral function is always positive definite. If this cut is present, then for even $N$ the theory is unviable because we cannot leave a cut at a non-zero angle unpaired with its complex conjugate. The cut at $\vp=\pi$ is also excluded, as argued above. Therefore, when $N=2$ we must exclude the $\vp=0$ cut and consider a pair of complex conjugate cuts. The first term in \Eqq{posicon1} is absent and so is the third. We have 
\be\label{psiN2}
\psi_i=\g_i\,{\rm Arg}_{i^*i} =\g_i\,{\rm Arg}\left(m^2-s\,\rme^{2\rmi\vp_i}\right),\qquad i=1,2\,.
\ee
Since $\vp_i\neq 0,\pi$, the only possibility to solve the stricter positivity condition \Eq{rhoiplusa} is to choose the two conjugate and opposite cuts on the imaginary line (Fig.~\ref{fig5}):
\be\label{pi2}
\vp_1=\frac{\pi}{2}\,,\qquad \vp_2=\frac{3\pi}{2}\,,
\ee
In fact, the condition $\psi_i=2\pi l$ must hold for all $s\geq m^2$ and, in particular, in the case when $s\to+\infty$ and
\be\label{psiN2b}
\psi_i=\g_i[2\vp_i-(2n+1)\pi]\,,\qquad n=0,1\,,
\ee
where the phase shift depends on the value of $\vp_i$. The right-hand side of \Eqq{psiN2b} vanishes for the values \Eq{pi2}.

The split kinetic term \Eq{m2B} corresponding to this configuration is
\be\label{m2BN2}
\cK_2(\B)= -\left(m^4+\B^2\right)^{\frac{\g}{2}},
\ee
with propagator \Eq{optNfinal} and spectral function \Eq{master} given by
\ba
\tilde G_2(-k^2) &=& \frac{1}{\left(k^4+m^4\right)^{\frac{\g}{2}}}=\int_{m^2}^{+\infty}\rmd s\,\frac{2s}{s^2+k^4}\,\rho_2(s)\,,\label{GN2}\\
\rho_2(s) &=& \frac{1}{|s^2-m^4|^{\frac{\g}{2}}}\frac{\sin\left(\frac{\pi\g}{2}\right)}{\pi}\,.\label{rhoN2}
\ea
Note that \Eqq{GN2} is still in the form of the usual Källén--Lehmann representation because there is a global spectral function $\rho_i(s)$ times an $(s,k)$-dependent factor. It is interesting to highlight that, in his book, Weinberg claimed that higher-derivative theories cannot be unitary because the Källén--Lehmann representation \Eq{kalegen2} implies that the propagator cannot vanish faster than $k^{-2}$ in the UV \cite[section 10.7]{Wei95}. We formalize Weinberg's argument in a no-go theorem presented in appendix \ref{appD}. Here we see that the hypotheses of the theorem do not hold for \Eqq{GN2} and its generalizations, since these multi-cut configurations, not envisaged before, are different from the single-cut expression \Eq{kalegen2} and do not suffer from the same limitations on the asymptotic behaviour.

\begin{figure}
\bc
\includegraphics[width=7cm]{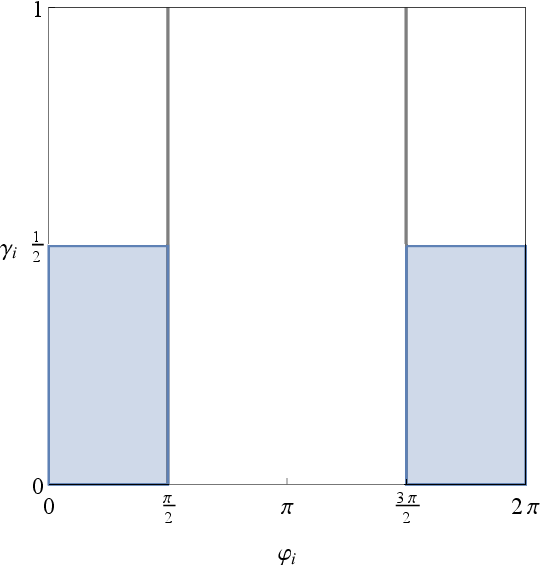}
\ec
\caption{\label{fig6} Angle $\vp_i$ and exponent $\g_i$ satisfying the positivity conditions \Eq{rhoiplus} with $\psi_i$ given by \Eqq{psiN2}, for $N=2$. Shaded areas and the two vertical lines are the regions in the parameter space where free-level unitarity is guaranteed.}
\end{figure}

If we abandon the stricter positivity condition \Eq{rhoiplusa} and limit ourselves to the inequalities \Eq{rhoiplus}, we have to ensure that they hold when $\psi_i$ is given by \Eqq{psiN2} for all $s\geq m^2$. This selects a certain region in the $(\vp_i,\g_i)$ plane (Fig.~\ref{fig6}):
\bs\ba
\hspace{-1.8cm}&&\text{non-renormalizable:\hspace{1.2cm} $0<\g_i< \frac12$}\,,\qquad \left\{0<\vp_i\leq \frac{\pi}{2}\right\}\cup \left\{\frac{3\pi}{2}\leq\vp_i<2\pi\right\},\label{N2cond1}\\
\hspace{-1.8cm}&&\text{strictly renormalizable:\hspace{1.3cm} $\g_i=\frac12$}\,,\qquad \left\{0<\vp_i\leq \frac{\pi}{2}\right\}\cup \left\{\frac{3\pi}{2}\leq\vp_i<2\pi\right\},\label{N2cond1b}\\
\hspace{-1.8cm}&&\text{super-renormalizable:\hspace{.8cm} $\frac12<\g_i<1$}\,,\qquad \vp_i=\frac{\pi}{2},\,\frac{3\pi}{2}\,.\label{N2cond2}
\ea\es
Thus, the special case \Eq{pi2} is the only one valid for any value of $\g_i$ in the interval $(0,1)$. Since $0<\g_i=\g/2<1$, the theory is non-renormalizable for $0<\g_i< 1/2$, strictly renormalizable for $\g_i=1/2$ and super-renormalizable for $1/2<\g_i<1$ (Fig.~\ref{fig5}).

One can check from \Eqqs{optNfinal}, \Eq{rhotot2} and \Eq{psiN2} that the propagator is indeed (here $\vp=\vp_i$)
\ba
\tilde{G}_2(-k^2)&=&\frac{1}{\left(m^4+k^4+2m^2k^2\cos\vp\right)^{\frac{\g}{2}}}\nn
&=&\frac{2}{\pi}\int_{m^{2}}^{+\infty}\!\rmd s\,\frac{s\cos\psi(s,\vp)+k^2\cos[\psi(s,\vp)-\vp]}{s^2+k^4+2sk^2\cos\vp}\nn
&&\qquad\qquad\times\frac{\sin\left(\frac{\pi\g}{2}\right)}{\left(s-m^2\right)^{\frac{\g}{2}}\left[s^2+m^4-2m^2s\cos(2\vp)\right]^{\frac{\g}{4}}}.
\ea


\section{Fractional gravity}\label{secgrav}

In the case of fractional gravity, the action is \cite{mf2} (we ignore the cosmological constant term)
\be\label{gravac}
S=\frac{1}{2\k^2}\int\rmd^4x\,\sqrt{|g|}\,\left[R+\ell_*^2 G_{\mu\nu}(-\ell_*^2\B)^{\g-2}\,R^{\mu\nu}\right],
\ee
where $\k^2=8\pi G$, $G$ is Newton's constant, $g$ is the determinant of the metric, $R$ is the Ricci scalar, $G_{\mu\nu}=R_{\mu\nu}-g_{\mu\nu}R/2$ is the Einstein tensor, $R_{\mu\nu}$ is the Ricci tensor and $\ell_*$ is a fundamental length scale. First, we will find the action for the graviton from \Eqq{gravac}. Then, we will discuss power-counting renormalizability and unitarity.


\subsection{Graviton action}

Expanding the action \Eq{gravac} around the flat spacetime background,
\be
g_{\mu\nu}=\eta_{\mu\nu}+h_{\mu\nu}\,,
\ee
at second order in the graviton we get the UV kinetic term
\be
S^{(2)}_\textsc{uv}=\int \rmd^{4}x\,h_{\mu\nu}\cO_\textsc{uv}^{\mu\nu,\rho\s}h_{\rho\s}\,,
\ee
where the operatorial kernel $\cO^{\mu\nu,\rho\s}$ is of the general form
\ba
\cO_\textsc{uv}^{\mu\nu,\rho\s}&=&\eta^{\mu\nu}\eta^{\rho\s}\B^{\g}+a_{2}\left(\eta^{\mu\rho}\eta^{\nu\s}+\eta^{\mu\s}\eta^{\nu\rho}\right)\B^{\g}+a_{3}\left(\eta^{\rho\s}\N^{\mu}\N^{\nu}+\eta^{\mu\nu}\N^{\rho}\N^{\s}\right)\B^{\g-1}\nn
&&+a_{4}\left(\eta^{\nu\s}\N^{\mu}\N^{\rho}+\eta^{\nu\rho}\N^{\mu}\N^{\s}+\eta^{\mu\s}\N^{\nu}\N^{\rho}+\eta^{\mu\rho}\N^{\nu}\N^{\s}\right)\B^{\g-1}\nn
&&+a_{5}\N^{\mu}\N^{\nu}\N^{\rho}\N^{\s}\B^{\g-2}\,,
\ea
where $a_i$ are numerical coefficients and $\N_\mu=\p_\mu$. A rigorous way to derive the action of the scalar modes of the graviton would be to work in the transverse-traceless gauge, decompose the spatial components into two polarization scalar modes, $h_{ij}=e_{ij}^+ h_++e_{ij}^\times h_\times$, and write the action for the scalars $h_+$ and $h_\times$. Here we just sketch the full derivation taking the transverse gauge $\N^\mu h_{\mu\nu}=0$. Then, only the first term in $\cO_\textsc{uv}^{\mu\nu,\rho\s}$ contributes and one gets the UV kinetic term between scalar fluctuations $h=\eta^{\mu\nu}h_{\mu\nu}$ of the form
\be\label{Suvh}
S^{(2)}_\textsc{uv}=\int \rmd^{4}x\,h\B^{\g}h\,.
\ee
At fourth order, it is not difficult to see that, in the UV,
\be\label{h4}
S^{(4)}_\textsc{uv}=\int \rmd^{4}x\left(\alpha_{1}h\B^{\g}h+\alpha_{2}h^{2}\B^{\g}h+\alpha_{3}h^{3}\B^{\g}h+\alpha_{4}h^{2}\B^{\g}h^{2}+\ldots+\alpha_{5}h\N^{\mu}h\N_{\mu}h\B^{\g-1}h\right),
\ee
where $\a_i$ are numerical coefficients.\footnote{Although one can always make a field redefinition to formally make the kinetic term local and canonical for a scalar field (with the risk of obtaining a physically different theory if the form factor is non-entire, as discussed in section \ref{sec1a}), this is not true in the case of gauge theories and gravity, as one can appreciate from the nonlinear actions \Eq{gauge} and \Eq{gravac}. In particular, even if one can redefine the graviton $h$ in \Eq{Suvh}, going back to the full gravitational action one is unable to find anything resembling a covariant metric theory like \Eq{gravac}. Therefore, the trade-off for this field redefinition is a complete loss of a fundamental non-linear formulation of the theory.} Regardless of the distribution of the derivatives on the scalars, we see that the total number of derivatives in all terms is always fixed to $2\g$. The conclusion is that we can never get the term of the form $\phi^{2}\B^{\g-1}\phi^{2}$ and the model \Eq{eq:model} does \emph{not} reproduce the action for the graviton starting from the non-linear theory \Eq{gravac}. 


\subsection{Generalized power counting}\label{genpowco}

Given that the quartic graviton UV interaction is not of the form of the model \Eq{suv}, we have to generalize the power-counting analysis of section \ref{powcou} to a different interaction. Consider then the $D=4$ scalar model
\be\label{genmod}
\cL=\frac{1}{2}a_\g\phi\B^{\g}\phi-\la_{\g-\a}\phi^{\frac{n}{2}}\B^{\g-\a}\phi^{\frac{n}{2}},
\ee
where we assume $\g>1$, $\g\geq\a\geq 0$ and $n\geq 4$. In the previous model \Eq{suv}, $n=4$ and $\a=1$, while in the case of the graviton in \Eq{h4} $n=4$ and $\a=0$. The distribution of the powers of $\phi$ in the interaction is immaterial for the purpose of power counting. Since $[\phi]=1$ in four dimensions, the dimensionality of the couplings is
\be
[a_\g]=2-2\g\,,\qquad [\la_{\g-\a}]=4-2(\g-\a)-n=[a_\g]+2+2\a-n\,.
\ee
Updating expression \Eq{sudedi} to the more general case \Eq{genmod}, the superficial degree of divergence is
\be
\om_\G\leq 4L+2(\g-\a)N_V-2\g N_I\,.
\ee
Combining the topological relation \Eq{pcrel1} and the generalization of \Eqq{pcrel2}
\be\label{pcrel3}
n N_V=N_E+2N_I\,,
\ee
we get
\be\label{pcrel4}
N_I=\frac{n(L-1)+N_E}{n-2}\,.
\ee
Then,
\ba
\om_\G&\stackrel{\text{\tiny\Eq{pcrel1}}}{\leq}&4L+2(\g-\a)(N_I+1-L)-2\g N_I\nn
&=&4L-2\a N_I-2(\g-\a)(L-1)\nn
&\stackrel{\text{\tiny\Eq{pcrel4}}}{=}&4L-2\a \frac{n(L-1)+N_E}{n-2}-2(\g-\a)(L-1)\nn
&=& L\left[4-2(\g-\a)-2\a\frac{n}{n-2}\right]+2\g-2\a\frac{N_E-2}{n-2}\nn
&=& 2L\left(2-\frac{2\a}{n-2}-\g\right)+2\g-2\a\frac{N_E-2}{n-2}\label{omifin1}\\
&=& 4-\frac{2\a N_E}{n-2}+2(L-1)\left(2-\frac{2\a}{n-2}-\g\right).\label{omifin2}
\ea
which generalizes inequality \Eq{omfin}. Take, as before, $N_E=2$. 

We can make a rough distinction between non-renormalizable and renormalizable theories looking at the sign of the coefficient in front of $L$ in \Eqq{omifin1}. If the sign is positive, then the superficial degree of divergence grows with the loop order and the theory is likely non-renormalizable. Otherwise, for a negative sign $\om_\G$ decreases in $L$, from some loop level on there are no divergences at all and the theory is renormalizable (strictly or even super-renormalizable). The limiting case when the coefficient is zero will not be discussed here. Thus, we distinguish three cases.
\begin{itemize}
\item If $[\la_{\g-\a}]>[a_\g]$, then $\a>(n-2)/2$ and the coefficient in front of $L$ in \Eq{omifin1} is always negative according to our assumptions:
\ben
2-\frac{2\a}{n-2}-\g<1-\g<0\,.
\een
Therefore, the theory is renormalizable for any $\g>1$.
\item If $[\la_{\g-\a}]=[a_\g]$, then $\a=(n-2)/2$, as for the scalar model \Eq{suv}. Equation \Eq{omifin1} becomes $\om_\G\leq 2L(1-\g)+2\g$ and the theory is renormalizable for $\g>1$, in agreement with the analysis of section \ref{powcou}.
\item If $[\la_{\g-\a}]<[a_\g]$, then $\a<(n-2)/2$, as for the graviton case \Eq{h4}, and the theory may or may not be renormalizable depending on the value of $\g$. In fact, since $0\leq 2\a/(n-2)<1$, we have
\ben
1-\g<2-\frac{2\a}{n-2}-\g\leq 2-\g\,,
\een
so that the theory is renormalizable either when $\g> 2$ or when $1<\g<2$ and $\a>(2-\g)(n-2)/2$. In particular, fractional gravity is renormalizable only for $\g> 2$.
\end{itemize}

We can be more specific and study the presence of logarithmic divergences order by order in perturbation theory using \Eqq{omifin2}. When $L=1$, there may be divergences. When $L>1$, there are at most logarithmic divergences provided $\om_\G\leq 0$, i.e., when
\be\label{gengmin}
\g \geq \frac{2L}{L-1}\left(1-\frac{\a}{n-2}\right).
\ee
This is the generalization of \Eq{gmin}, which is recovered when $1-\a/(n-2)=1-1/2=1/2$. For $\a>n-2$, condition \Eq{gengmin} is always satisfied, since the right-hand side would be negative definite. The systems of interest \Eq{suv} and \Eq{h4} correspond to $\a<n-2$ ($0,1=\a<2$), in which case the range \Eq{lmax} becomes
\be\label{lmax2}
\frac{2(L_{\rm max}+1)}{L_{\rm max}}\left(1-\frac{\a}{n-2}\right)<\g\leq\frac{2L_{\rm max}}{L_{\rm max}-1}\left(1-\frac{\a}{n-2}\right).
\ee
If $\g$ falls within this interval, then the model has divergences for $L=1,\ldots,L_{\rm max}$. In particular, when $\a=0$ (fractional gravity):
\begin{itemize}
\item For $\g>4$, the theory has divergences only for $L=1$ and is one-loop super-renormalizable.
\item For $3<\g\leq 4$, the theory has divergences at $L=1,2$ loops.
\item For $\frac{8}{3}<\g\leq 3$, the theory has divergences at $L=1,2,3$ loops.
\item For $\frac{5}{2}<\g\leq\frac{8}{3}$, the theory has divergences at $L=1,2,3,4$ loops.
\end{itemize}
This result is independent on the interaction order $n$ and applies to all graviton interactions. Therefore, fractional gravity is
\bs\label{reno}\ba
&&\text{non-renormalizable:\hspace{2.85cm} $\g<2$}\,,\\
&&\text{strictly renormalizable:\hspace{2.3cm} $\g=2$}\,,\\
&&\text{super-renormalizable:\hspace{1.9cm} $2<\g\leq 4$}\,,\\
&&\text{one-loop super-renormalizable:\hspace{.3cm} $4<\g$}\,.
\ea\es


\subsection{Finiteness}

Having found a range of $\g$ where fractional quantum gravity is strictly or super-renormalizable, we proceed by constructing killer operators for the gravitational action to make the quantum theory finite.

The UV part of the action \Eq{gravac} is
\be\label{killgrav}
S_\textsc{uv}\propto \int\rmd^{4}x\,\sqrt{|g|}\,G_{\mu\nu}\B^{\g-2}R^{\mu\nu},
\ee
and the UV graviton propagator scales as $k^{-2\g}$ in momentum space. For the UV-divergent terms at the one-loop level in the super-renormalizable ($\g>4$) gravitational theory, we can choose the basis
of irreducible terms, namely
\be
S_{{\rm div}}=\!\int\!d^{4}x\sqrt{|g|}\left(\beta_{R}R^{2}+\beta_{U}U^{2}+\beta_{C}C^{2}\right),
\ee
where $U_{\mu\nu}$ is the trace-free Ricci tensor in $D=4$ spacetime dimensions
\be
U_{\mu\nu}=R_{\mu\nu}-\frac{1}{4}g_{\mu\nu}R\,,
\ee
and $C_{\mu\nu\s\tau}$ is the standard Weyl tensor of Riemannian geometry in four dimensions:
\be
C_{\mu\nu\s\tau}\coloneqq R_{\mu\nu\s\tau}+\left(g_{\nu[\s}R_{\tau]\mu}-g_{\mu[\s}R_{\tau]\nu}\right)+\frac{1}{3}\,Rg_{\mu[\s}g_{\tau]\nu}\,.
\ee
By the squares above, we denote their natural quadratic contractions: $U^{2}=U_{\mu\nu}U^{\mu\nu}$ and $C^{2}=C_{\mu\nu\s\tau}C^{\mu\nu\s\tau}$. Although by a slight change of basis one term of the three above can be eliminated by the Gauss--Bonnet theorem, we prefer to use this (redundant) basis since the contributions to the beta functions are very clear and direct in this way. All three coefficients $\beta_{R},$ $\beta_{U}$ and $\beta_{C}$ contain divergences of the theory. To cancel them, we employ the killing procedure of section \ref{killers}. For this, we can use three killer operators in the form
\be
S_{{\rm K}}=\int\rmd^{4}x\,\sqrt{|g|}\left(s_{R}R^{2}\B^{\g-4}R^{2}+s_{U}U^{2}\B^{\g-4}U^{2}+s_{C}C^{2}\B^{\g-4}C^{2}\right).\label{eq:gravitational_killers}
\ee
Thanks to the tracelessness property of the tensors $U_{\mu\nu}$ and $C_{\mu\nu\s\tau}$, we are sure that dependence of the coefficients $\beta_{R}$, $\beta_{U}$ and $\beta_{C}$ will generally be affine and diagonal functions in the non-running killer coefficients $s_{R}$, $s_{U}$ and $s_{C}$. (This is a consequence of the fact that, for example, expressions such as $U_{\mu\nu}U_{\rho\sigma}$ resulting from second variations can be completed to scalars, i.e., traced, in one unique way, giving us the invariant square $U^{2}$.) Namely, we have
\ba
\beta_{R}&=&a_{R}s_{R}+\tilde{a}_{R},\\
\beta_{U}&=&a_{U}s_{U}+\tilde{a}_{U},\\
\beta_{C}&=&a_{C}s_{C}+\tilde{a}_{C},
\ea
where the constant numbers $a_{R}$, $a_{U}$ and $a_{C}$ are non-zero (because the operators in \Eqq{eq:gravitational_killers} are non-trivial). This implies that, by choosing the special values for the killer coefficients
\be
s_{R}^{*}=-\frac{\tilde{a}_{R}}{a_{R}},\qquad
s_{U}^{*}=-\frac{\tilde{a}_{U}}{a_{U}},\qquad
s_{C}^{*}=-\frac{\tilde{a}_{C}}{a_{C}},
\ee
we realize the condition that all perturbative beta functions of the theory cancel at the one-loop level ($\beta_{R}=\beta_{U}=\beta_{C}=0$). Hence the quantum theory is finite. After applying the Gauss--Bonnet theorem, the $D=4$ total classical action is \Eq{gravac} augmented by \Eqq{eq:gravitational_killers} where the derivative operators are modified similarly to \Eq{killL}, such that they interpolates between the UV and the IR regime with a well-defined IR limit:
\ba
S&=&\frac{1}{2\k^2}\int\rmd^4x\,\sqrt{|g|}\,\left[R+\ell_*^2 G_{\mu\nu}(-\ell_*^2\B)^{\g-2}\,R^{\mu\nu}\vphantom{\frac{(-\B)^\g}{\B^4+\Lambda_\textsc{ir}^4}}\right.\nn
&&\qquad\quad\left.+\a_0 R^{2}\frac{(-\B)^\g}{\B^4+\Lambda_\textsc{ir}^4}R^{2}+\a_2 R_{\mu\nu}R^{\mu\nu}\frac{(-\B)^\g}{\B^4+\Lambda_\textsc{ir}^4}(R_{\mu\nu}R^{\mu\nu})\right],\label{gravactot}
\ea
where $\g>4$, we made explicit the signs attached to $\B$ operators and the coefficients $\a_0$ and $\a_2$ depend on $s_{R}^{*}$, $s_{U}^{*}$ and $s_{C}^{*}$. A similar construction, but with more killers, holds if one starts from the strictly renormalizable version of the theory ($2<\g<4$).


\subsection{Double splitting and unitarity}\label{douspl}

The renormalizability range $\g\geq 2$ for fractional quantum gravity is incompatible with the unitarity range $0<\g<2$ found with the splitting procedure of section \ref{sec4}. Since we cannot change the power counting, the only possibility to make the theory work is to modify the splitting procedure in such a way as to extend the range of $\g$ to values greater than 2. We illustrate this with a scalar field model.

Revisiting the whole splitting apparatus, one can see that there are two main sources of constraints on $\g$. One is the contribution of the branch points, which gives the bound \Eq{gN}. The other is the contribution of the branch cuts, which yields the restrictive upper bound \Eq{empibou}. When $N=2$, the bound \Eq{empibou} can be over-crossed when the cuts are chosen on the imaginary axis, but in that case \Eq{gN} limits the fractional exponent to the range $0<\g<2$. Therefore, a natural first step in our strategy is to lift the bound \Eq{gN}. This is indeed possible if we recall that the branch points of the split propagator \Eq{GN} are also singularity points. If we redefined the propagator so that it were non-singular at branch points, then the contribution of the arcs around these points would vanish and the bound \Eq{gN} would be completely lifted.

To achieve this result, we propose a double-splitting procedure. The branch cut of the original UV propagator \Eq{Gz} is, so to speak, split twice, first into $N$ cuts and then each of these is further split into $\tilde N$ new cuts, possibly in a degenerate way:
\be\label{ww}
\tilde G(z) = \frac{1}{\left\{w_{\tilde N}[w_N(z)]\right\}^\g}\,,\qquad w_N(z)=\prod_{i=1}^N(m^2-\rme^{-\rmi\vp_i}z)^\frac{1}{N},
\ee
where $w_N(z)\to m^2-z$ in the limit $\vp_i\to 0$ and the angles in $w_N$ are chosen in such a way as to get a real-valued expression, as in section \ref{sec4a}. For example,
\ba
w_2(z)&=& \left(m^4+z^2\right)^\frac12,\label{w2}\\
w_3(z)&=& \left(m^2-z\right)^\frac13\left(m^4-2m^2z\cos\vp+z^2\right)^\frac13,\qquad 0<\vp<\frac{\pi}{2}\,.\label{w3}
\ea

As in section \ref{sec4}, we must consider the contributions of arcs at infinity, branch points and branch cuts separately. The contribution of arcs at infinity is obvious and still yields the lower bound \Eq{g0}, $\g>0$. Next, we consider branch points. The function $w_N(z)$ in \Eq{ww} has $N$ branch points $z_i=m^2\exp(\rmi\vp_i)$ given by the solutions of the equation
\be
w_N(z_i)=0\,.
\ee
Since a non-zero mass $m^2\neq 0$ implies that $w_{N}(0)\neq 0\neq w_{\tilde N}(0)$, one has
\be
w_{\tilde N}[w_N(z_i)]=w_{\tilde N}(0)\neq 0\,,
\ee
which means that the branch points $z_i$ are not singularities of the propagator. This bears an important consequence: the contribution of the mini-contours around the $z_i$ is zero. In fact, consider the contour $C_\ve^i$ around the branch point $z_i$ and parametrize it as $z=z_i+\ve\exp(\rmi\t)$. Then,
\ba
w_N(z_i+\ve\,\rme^{\rmi\t})&=&(-\ve\,\rme^{\rmi\t}\,\rme^{-\rmi\vp_i})^\frac{1}{N}\prod_{j\neq i}\left(m^2-m^2\rme^{\rmi\De_{ij}}\right)^\frac{1}{N}+\dots\nn
&\eqqcolon& \ve^\frac{1}{N}\,\rme^{\frac{\rmi\t}{N}} f_N(\vp_i,\vp_{j\neq i})\,,\label{wNpar1}
\ea
where $\De_{ij}$ is defined by \Eq{Deij} and $f_N(\vp_i,\vp_{j\neq i})\neq 0$. Furthermore,
\ba
w_{\tilde N}[w_N(z)]&=&\prod_{l=1}^{\tilde N}\left[m^2-\rme^{-\rmi\vp_l}w_N(z)\right]^\frac{1}{{\tilde N}}\nn
&\stackrel{\text{\tiny\Eq{wNpar1}}}{=}& \prod_{l=1}^{\tilde N}\left(m^2-\rme^{-\rmi\vp_l}\ve^\frac{1}{N}\,\rme^{\frac{\rmi\t}{N}} f_N\right)^\frac{1}{{\tilde N}}\nn
&=& m^2+O(\ve^\frac{1}{N})\,,\label{wNpar2}
\ea
so that there is a regular contribution from the branch point in the limit $\ve\to0$, contrary to \Eqq{zizj}. We can easily adapt \Eqq{ceps} and get the generalization of \Eqq{zizj} for a doubly split propagator. The leading term is
\ba
\frac{1}{2\pi\rmi}\int_{C_\ve^i}\rmd z\,\frac{\tilde G(z)}{z+k^2} &=&\frac{\ve}{z_i+k^2}\int_{\t_1}^{\t_1+2\pi}\frac{\rmd\t}{2\pi}\,\rme^{\rmi\t}\tilde G(z_i+\ve\,\rme^{\rmi\t})+\dots\nn
&\stackrel[\text{\tiny\Eq{wNpar2}}]{\text{\tiny\Eq{ww}}}{=}& \frac{\ve}{z_i+k^2}\frac{1}{m^{2\g}}\frac{\rme^{\rmi(\t_1+2\pi)}-\rme^{\rmi\t_1}}{2\pi\rmi}+O(\ve^2)\nn
&=&O(\ve^2)\,,
\ea
which vanishes in the limit $\ve\to 0$ for any value of $\g$. (Once again, we stress that the limit $m\to 0$ is taken at the end of any other calculation.) In particular, the bound \Eq{gN} on $\g$ disappears.

For certain choices of the splitting, the set of $z_i$ does not exhaust all the branch points of the propagator but one can show that any extra branch point $\tilde z_i$ should lie along some branch cut. This can occur, for instance, for the $(\tilde N,N)=(4,3)$ splitting. We do not consider such configurations further.\footnote{They are harmless anyway, because the contribution of all extra branch points is already included in the contribution of the cuts they branch from, which we treat in the most general way.} A final restriction to the types of double splitting is the exclusion of configurations admitting extra branch points $\tilde z_i$ which are also singularities of the propagator. Therefore, we will choose $(\tilde N,N)$ so that
\be\label{wwnosing}
w_{\tilde N}[w_N(\tilde z_i)]\neq 0\,.
\ee
A fully calculable, worked-out example reported in appendix \ref{appF} is $(\tilde N,N)=(2,3)$.

This leaves us to deal with the paths along branch cuts, which are by far the hardest part of the contour to compute. In fact, the general expression \Eq{ww} gives rise to a pattern of curved cuts in the complex plane. The explicit calculation of the contribution of each cut in \Eq{ww} becomes rapidly intractable with the increase of $\tilde N, N$ and, even when it is doable, it gives a negative result: although by construction the spectral function of the cuts is real, the positivity requirement not only reinstates the bound \Eq{gN}, but it further restricts the range of $\g$ to values $\g<2$, or it even shrinks it to an empty set. We checked this in the simplest example $(\tilde N,N)=(2,3)$, where all the cuts are straight lines and their contributions can be calculated explicitly (appendix \ref{appF}).

It is at this point that we introduce the modified Anselmi--Piva prescription \Eq{optAP} as the second step of our solution to the unitarity problem. Indeed, \Eqq{optAP}, applies to the UV part of the propagator and guarantees that, no matter how intricate the net of branch cuts is in the complex plane, their contribution will always multiply the principal value PV and will not give rise to degrees of freedom propagating on external legs. Also, this contribution will be real and finite for any $\g$ thanks to \Eq{wwnosing}, so that it is well defined mathematically. The same conclusion is reached if we decompose the metric into $g_{\mu\nu}=\bar g_{\mu\nu}+2\chi_{\mu\nu}$ and, instead of \Eq{optAP}, we implement the Anselmi--Piva procedure on the spin-2 field $\chi_{\mu\nu}$ as done in \cite{Anselmi:2018tmf,Anselmi:2018bra} for quadratic gravity.

This observation is a game changer, because it makes the calculation of the cuts in \Eq{ww} unnecessary. To summarize, iterating the splitting of the propagator we remove the constraints on $\g$ previously coming from the branch points but this does not solve the problem completely because there are constraints also from the branch cuts, if we force the positivity condition thereon. The ultimate solution is to impose the Anselmi--Piva procedure, which removes all UV unstable modes as well as the need of the positivity condition on the spectral function of the cuts. This leads to a unitary theory for any $\g>1$ and, in particular, to renormalizable fractional quantum gravity when $\g>2$.


\section{Discussion}\label{sec5}

In this work, we studied the power-counting renormalizability and free-level unitarity of a multi-fractional scalar field theory with fractional d'Alembertian operators. We refined and extended the results of \cite{mf1} in the case of derivative interactions and found that the theory is higher-loop super-renormalizable for $1<\g\leq 2$ and one-loop super-renormalizable for $\g>2$. The same conclusion holds for fractional gauge theories (section \ref{sec1b}). 

Regarding the prescription on the propagator of the theory, we considered the Feynman (causal) and the Dyson (anti-causal) prescription, as well as the Anselmi--Piva prescription which leads to the conclusion that the gas of quasi-particles associated with the fractional operator are fakeons, i.e., they are always virtual and never appear in external legs. This first-timer result is attractive for its simplicity and for generalizing the application of the Anselmi--Piva procedure to a class of theories much wider than higher-derivative local theories. However, the ensuing theory is still confined to a corner of parameter space where it is non-renormalizable.

Taking a novel representation of the $\B^\g$ operator as the product of two or more copies $\B^{\g_i}$ with smaller fractional exponents $\g_i$, we were able to achieve unitarity and higher-loop super-renormalizability (extendable to finiteness by adding killer operators) in the case of two copies and $1<\g<2$. Non-zero masses in the copies are necessary for computational purposes but some or all of them can be sent to zero at the end. This splitting procedure is a sort of regularization and is new as an application to branch cuts but, by itself, the introduction of mass scales eventually set to vanish is not a novelty in QFT with ordinary poles. For example, a class of theories of conformal supergravity can be built as the massless limit of Einstein--Weyl supergravity \cite{Ferrara:2020zef}.

We also found a simple explanation of why branch cuts do not affect unitarity if they appear in conjugate pairs: in that case, the spectral function is real-valued, otherwise any individual branch cut would carry a phase that could not be reabsorbed in the absence of a conjugate counter-part. This generalizes a similar conclusion obtained for complex conjugate poles in higher-derivative theories \cite{Liu:2022gun}. We showed that each splitting leads to a unique result when $\g_i=\g/N$, while different splittings can lead to theories with the same renormalization properties. In particular, we obtain a strictly renormalizable theory only for $N=2$, while in all the other cases the theory is non-renormalizable. This scheme confirms what has become increasingly apparent since the proposal of the Anselmi--Piva procedure: To a certain degree, the issue of unitarity lies more in the way the theory is technically defined rather than on physical considerations, although the latter do provide a strong guidance on how to deal with such technicalities. This conclusion implies the somewhat radical perspective that the physics of quantum fields, including quantum gravity, might be described more efficiently not by individual theories, but in terms of universality classes \cite{Liu:2022gun}. Non-local theories with fractional operators form a class on their own that join higher-derivative and asymptotically polynomial theories in the fray.

These findings have an immediate application to other quantum theories with fractional d'Alembertian, as we have seen for gauge theories. In the case of fractional quantum gravity (section \ref{secgrav}), we met some extra difficulties because renormalizability can be obtained only for $\g\geq 2$, which lies outside the unitarity range in the splitting procedure. We circumvented this problem by implementing a two-step technical trick, first a double splitting of the kinetic operator and then imposition of the Anselmi--Piva prescription in the UV sector of the theory. Thus, fractional gravity is unitary for $\g>1$, non-renormalizable for $\g<2$, strictly renormalizable for $\g=2$ (Stelle gravity), super-renormalizable for $2<\g\leq 4$ and one-loop super-renormalizable for $\g>4$.

Thanks to the results of the present work, the possibility to have a consistent theory of quantum gravity with fractional d'Alembertian has become more concrete. There are still several important aspects to investigate. Unitarity has been shown here at the tree level and this is an encouraging result but, in order to conclude that the theory is free from quantum instabilities and that probability is conserved, one should check full perturbative unitarity using Cutkosky rules \cite{Cutkosky:1960sp}, which have already been employed successfully in other theories with non-local operators \cite{Pius:2016jsl,Briscese:2018oyx,Chin:2018puw}. 

Related to perturbative unitarity is the topic of scattering amplitudes, which require the computation of terms in the finite part of the effective action. As a preliminary consideration, we note that this other chapter of the fractional quantum theory, to be written in the future, will be in principle independent of the properties of renormalizability and unitarity presented here. 
 In the golden age of QFT, it was believed that a ``good'' energy behaviour of scattering amplitudes is related to, or even necessary for, renormalizability \cite{LlewellynSmith:1973yud}. A growth of tree-level scattering amplitudes at high energies $E$ in non-renormalizable theories makes these amplitudes exceed the so-called unitarity bound \cite{Abe:2018rwb}. This fact, together with the finding that in some renormalizable gauge theories the violation of the unitarity bound does not happen \cite{Cornwall:1973tb,Bell:1973ex}, led to the conjecture that all renormalizable theories should respect the unitarity bound \cite{LlewellynSmith:1973yud}. An added argument advanced in \cite{LlewellynSmith:1973yud} was that such a violation would necessarily signal a breaking of perturbation theory above some energy. Later on, the case of Einstein's gravity supported the conjecture, since the theory is non-renormalizable and, indeed, its tree-level scattering amplitudes diverge as $E^2$ at high energies \cite{DeWitt:1967uc,Berends:1974gk,Grisaru:1975bx}. Thus, the issues of renormalizability, unitarity and validity of the perturbative expansion all mix up in the question of how scattering amplitudes behave at high energies. However, counter-arguments to \cite{LlewellynSmith:1973yud} were found quite soon \cite{Berends:1974gk} and, nowadays, we have ample proof that these three aspects are unrelated to the high-energy behaviour of scattering amplitudes. In general, it can happen in any theory that some scattering amplitudes diverge when the scattering energy tends to infinity. This could be taken as a signal that we enter a strongly coupled regime of the theory in the UV, where the coupling is defined operationally from physical cross sections as for example done in \cite{Weinberg:1980gg}. However, this phenomenon has nothing to do with renormalizability, UV-finiteness or unitarity. Two explicit examples with different renormalizability and unitarity properties should convince the reader about this fact. The first example is quadratic Stelle gravity in four dimensions, which is a renormalizable theory and, yet, some of its tree-level scattering amplitudes diverge at high energies as $E^2$ \cite{Dona:2015tra,Abe:2020ikj,Abe:2022spe}, exactly as in Einstein's gravity, which is clearly a unitary theory. Summing also the contributions of the decay of two spin-2 ghosts and of one ghost and one graviton, the total result for the minimally inclusive cross section is $O(E^0)$ and does respect the unitarity bound \cite{Holdom:2021hlo}, despite Stelle's theory being non-unitary. Also, this does not cancel the fact that not all the amplitudes of this theory fulfill the bound. The second example is non-local quantum gravity with asymptotically polynomial operators, which is super-renormalizable or even finite depending on whether one adds killer operators or not. Also in this case the graviton tree-level scattering amplitudes diverge as $E^2$ at $E \to +\infty$ \cite{Dona:2015tra,Modesto:2021soh} but this theory is known to be perturbatively unitary at all orders \cite{Briscese:2018oyx}. The energy divergence is a non-issue because, once perturbative unitarity is proven, it must be preserved also at the level of the fully resummed scattering amplitudes. A violation of unitarity at any given loop order must be compensated upon resummation by higher-order contributions \cite{Abe:2018rwb}. Although we have not computed either graviton amplitudes or the Cutkosky rules in the fractional theory, the same conclusions will apply here.

On a separate subject, we foresee violations of micro-causality, both because this can happen in the presence of non-integer powers of the d'Alembertian operator \cite{BG,Belenchia:2014fda} and also due to the presence of fakeons \cite{Anselmi:2018tmf,Anselmi:2018bra}. This may be an asset rather than a liability, since violations of micro-causality can give rise to interesting predictions of new phenomena while respecting other observations, in contrast with the more dangerous breaking of macro-causality not happening here. The topic of the classical limit is related. Another consequence of having fake particles is that the classical limit of the theory is not \Eq{gravac} or \Eq{gravactot} but is obtained by projecting out fakeons perturbatively according to a \emph{classicization} procedure \cite{Anselmi:2018bra,Anselmi:2019rxg}. The resulting classical equations of motion are expected to deviate from the standard Einstein equations and give rise to new phenomenology, such as the above-mentioned violation of micro-causality or a modified cosmological evolution. These and other physical consequences of \Eqqs{gravac} and \Eq{gravactot}, some of which were pointed out in \cite{mf2}, will require further study.


\section*{Acknowledgments}

G.C.\ thanks D.~Anselmi and G.~Nardelli for useful discussions. L.R.\ thanks the Department of Physics, UFJF, for the warm hospitality. The authors are supported by grant PID2020-118159GB-C41 funded by MCIN/AEI/10.13039/501100011033.


\appendix


\section{BPHZ renormalization}\label{secbphz}

It is known that the BPHZ renormalization procedure can be applied to some non-local theories. The simplest case is non-local quantum gravity with asymptotically polynomial form factors \cite{BasiBeneito:2022wux}. The UV limit of that theory is local, since those particular form factors reduce to higher-derivative polynomials at high energy. Then, the BPHZ scheme applies without any change, and so does the power counting. More caution should be exercised in theories with mixed UV/IR divergences, which require a modification of the BPHZ scheme. Such is the case of matrix field theory \cite{Grosse:2016pob,Hock:2020rje} and non-commutative field theory \cite{Blaschke:2012ex,Blaschke:2013cba}. Non-commutative and matrix theories are examples of combinatorially non-local field theories, where field arguments are convoluted pair-wise in the interactions instead of all together at one point as in local theories. An extension of the BPHZ scheme is possible also in this more general class \cite{Thurigen:2021ntr}. Different from the above examples and less explored is the case of theories, such as fractional ones, with a non-local UV limit and without IR divergences. In this sub-section, we show that the BPHZ scheme applies to fractional QFTs without modifications.

The BPHZ procedure essentially consists of three ingredients (see, e.g., \cite{Blaschke:2013cba} for a review):
\begin{enumerate}
\item A regularization-independent subtraction scheme systematically subtracting any overall UV divergences as well as sub-divergences from any Feynman integral of all one-particle-irreducible graphs in the theory. The aim here is to get a convergent expression for these integrals.
\item Locality of counter-terms, as established by Weinberg's theorem \cite{Weinberg:1959nj}. The terms subtracted in the procedure in the previous point have the same form of the UV action, which is constrained by a power-counting analysis of UV divergences. Hence, one proves additive renormalizability if the UV action absorbs all divergences. With a redefinition of the constants $Z_{i}$ used for the wave-function normalization of all fields, masses and couplings present in the UV action, one shows that additive renormalizability is equivalent to multiplicative renormalizability, since all bare quantities are successfully reabsorbed in the renormalization constants $Z_{i}$.
\item Conditions for renormalization of the theory that can be placed at any energy scale. A change of the scale corresponds to a finite renormalization of couplings (addition of finite counter-terms).
\end{enumerate}
Point 1 can be performed using momentum-space subtractions in the integrands of Feynman integrals. The subtraction of divergences from the Feynman integral and the consequent extraction of its finite part is known as the Bogoliubov $R$-operation on the integral. Due to the possible presence of UV sub-divergences, one needs recursion relations for the $R$-operation. These are obtained by considering Zimmermann's forest formula which yields finite (i.e., renormalized) total Feynman integrals. This also allows one to inductively prove the locality of the subtractions done by counter-terms. The needed counter-terms are local, which means that, if they feature $\B^\g$ operators, the exponents $\g$ must be positive integers. However, when $\g>1$ in the original theory, nowhere in the proof of locality of UV counter-terms does one have to assume that the exponents $\g$ in the initial action be integers. Compare also a more detailed discussion on the structure of higher-loop beta functions in section \ref{betas}.

For the subtraction scheme, we can use the following notation. Let $I_\G(k,p)$ be the integrand expression of a Feynman diagram $\Gamma$, where $p$ collectively denotes external momenta $p_i$ ($i=1,2,\dots$) and $k$ is the set of all internal momenta $k_i$ over which we have to integrate in the whole Euclidean range (from the IR regime $|k|=0$ towards the UV limit $|k|\to+\infty$). Let us consider the case in which the integral $ \int_k 
I_{\Gamma}(k,p)$ is divergent in the UV limit of $k$ integration. The Bogoliubov $R$-operation is defined as producing the finite integral
\be\label{Ifin}
I_{\Gamma}^{{\rm fin}}(p)\coloneqq\int_k 
R_{\Gamma}I_{\Gamma}(k,p)\,,\qquad R_{\Gamma}\coloneqq\mathbbm{1}-t_{p}^{N}\,,
\ee
where $N\in\mathbb{N}_+$ is related to the superficial degree of divergence $\om_\G$ of the diagram $\G$, obtained from a naive power-counting analysis, and the Taylor multi-expansion operator acting on integrands with several external momenta is given by
\be
\left(t_{p}^{N}I_{\Gamma}\right)(k,p)\coloneqq\sum_{l=0}^{N}\frac{1}{l!}p_{i_{1}}^{\mu_{1}}\cdots p_{i_{l}}^{\mu_{l}}\frac{\p^{l}I_{\Gamma}}{\partial p_{i_{1}}^{\mu_{1}}\cdots\p p_{i_{l}}^{\mu_{l}}}(k,p=0)\,.\label{eq:taylorexp}
\ee
If $\om_\G<0$, then $\left(t_{p}^{\om_\G}I_{\Gamma}\right)(k,p)=0$. Here, in particular, we focus on theories 
 where the integrand $I_\G(k,p)$ takes the form of a generalized polynomial in the UV regime,
\be
 I_\G\sim\mathfrak{p}(K)=\sum_{\gamma'}a_{\gamma'}K^{\gamma'},
\label{eq:genpoly}
\ee
where $K$ is a linear combination of internal and external momenta $k,p$ and the values of $\gamma'\in\mathbb{R}$ are ordered and not necessarily integer. Expression \Eq{eq:genpoly} is a generalized polynomial in momentum $K$, i.e., the exponents $\gamma'$ can be non-integer or even irrational; this is an example of a generalized Laurent or Puiseux expansion around the formal point $K=+\infty$. 


In order to understand divergences, the integral $\int_k I_\G$ can be treated as an indefinite integral over a set of variables $k$ of loop integrations. Based on \Eqq{eq:genpoly}, the integral defined with the UV cut-off related to the momentum $K$ has the following UV asymptotics:
\be
\int_k I_\G\sim\int_k \mathfrak{p}(K)=\sum_{\gamma'\neq0}b_{\gamma'}K^{\gamma'}+ b_0\ln K,
\label{eq:genpoly2}
\ee
with the same qualifications as for the generalized Laurent expansion in (\ref{eq:genpoly}). Moreover, for \Eq{eq:genpoly2} we require that $\gamma'_{\max}=\om_\G$ be finite. Also $\gamma'_{\min}$ is finite if the generalized polynomial contains finitely many terms, while $\gamma'_{\min}\to -\infty$ when the number of terms in \Eq{eq:genpoly} is infinite (this is to avoid the focusing point at some finite $\gamma'$). In the case of the kinetic term in \Eqq{eq:model}, $\gamma'_{\max}=\gamma'_{\max}(\g)$ is an affine function of the fractional exponent $\g$:
\be
\gamma'_{\max}(\g)=B_0+B_1\g\,,\qquad B_{0,1}\in\mathbb{R}\,,
\ee
where $B_0$ and $B_1$ are constants. Note that the series \Eq{eq:genpoly2} is ``doubly'' generalized due to the appearance of the $b_0\ln K$ term instead of a constant term $b_0$. This logarithm comes from the indefinite integral $\int\rmd k/k$  of the generalized monomial $k^{-1}$ term possibly present in the expansion of $I_\G$ in \Eq{eq:genpoly}. Since we are only interested in the divergent part of $\int_k I_\G$ and inverse powers of $k$ in \Eq{eq:genpoly2} produce convergent terms, the maximal exponent $\gamma'_{\max}$ can be effectively set to zero if there are no $\gamma'>0$ and $b_0\neq 0$ in \Eq{eq:genpoly2}. 

In such circumstances, one can immediately check that the $R$-operation applied to the generalized polynomial of degree $\gamma'_{\max}$ in $\int_k I_\G$ renders the integral finite, provided that the upper limit $N$ in \Eqq{eq:taylorexp} be taken as the largest integer smaller than or equal to $\gamma'_{\max}+1$:
\be\label{gmax}
N=\lfloor \gamma'_{\max}+1 \rfloor=\begin{cases}
\lceil\gamma'_{\max}\rceil & {\rm if}\quad\ensuremath{\gamma'_{\max}}\notin\mathbb{Z}\\
\gamma'_{\max}+1 & {\rm if}\quad\ensuremath{\gamma'_{\max}}\in\mathbb{Z}
\end{cases}\,.
\ee
Let us consider four applications of \Eqqs{eq:taylorexp} and \Eq{gmax}. 
\begin{itemize}
\item The first is a local higher-derivative theory where $\int_k\mathfrak{p}(K)$ is an ordinary polynomial with all exponents $\gamma'$ being positive natural numbers, so that its degree is $\gamma'_{\max}=N-1\in\mathbb{N}_+$. Then, we need to take $N=\gamma'_{\max}+1$ derivatives of the integrand $I_\G$ to arrive at a convergent integral, because after taking only $\gamma'_{\max}$ derivatives with respect to external momenta $p$ the integrand might be still (logarithmically) divergent in the UV. This condition in \Eq{gmax} is determined by the fact that each differentiation with respect to $p^{\mu}$ lowers the degree of divergence of the propagator and thereby of the integral. Actually, due to how the subtraction mechanism works, the divergent part of the integrand $I_{\Gamma}(k,p)$ is always and necessarily a standard (not generalized) polynomial (possibly with a logarithmic term $\ln K$ instead of a constant) in the external momenta $p$, whose degree $\lfloor \gamma'_{\max} \rfloor$ coincides with, or is the integer floor of, the degree of divergence $\om_\G$ of this diagram. 
\item The second application is a plain Laurent series
\be\label{IGlau}
I_\G = a_1 (k+p)+a_0 +a_{-1}(k+p)^{-1}+a_{-2}(k+p)^{-2}\,,
\ee
which mimics non-local models with inverse powers of the d'Alembertian. Here we consider the integrand $I_\G$ written schematically and dependent only over the magnitudes of the one-dimensional loop integration momentum $k$ and of the external momentum $p$. Integrating in $k$, we get $\int \rmd k I_\G = (a_1/2) k^2 +(a_0+a_1p) k +a_{-1}\ln (k+p) - a_{-2}(k+p)^{-1}$, where the highest exponent is $\gamma'_{\max}=2$. Of course, this expression is power-law UV divergent (limit $k\to+\infty$). Apply now the BPHZ scheme \Eq{Ifin} and \Eqq{eq:taylorexp} as an iteration of $p$ derivatives. Let us accordingly define $I_\G^{(l)}=\rmd^l I_\G/\rmd p^l$, where $l=0,1,\dots,N$ and $N=3$ as established by \Eq{gmax}. 
 Evaluating at vanishing external momentum $p$, we have
\ba
&&\left.I_\G^{(0)}\right|_{p=0} = a_1 k+a_0 +a_{-1}k^{-1}+a_{-2}k^{-2},\quad \left.I_\G^{(1)}\right|_{p=0} = a_1-a_{-1}k^{-2}-2a_{-2} k^{-3}\nn
&& \left.I_\G^{(2)}\right|_{p=0} = 2a_{-1}k^{-3}+6a_{-2}k^{-4},\quad \left.I_\G^{(3)}\right|_{p=0} = -6a_{-1}k^{-4}-24a_{-2}k^{-5}\,.
\ea
Then, the Bogoliubov $R$-operation is
\ba
R_\G I_\G &=& I_\G -\left[\left.I_\G^{(0)}\right|_{p=0}+p \left.I_\G^{(1)}\right|_{p=0}+\frac12 p^2\left.I_\G^{(2)}\right|_{p=0}+\frac16 p^3\left.I_\G^{(3)}\right|_{p=0}\right]\nn
&=&\frac{4a_{-2} p^3}{k^5}+\frac{(a_{-1} p-3 a_{-2}) p^2}{k^4}-\frac{(a_{-1}p-2a_{-2}) p}{k^3}+\frac{a_{-1}p-a_{-2}}{k^2}\nn
&&+\frac{a_{-2}}{k^2+2 kp+p^2}+\frac{a_{-1}}{k+p}-\frac{a_{-1}}{k}\,.
\ea
The integral of this, $\int \rmd k\, R_\G I_\G$, which corresponds to $I_{\Gamma}^{{\rm fin}}(p)$ in \Eqq{Ifin}, is convergent in the UV regime of the variable $k$ (while at the same time being IR-divergent due to the form of the Laurent series; this is a typical feature of non-local models with inverse d'Alembertian). In particular, the two dangerous terms $a_{-1}/(k+p)$ and $-a_{-1}/k$ cancel each other in the UV regime $k\gg p$. Explicitly, given the IR cut-off $k_{\textsc{ir}}>0$, \Eqq{Ifin} is given by
\ba
I_{\Gamma}^{{\rm fin}}(p)&=&\int_{k_{\textsc{ir}}}^{+\infty} \rmd k\, R_\G I_\G\nn
&=&\frac{a_{-1} p^3}{3 k_{\textsc{ir}}^3}-\frac{a_{-1} p^2}{2k_{\textsc{ir}}^2}+\frac{a_{-1} p}{k_{\textsc{ir}}}+\frac{a_{-2}p^4}{k_{\textsc{ir}}^4 (k_{\textsc{ir}}+p)}+a_{-1}\ln\left(\frac{k_{\textsc{ir}}}{k_{\textsc{ir}}+p}\right),\label{ex1fin}
\ea
which is UV convergent. Actually, one can easily verify that the subtraction procedure is successful in removing the UV divergences already for a smaller $N=2$ or $N=1$. This is a manifestation of the fact that power-law divergences are non-universal and can be removed in certain regularization schemes such as the dimensional one. Consistently, we would have obtained the same result \Eq{ex1fin} when setting $a_1=0=a_0$ from the start. 
\item As a third check of \Eq{gmax}, we notice that, for the generalized polynomial in \Eq{eq:genpoly2} with $b_0\neq 0$ and all other $\g'<0$, we have $\gamma'_{\max}=0$ and logarithmic UV divergences, which are universal in the sense of appearing in any regularization scheme. To remove them with the subtraction procedure, one needs to take one derivative, hence the bottom case $N=\gamma'_{\max}+1=1$ in \Eqq{gmax}. 
\item The last two cases above are somewhat pathological for the presence of IR divergences in the limit $k_{\textsc{ir}}\to 0$, which would require \emph{ad hoc} adulterations of the original BPHZ scheme \cite{Grosse:2016pob,Hock:2020rje,Blaschke:2012ex,Blaschke:2013cba,Thurigen:2021ntr} (notice that \Eq{ex1fin} is regularization-dependent). For this reason, we turn to a more interesting and non-trivial example with non-integer exponents. Consider a scalar fractional QFT with quartic interaction and, in particular, the two-loop correction to the two-point function coming from the ``cactus'' diagram $\vcenter{\hbox{\includegraphics[height=.8cm]{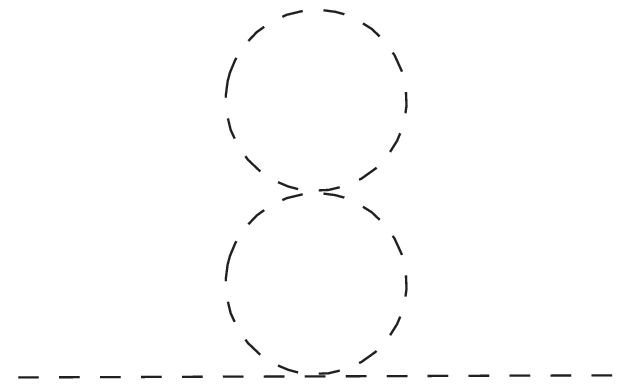}}}$ (two bubbles on top of each other and the two external legs tangent to one of the bubbles). The full integrand for this diagram in the model \Eq{eq:model} is
\ba
I_\G(p) &=& \int\frac{\rmd^{4}k}{(2\pi)^{4}}\int\frac{\rmd^{4}q}{(2\pi)^{4}} \frac{\rmi}{\left[a_{\gamma}(k^{2})^{\gamma}+k^{2}\right]^2\left[a_{\gamma}(q^{2})^{\gamma}+q^{2}\right]}\nn
&&\qquad\times\left(24\lambda_{0}+8\lambda_{\gamma-1}\left\{ \left[\left(p+k\right)^{2}\right]^{\gamma-1}+\left[\left(p-k\right)^{2}\right]^{\gamma-1}\right\} \right)\nn
&&\qquad\times\left(24\lambda_{0}+8\lambda_{\gamma-1}\left\{ \left[\left(k+q\right)^{2}\right]^{\gamma-1}+\left[\left(k-q\right)^{2}\right]^{\gamma-1}\right\} \right).\label{2loopint}
\ea
The UV asymptotic form of this integrand simultaneously in the $k$ and $q$ integration variables reads
\be
I_\Gamma(p)\sim \sum_c k^{\gamma'_{\max}-c}q^{c},
\label{uvasym2poly}
\ee
where $\g'_{\max}=\om_\G=4-2\g$ (this is obtained from standard power counting of both integrals over $k$ and $q$ momenta, where the contribution from the measure $\rmd^4k\,\rmd^4q$ is also included). One would get the same asymptotics for the melonic two-loop diagram $\vcenter{\hbox{\includegraphics[height=.4cm]{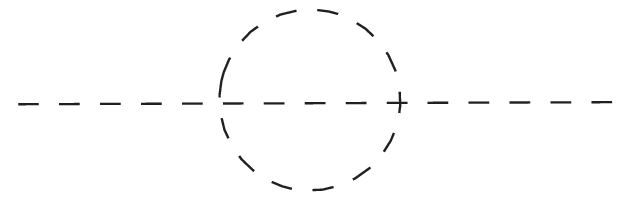}}}$. To keep our argument simple, here we can just analyze the leading term in the expansion of \Eq{2loopint} for large values of one of the integration momenta, i.e., when $k\to+\infty$; effectively, we take the $c=0$ term of the sum in \Eq{uvasym2poly} and study the term $k^{\g'_{\max}}$. Moreover, due to the structure of the integrand, we can substitute this term with $(k+p)^{\gamma'_{\max}}$, where $p$ is the external momentum of the diagram. Now, by taking $N= \lfloor \gamma'_{\max}+1 \rfloor$ derivatives with respect to $p$ of this expression, we arrive at an integral completely convergent in the $k\to+\infty$ limits. This is because, due to the integrand being symmetric with respect to exchange of $p$ with $k$, derivatives with respect to $p$ have the same effect as those in $k$, and the latter decrease the level of divergence in $k$ integrals. Therefore, after exhausting the subtraction procedure, the final integral is convergent in the $k$ variable. We also note that the UV divergences of two-loop integrals such as \Eq{2loopint} disappear completely when $\g'_{\max}<0$, leading to the condition $\g>2$ (that is, in one-loop super-renormalizable scalar theories, as we see in section \ref{powcou}).
\end{itemize}
In this way, in any local (higher-derivative) or fractional covariant QFT treated with the BPHZ procedure we deal only with convergent integrals and never have to write UV-divergent expressions explicitly. The universal regularization for a given diagram $\Gamma$ is done by means of a subtraction procedure which is universal given the set of external momenta $p$. If there were only one diagram $\Gamma$ to consider in the theory, this universal regularization of one integral would be enough. However, in many cases, and especially in gauge theory, we need to consider many diagrams at the same time at the same loop level and to renormalize them consistently (i.e., to have the same renormalization conditions everywhere according to point 3). In this case, subtraction of divergences by differentiation would lead to incoherent results for the whole theory since different integrals would be regularized separately and independently. However, the presence of gauge symmetry in the quantum theory necessarily relates different diagrams and the quantum results for different loop integrals. Therefore, regularization by ordinary methods with the absorption of UV divergences by gauge-covariant counter-terms must be applied. These counter-terms successfully renormalize the gauge theory additively because they take care of all divergent diagrams at one stroke and they preserve the gauge symmetry on the quantum level.

Next, in fractional QFTs we can reproduce the standard analysis of overlapping divergences, which arise due to superficially divergent sub-graphs of a given diagram $\Gamma$. Again, the $R$-operation works in a recursive manner. Here, the so-called Zimmermann forest formula circumvents the problems with overlapping divergences since it only requires to deal with divergences of \emph{non-overlapping} sub-graphs. In the case where the diagram contains divergent sub-graphs, the total finite Feynman integral can be obtained by the following version of the Bogoliubov $R-$operation:
\be
R_{\Gamma}I_{\Gamma}=\left(\mathbbm{1}-t_{\Gamma}\right)\left[\prod_{\kappa\in{\cal F}}(\mathbbm{1}-t_{\kappa})\right]I_{\Gamma}\,,\label{eq:bogo}
\ee
where the set ${\cal F}$ consists of all renormalization parts $\kappa\neq\Gamma$ of the total diagram $\Gamma$. This allows one to eliminate an overall divergence from the graph $\Gamma$. Moreover, here it is understood that the subtractions are performed starting from the innermost nested sub-graphs and proceeding outwards. This algorithm can always be employed to remove potentially divergent contributions from a given Feynman graph $\Gamma$ and to get, therefore, absolutely convergent integrals (according to the powerful Weinberg's power-counting theorem \cite{Weinberg:1959nj}). None of this depends on whether the kinetic terms (or interactions) are higher-derivative or fractional in nature.

Overlapping sub-graphs can be dropped systematically from \Eqq{eq:bogo} \cite{Bergere:1974zh}. In the renormalization procedure using counter-terms, this implies the removal of a possible source of non-locality, namely, divergences of overlapping type, which are always eliminated by lower-order counter-terms \cite{Blaschke:2013cba}. Therefore, no new higher-loop counter-term is needed for this purpose and one can simplify \Eqq{eq:bogo} to Zimmermann forest formula
\be
R_{\Gamma}I_{\Gamma}=\left(\mathbbm{1}-t_{\Gamma}\right)\sum_{\alpha}\left[\prod_{\kappa\in\cF_{\alpha}}(-t_{\kappa})\right]I_{\Gamma}\,,\label{eq:zimmer}
\ee
where $\alpha=0,1,\dots$ labels all sets (forests) $\cF_{\alpha}$ of renormalization parts $\kappa\subset\Gamma$ containing only non-overlapping sub-graphs. The operator associated with the empty forest $\left\{ \emptyset\right\}$ is defined as $t_{\emptyset}=-\mathbbm{1}$. 
It is also convenient to define a special subtraction operator $\bar{R}_{\Gamma}$ for a graph $\Gamma$ different from the $R_{\Gamma}$ operation, according to
\be
R_{\Gamma}I_{\Gamma}\eqqcolon (\mathbbm{1}-t_{\Gamma})\bar{R}_{\Gamma}I_{\Gamma}\,.
\ee
Then, the forest formula for the operation $\bar{R}_{\Gamma}$ is
\be
\bar{R}_{\Gamma}I_{\Gamma}=\sum_{\alpha}\left[\prod_{\kappa\in{\cal F}_{\alpha}}(-t_{\kappa})\right]I_{\Gamma}\,,\label{eq:zimmerbar}
\ee
where now the index $\alpha$ runs over all forests ${\cal F}_{\alpha}$ containing only non-overlapping sub-graphs. Finally, this modified
subtraction operator $\bar{R}_{\Gamma}$ satisfies the BPHZ recursion relation
\be
\bar{R}_{\Gamma}I_{\Gamma}=\sum_{\psi}\left[\prod_{\kappa\in\psi}(-t_{\kappa})\bar{R}_{\kappa}\right]I_{\Gamma}=\sum_{\psi}I_{\Gamma/\psi}\prod_{\kappa\in\psi}(-t_{\kappa})\bar{R}_{\kappa}I_{\kappa}\,,\label{eq:recursion}
\ee
where $\psi$ labels all possible sets of disjoint renormalization parts of $\Gamma$, including the empty set. The operator $\bar{R}_{\Gamma}$ is determined by the recursive relation \Eq{eq:recursion} in terms of the operators $\bar{R}_{\kappa}$ associated with lower-order sub-graphs $\kappa$ of $\Gamma$. While \Eq{eq:recursion} only involves disjoint sub-graphs, the forest formula \Eq{eq:zimmer} applies to non-overlapping sub-graphs, which can be both disjoint and nested. 

Equation \Eq{eq:zimmerbar} solves the BPHZ recurrence formul\ae\ and, at the same time, avoids the problem of overlapping sub-graphs. This concludes the proof that the BPHZ procedure is applicable to fractional field theories. In none of the arguments above did we have to require that $\gamma$ (appearing inside $\g'_{\max}$) be an integer. The only constraint was that the order $N$ of the Taylor expansion in the subtraction method be integer. Thanks to \Eqq{gmax}, the proof of multiplicative renormalizability using the BPHZ procedure works also for theories with fractional values of $\g'_{\max}(\g)$, provided $\g\geq 2$. In conclusion, in theories defined by generalized polynomials in the UV regime (i.e., non-integer powers when $|k|\to+\infty$), \emph{the counter-terms to be added in the subtraction procedure in point 1 are local operators}. Therefore, we can consistently apply the power counting as in sections \ref{powcou} and \ref{genpowco}.


\section{Leibniz rule for powers of the d'Alembertian}\label{appA}

The Leibniz rule is one of the most definitory properties of fractional operators. Due to its importance, we dedicate this section to it and derive two novel results: the composition rule for integer and for fractional power of the $\B$ operator.

For ordinary derivatives in one dimension, the Leibniz rule for two functions $A$ and $B$ is
\be\label{pnAB}
\p^n(AB)=\sum_{j=0}^{n}\frac{n!}{j!(n-j)!} (\p^{n-j}A)(\p^j B)=\sum_{j=0}^{n}\frac{\G(n+1)}{\G(j+1)\G(n-j+1)} (\p^{n-j}A)\p^j B\,.
\ee
In fractional calculus, there exist different derivatives of non-integer order but, in general, they all admit a generalization of the Leibniz rule that can be proved rigorously \cite{SKM}. For instance, for the Liouville derivative of order $\g$ one has \cite{mf1}
\be\label{leru}
{}_\infty\p^\g(AB)=\sum_{j=0}^{+\infty}\frac{\G(\g+1)}{\G(j+1)\G(\g-j+1)} ({}_\infty\p^{\g-j}A)\p^j B\,,
\ee
where it is easy to recognize a strong parallelism with \Eq{pnAB}. In this paper, we need a generalized Leibniz rule for non-integer powers of the covariant d'Alembertian. The fact that we act with two $D$-dimensional derivatives at each iteration makes the Leibniz rule already complicated in the case of integer $\g=n$. Since, to the best of our knowledge, even this case has not been discussed much in the literature, let us start from the calculation of $\B^n(AB)$ in Minkowski spacetime.


\subsection{Leibniz rule for integer powers \texorpdfstring{$\B^n$}{boxn}}\label{Leiru1}

For $n=1$, we have three terms:
\be
\B(AB)=(\B A)B+2\N_{\mu}A\N^{\mu}B+A\B B\,.
\ee
For $n=2$, the number of terms doubles:
\ba
\B^{2}(AB)&=&
(\B^2A)B+2\B A\B B+4(\N_{\mu}\N_{\nu}A)\N^{\mu}\N^{\nu}B+4(\B\N_{\mu}A)\N^{\mu}B\nn
&&+4(\N_{\mu}A)(\B\N^{\mu}B)+A\B^2B\,,
\ea
where we used the fact that $\N$ and $\B$ commute on Minkowski spacetime. If we work in a $\phi\B^2\phi$ theory, when computing the one-loop effective action we can throw away the last two terms, which are $O(\N^3B)$.

For arbitrary $n\geq 1$, noting that $\B^n=\prod_{i=1}^n g_{\mu_i\nu_i}\N^{\mu_i}\N^{\nu_i}$, we were able to derive the Leibniz rule in Minkowski spacetime:
\bs\label{BnAB}\ba
\B^{n}(AB) &=& \sum_{j=0}^{n}\sum_{l=0}^{j}c_{n,j,l}\left(\B^{n-j}\prod_{i=1}^{l}\N_{\mu_i}A\right)\left(\B^{j-l}\prod_{i=1}^{l}\N^{\mu_i}B\right),\\
c_{n,j,l} &=& \frac{n!}{(n-j)!}\frac{2^{l}}{l!(j-l)!}\,,\label{ccoef}
\ea\es
where $\prod_{i=1}^{l}\N_{\mu_i}A=\N_{\mu_1}\cdots\N_{\mu_l}A$. In particular, there are only four terms which are at most of second order in the derivatives of $B$ ($j=0$, $l=0$; $j=1$, $l=0,1$; $j=2$, $l=2$):
\ba
\B^{n}(AB)&=&(\B^{n}A)B+2n(\B^{n-1}\N_{\mu}A)\N^{\mu}B+n(\B^{n-1}A)\B B\nn
&&+2n(n-1)(\B^{n-2}\N_{\mu}\N_{\nu}A)\N^{\mu}\N^{\nu}B+O\left(\N^3B\right).\label{BnAB2}
\ea

To prove \Eqq{BnAB} by induction, assume that it holds for $\B^n$ and apply it to the left- and right-hand side of the identity $\B^{n+1}(AB)=\B\B^n(AB)$. Redefining the summation indices to match all terms, one finds a recurrence equation for the coefficient $c_{n,j,l}$ with initial conditions, which are solved by \Eqq{ccoef}.

A more instructive proof makes use of the generating functional of the powers of the d'Alembertian. We remind the reader that the basic binomial expansion for the case of natural finite $n$ is
\be\label{binexp}
(a+d)^{n}=\sum_{j=0}^{n}\frac{n!}{j!(n-j)!}\,a^{j}d^{n-j}.
\ee
Applying this formula twice on the generating functional $(a+2b+c)^{n}$, we obtain 
\ba
(a+2b+c)^{n}&=&[a+(2b+c)]^{n}=\sum_{j=0}^{n}\frac{n!}{j!(n-j)!}(2b+c)^{j}a^{n-j}\nn
&=&\sum_{j=0}^{n}\sum_{l=0}^{j}\frac{n!}{j!(n-j)!}\frac{j!}{l!(k-l)!}2^{l}b^{l}c^{j-l}a^{n-j}\nn
&=&\sum_{j=0}^{n}\sum_{l=0}^{j}c_{n,j,l} b^{l}c^{j-l}a^{n-j}.\label{genfun}
\ea
This double Taylor expansion in the variables $a$, $b$ and $c$ is done in the order marked by the brackets in the expression for the generating function $(a+2b+c)^{n}=[a+(2b+c)]^{n}$. Of course, at the level of the generating functional the change of the order of summation between $a$, $b$ and $c$ variables is possible and we will use this property also later. 

Now we identify the variables $a,b,c$ with the operators
\be
a=\B_{A},\quad b=\N_{\mu,A}\N_{B}^{\mu},\quad c=\B_{B},
\ee
where the subscripts $A$ and $B$ denote the object on which the differential operator acts. Therefore, in \Eqq{genfun} we have
\be
b^{l}c^{j-l}a^{n-j}(AB)=\big(\N_{\mu_{1}}\ldots\N_{\mu_{l}}\B^{n-j}A\big)\big(\N^{\mu_{1}}\ldots\N^{\mu_{l}}\B^{j-l}B\big),
\ee
and we obtain \Eq{BnAB}. In general, the operators $\B_{A}$ and $\B_{B}$ commute in any spacetime, while on a spacetime with zero Riemann curvature we have moreover that they also commute with the mixed operator $\N_{\mu,A}\N_{B}^{\mu}$. This full commutativity between $a$, $b$ and $c$ variables in the generating functional is essential to perform the standard sequential binomial expansions.



\subsection{Leibniz rule for arbitrary powers \texorpdfstring{$\B^\g$}{boxgamma}}\label{Leiru2}

The generalization of \Eqq{BnAB} to a non-integer power is
\bs\label{BgAB}\ba
\B^{\g}(AB) &=& \sum_{j=0}^{+\infty}\sum_{l=0}^{j}c_{\g,j,l}\left(\B^{\g-j}\prod_{i=1}^{l}\N_{\mu_i}A\right)\left(\B^{j-l}\prod_{i=1}^{l}\N^{\mu_i}B\right),\\
c_{\g,j,l} &=& \frac{\G(\g+1)}{\G(\g-j+1)}\frac{2^{l}}{l!(j-l)!}\,,\label{cgcoef}
\ea\es
which is the analogue of \Eqq{BnAB} with the substitution $n\to\g$. In particular, the four terms at most of second order in the derivatives of $B$ are the same as in \Eq{BnAB2} with $n\to \g$:
\ba
\B^{\g}(AB)&=&(\B^{\g}A)B+2\g(\B^{\g-1}\N_{\mu}A)\N^{\mu}B+\g(\B^{\g-1}A)\B B\nn
&&+2\g(\g-1)(\B^{\g-2}\N_{\mu}\N_{\nu}A)\N^{\mu}\N^{\nu}B+O\left(\N^3B\right).\label{BgAB2}
\ea
Formula \Eq{BgAB} can be guessed by analogy with \Eqq{leru} but here we also give a formal proof of it along steps similar to those followed in \cite[section 15]{SKM} in the case of fractional derivatives. 

Using the generalized binomial expansion extending \Eq{binexp} to an arbitrary real $\g$,
\be
(a+d)^{\g}=\sum_{n=0}^{+\infty}\frac{\G(\g+1)}{n!\G(\g-n+1)}a^{\g-n}d^{n},\label{eq:binomial_gamma}
\ee
we recast the operator $\B^\g$ in a formal series representation:
\be
\B^\g=\lim_{M\to 0}(M^2+\B)^\g=\lim_{M\to 0}\sum_{n=0}^{+\infty}b_{\g,n}^M\B^n\,,\qquad
b_{\g,n}^M\coloneqq\frac{\G(\g+1)}{n!\G(\g-n+1)}(M^2)^{\g-n}.\label{BgM}
\ee
Here we have omitted a $-$ sign in front of each and every $\B$ symbol in our $(-,+,\dots,+)$ signature, while the formula is correct as it is in the $(+,-,\dots,-)$ signature. Applying \Eq{BgM} to $AB$ and using \Eqq{BnAB}, we get
\ba
\B^\g(AB)&=&\lim_{M\to 0}\sum_{n=0}^{+\infty}\frac{\G(\g+1)}{n!\G(\g-n+1)}(M^2)^{\g-n}\B^n(AB)\nn
&=&\lim_{M\to 0}\sum_{n=0}^{+\infty}\sum_{j=0}^{n}\sum_{l=0}^{j}b_{\g,n}^M c_{n,j,l}\big(\B^{n-j}A_l\big)\big(\B^{j-l}B_l\big),
\ea
where we used the shorthand notation $A_l=\prod_{i=1}^{l}\N_{\mu_i}A$ and $B_l=\prod_{i=1}^{l}\N^{\mu_i} B$. Now we apply the general rule for changing the order of a finite and an infinite sum:
\be
\sum_{n=0}^{+\infty}\sum_{j=0}^{n} f_{n,j}=\sum_{j=0}^{+\infty}\sum_{n=0}^{+\infty} f_{n+j,j}\,,
\ee
so that
\ba
\B^{\g}(AB)&=&\lim_{M\to0}\sum_{j=0}^{+\infty}\sum_{n=0}^{+\infty}\sum_{l=0}^{j}b_{\g,n+j}^{M}c_{n+j,j,l}\big(\B^{n}A_l\big)\big(\B^{j-l}B_l\big)\nn
&=&\lim_{M\to0}\sum_{j=0}^{+\infty}\sum_{l=0}^{j}\sum_{n=0}^{+\infty}b_{\g,n+j}^{M}c_{n+j,j,l}\big(\B^{n}A_l\big)\big(\B^{j-l}B_l\big)\nn
&=&\lim_{M\to0}\sum_{j=0}^{+\infty}\sum_{l=0}^{j}\sum_{n=0}^{+\infty}b_{\g-j,n}^{M}c_{\g,j,l}\big(\B^{n}A_l\big)\big(\B^{j-l}B_l\big)\nn
&=&\lim_{M\to0}\sum_{j=0}^{+\infty}\sum_{l=0}^{j}c_{\g,j,l}\big[(M^2+\B)^{\g-j}A_l\big]\big(\B^{j-l}B_l\big)\nn
&=&\sum_{j=0}^{+\infty}\sum_{l=0}^{j}c_{\g,j,l}\big(\B^{\g-l}A_l\big)\big(\B^{j-l}B_l\big)\,,\label{eq:proof}
\ea
which yields \Eqq{BgAB}. In the third line, we took advantage of the relation
\ba
b_{\g,n+j}^{M}c_{n+j,j,l}&=&\frac{\G(\g+1)}{(n+j)!\G(\g-n-j+1)}(M^2)^{\g-n-j}\frac{(n+j)!2^{l}}{n!l!(j-l)!}\nn
&=&\frac{\G(\g+1)}{\G(\g-n-j+1)}(M^2)^{\g-n-j}\frac{2^{l}}{n!l!(j-l)!}\nn
&=&\frac{\G(\g+1)}{n!\G(\g-n-j+1)}(M^2)^{\g-j-n}\frac{2^{l}}{l!(j-l)!}\nn
&=&\frac{\G(\g-j+1)}{n!\G(\g-j-n+1)}(M^2)^{\g-j-n}\frac{\G(\g+1)2^{l}}{\G(\g-j+1)l!(j-l)!}\nn
&=&b_{\g-j,n}^{M}c_{\g,j,l}\,.\label{eq:relationbinomials}
\ea


\section{One-loop effective action in BV formalism}\label{appB}

In section \ref{appB1}, we compute the one-loop effective action and derive the same beta functions \Eq{beta1} and \Eq{beta2} obtained using Feynman diagrams. Sections \ref{appB2}--\ref{appB4} contain some technical intermediate steps to reach this result. In particular, we show that the first two terms in the operator \Eq{hesred} do not contribute to the divergent part of the one-loop effective action. We also discuss the limit $\g\to 1$.


\subsection{One-loop effective action: BV technique}\label{appB1}

The beta function of the quartic coupling turns out to be $\g$-independent. To show this, we consider the action of the theory for $\g>1$ and the terms with the highest number of derivatives, which are the only ones contributing to UV divergences. The leading term in the number of derivatives giving a non-trivial kinetic term is
\be
\cL_{\rm kin}=\frac{1}{2}a_\g\phi^{}\B^{\g}\phi^{},\label{Xkin}
\ee
while the interaction term containing vertices is
\be
\cL_{\rm int}=-\lambda_{\g-1}\phi^{2}\B^{\g-1}\phi^{2}.\label{Xint}
\ee
In the BV formalism, one first computes the Hessian $\Hes_{ij}\coloneqq \de^2 S/\de\Phi_i\de\Phi_j$, where $\Phi_i$ are the fields in the action. In the present case, $\Hes$ is a rank-$0$ operator and $\Phi_i=\phi$. From the kinetic part \Eq{Xkin}, we get the contribution to the Hessian (acting between fluctuations $\delta\phi$) in the simple form
\be\label{heskin}
\Hes_{\rm kin} =a_\g\B^{\g}\,.
\ee
The Hessian from the interaction requires more work. The first-order variation of \Eq{Xint} is
\ben
\delta \cL_{\rm int}=-\lambda_{\g-1}\left[2\phi\delta\phi\B^{\g-1}\phi^{2}+2\phi^{2}\B^{\g-1}\left(\phi\delta\phi\right)\right],
\een
so that the second-order variation is
\be\label{de2X}
\delta^{2}\cL_{\rm int}=-\lambda_{\g-1}\left[2\delta\phi\delta\phi\B^{\g-1}\phi^{2}+4\phi\delta\phi\B^{\g-1}\left(\phi\delta\phi\right)+4\phi\delta\phi\B^{\g-1}\left(\phi\delta\phi\right)+2\phi^{2}\B^{\g-1}\left(\delta\phi\delta\phi\right)\right].
\ee

Thanks to the results of appendix \ref{appA} and the Leibniz rule \Eq{BgAB}, integration by parts is carried in the standard way for the fractional d'Alembertian, $A\B^\g B =B\B^\g A+O(\N)$, where $O(\N)$ is a total derivative. This can also be shown from the formal series \Eq{BgM} or from the Schwinger representation of $\B^\g$ \cite{mf1,Bal60}. Then, after integrating by parts, we find that the first and the last terms in \Eq{de2X} combine and we are left with
\be
\delta^{2}\cL_{\rm int}=-\lambda_{\g-1}\left[4\delta\phi\delta\phi\B^{\g-1}\phi^{2}+8(\phi\delta\phi)\B^{\g-1}\left(\phi\delta\phi\right)\right]+O(\N)\,.\label{sym2var}
\ee
The first term is a total derivative except when $\g=1$; here the $\B^{\g-1}$ operator acts on the square of the background field $\phi$. In the second term, the same operator acts on the product $\phi\delta\phi$. This term is symmetric with respect to the product $\phi\delta\phi$, meaning that its contribution to the Hessian will be manifestly self-adjoint. 

The Hessian is a differential operator acting between two fluctuation fields $\delta\phi$, where $\phi$ factors play the role of constant background coefficients. As gathered from the above power-counting analysis of divergences, we expect these coefficients to appear in the UV regime of the theory only in the form of two simple structures: $\phi\B\phi$ (equivalent by integration by parts to $-\N_\mu\phi\N^\mu\phi$) and $\phi^4$ (without derivatives). This information is important to select relevant terms from \Eqq{sym2var}. In conclusion, having in mind the computation of UV divergences, we focus on terms with at most two derivatives between background fields. Therefore, we must expand the action of the $\B^{\g-1}$ operator in the second term in \Eq{sym2var} and pick terms with up to two derivatives acting on the background field $\phi$. For this purpose, we use the generalized Leibniz rule for the powers of the covariant d'Alembertian derived in section \ref{Leiru2}, in particular, in its truncated form \Eq{BgAB2} which contains all the relevant terms, where $A=\delta\phi$ (on which we can have any number of derivatives) and $B=\phi$ (with at most two derivatives). We get the contribution
\ban
\delta^{2}\cL_{\rm int}&=&-\lambda_{\g-1}\!\left[\vphantom{\frac{1}{2}}4\delta\phi\delta\phi\B^{\g-1}\phi^{2}+8\phi^2\delta\phi\B^{\g-1}\delta\phi+16(\g-1)\phi(\N^\mu\phi)\delta\phi\N_\mu\B^{\g-2}\delta\phi\right.\\
&&\left.+8(\g-1)\phi(\B\phi)\delta\phi\B^{\g-2}\delta\phi+16(\g-1)(\g-2)\phi(\N^{\mu}\N^{\nu}\phi)\delta\phi\N_{\mu}\N_{\nu}\B^{\g-3}\delta\phi\vphantom{\frac{1}{2}}\right]\!+\dots.
\ean
The Hessian from the interaction part \Eq{Xint} is then
\ba
\Hes_{\rm int} &=& -\lambda_{\g-1}\!\left[4(\B^{\g-1}\phi^{2})+8\phi^2\B^{\g-1}+16(\g-1)\phi(\N^\mu\phi)\N_\mu\B^{\g-2}+8(\g-1)\phi(\B\phi)\B^{\g-2}\right.\nn
&&\left.+16(\g-1)(\g-2)\phi(\N^{\mu}\N^{\nu}\phi)\N_{\mu}\N_{\nu}\B^{\g-3}\right],\label{Hint}
\ea
where the brackets in the first term indicate that the operator acts only on the background quantities. As we said above, we keep this term only for the $\g=1$ case. In this formula, we pay special attention to the first two terms, which have no derivatives between the background fields $\phi\phi$ and are the background contribution in the divergent part of the effective action. One sees that the coefficients of these terms are completely independent of the fractional exponent $\g$.

It is reassuring that, since in the UV kinetic term of the theory we have $2\g$ derivatives between fluctuations, then in the relevant part of the interaction Hessian $\Hes_{\rm int}$ we have terms at least with $2\g-4$ derivatives between fluctuations. After effectively dividing by the leading part of the Hessian $\Hes_{\rm kin}$ from the kinetic operator, we will obtain terms which certainly give rise to UV divergences in $D=4$ dimensions, since they have the number of derivatives in the correct range (and, for example, we know that the functional trace of operators like $\B^{-3}$ are convergent in $D=4$ dimensions \cite{Barvinsky:1985an}). We proceed further by expanding and extracting the total Hessian according to
\be
\Hes=\Hes_{\rm kin} + \Hes_{\rm int}=\Hes_{\rm kin}\left(1 + \Hes_{\rm red}\right),
\ee
where we used \Eq{heskin} and defined the reduced operator
\ba
\Hes_{\rm red} &\coloneqq& \Hes_{\rm kin}^{-1}\Hes_{\rm int}\nn
&=&-\frac{8\lambda_{\g-1}}{a_\g}\left[\frac12(\B^{\g-1}\phi^{2})\B^{-\g}+\phi^2\B^{-1}+2(\g-1)\phi(\N^\mu\phi)\N_\mu\B^{-2}\right.\nn
&&\left.\vphantom{\frac12}+(\g-1)\phi(\B\phi)\B^{-2}+2(\g-1)(\g-2)\phi(\N^{\mu}\N^{\nu}\phi)\N_{\mu}\N_{\nu}\B^{-3}\right].\label{Hred}
\ea
Notice that the exponents on operators acting between fluctuations are $\g$-independent. 

Next, we use the standard formula
\be
\Gamma^{(1)}=\frac{\rmi}{2} {\rm Tr} \ln \Hes\,,
\ee
where ${\rm Tr}$ with capital T is the functional trace over position space, spacetime indices and internal field indices, and we perform the expansion of the logarithm into a series separating the unity and its proper argument, that is
\ben
\ln(1+z)=\sum_{j=1}^{+\infty}(-1)^{j-1}\frac{z^j}{j}=z-\frac{z^2}{2}+O(z^3)\,.
\een
We need to truncate this expansion at the second order in the argument $z=\Hes_{\rm red}$. This is because we already have terms quadratic in the background scalar field in the reduced Hessian operator $\Hes_{\rm red}$. Expected divergences are at most quartic in the background field $\phi$, therefore, we need at most the $O(z^2)$ term while we can neglect terms of order of $z^3$ and higher. Then, for $\g\neq1$,
\ba
\ln \Hes &=&  \ln\Hes_{\rm kin}+\ln\left(1 + \Hes_{\rm red}\right)\nn
&=&\Hes_{\rm red}-\frac12\Hes_{\rm red}^2+\dots\nn
&=&\ln(a_\g\B^\g)-\frac{4\lambda_{\g-1}}{a_\g}(\B^{\g-1}\phi^{2})\B^{-\g}-\frac{8\lambda_{\g-1}}{a_\g}\left[\phi^2\B^2+2(\g-1)\phi(\N^\mu\phi)\N_\mu\B\right.\nn
&&\left.+(\g-1)\phi(\B\phi)\B+2(\g-1)(\g-2)\phi(\N^{\mu}\N^{\nu}\phi)\N_{\mu}\N_{\nu}\right]\B^{-3}\nn
&&-32\left(\frac{\lambda_{\g-1}}{a_{\g}}\right)^{2}\phi^{4}\B^{-2}+\dots\,,\label{hesred}
\ea
where we neglected terms not contributing to UV divergences. The first term $\sim\ln\B^\g$ can be thrown away because its trace is proportional to powers of the curvature and, therefore, it does not contribute to the divergent part in Minkowski spacetime (section \ref{appB2}). As argued in section \ref{appB3}, the second term does not contribute either, since $\g\neq 1$.

Performing all functional traces explicitly,
\ba
\left.{\rm Tr}\ln\Hes\right|_{{\rm div}}&=&-\frac{2}{\pi^{2}}\left(\frac{\lambda_{\g-1}}{a_{\g}}\right)^{2}\phi^{4}-\frac{\g-1}{2\pi^{2}}\frac{\lambda_{\g-1}}{a_{\g}}\phi\B\phi\nn
&&-\frac{1}{\pi^{2}}\frac{\lambda_{\g-1}}{a_{\g}}(\g-1)(\g-2)\phi\N^{\mu}\N^{\nu}\phi\,{\rm tr}\left(\N_{\mu}\N_{\nu}\B^{-3}\mathbbm{1}\right),
\ea
where ${\rm tr}$ is the trace only on Lorentz and internal field indices and we used \cite{Barvinsky:1985an}
\be
 {\rm tr}(\B^{-2}\mathbbm{1}) = 1\,, \qquad {\rm tr}(\N^{\mu}\mathbbm{1}) = 0\,.
\ee
Notice that the functional trace of the differential operator $\N^\mu\B^{-2}$ vanishes, since in $D=4$ dimensions there are no background quantities (such as curvature tensors) with which one could construct a non-zero expression for such a trace with energy dimensionality 1.\footnote{Here we cannot use the background scalar field $\phi$ since we assume the $\mathbb{Z}_2$ symmetry of the theory.}

We also integrate by parts under the spacetime volume trace $\int \rmd^4 x$ to produce the effective action. In this way, up to total derivatives we get
\ba
\left.{\rm Tr}\ln\Hes\right|_{{\rm div}}&=&-\frac{2}{\pi^{2}}\left(\frac{\lambda_{\g-1}}{a_{\g}}\right)^{2}\phi^{4}-\frac{\g-1}{2\pi^{2}}\frac{\lambda_{\g-1}}{a_{\g}}\phi\B\phi-\frac{1}{\pi^{2}}\frac{\lambda_{\g-1}}{a_{\g}}(\g-1)(\g-2)\phi\N_{\mu}\N_{\nu}\phi\frac{1}{4}\eta^{\mu\nu}\nn
&=&-\frac{2}{\pi^{2}}\left(\frac{\lambda_{\g-1}}{a_{\g}}\right)^{2}\phi^{4}-\frac{\g-1}{2\pi^{2}}\frac{\lambda_{\g-1}}{a_{\g}}\phi\B\phi-\frac{1}{4\pi^{2}}\frac{\lambda_{\g-1}}{a_{\g}}(\g-1)(\g-2)\phi\B\phi\nn
&=&-\frac{2}{\pi^{2}}\left(\frac{\lambda_{\g-1}}{a_{\g}}\right)^{2}\phi^{4}-\frac{1}{4\pi^{2}}\frac{\lambda_{\g-1}}{a_{\g}}\g(\g-1)\phi\B\phi\,.
\ea
Compared with the action \Eq{SUVaction}, which is the integer part of the total action \Eq{eq:model}, we obtain 
\ba
\beta_1&=&-\frac{\g(\g-1)}{2\pi^2}\frac{\la_{\g-1}}{a_{\g}}\,,\label{beta1app}\\
\beta_{\la_0}&=&\frac{48}{\pi^{2}}\left(\frac{\la_{\g-1}}{a_{\g}}\right)^2,\label{beta2app}
\ea
in perfect agreement with the results \Eq{beta1} and \Eq{beta2} obtained via the calculation with Feynman diagrams.

The first expression depends on the $\g$ parameter but it vanishes continuously in the limit $\g\to1$, giving no wave-function renormalization in the case of standard $\lambda\phi^{4}$ theory. This limiting case is further discussed in section \ref{appB4}.


\subsection{Traces of fractional operators: first term in \Eqq{hesred}}\label{appB2}

Given the function
\be
f(z)=\ln\left[(1+z)^{\g}\right]\,,
\ee
its Taylor expansion around the point $z=0$ is
\be
f(z)=\sum_{n=1}^{+\infty}\frac{f^{(n)}(0)}{n!}z^{n}\,,
\ee
where we used $f(0)=0$. The trace is
\be
{\rm Tr}\,f(z)=\sum_{n=1}^{+\infty}{\rm Tr}\left[\frac{f^{(n)}(0)}{n!}z^{n}\right]\,.
\ee
We find that
\ban
f^{(1)}(z)&=&\frac{\rmd}{\rmd z}\ln\left[(1+z)^{\g}\right]=\left[(1+z)^{\g}\right]^{-1}\frac{\rmd}{\rmd z}(1+z)^{\g}\\
&=&\left[(1+z)^{\g}\right]^{-1}\g(1+z)^{\g-1}=\frac{1}{(1+z)^{\g}}\g(1+z)^{\g-1}=\frac{\g}{1+z}\,,
\ean
which yields under the trace
\ben
{\rm Tr}\left[f^{(1)}(z)z\right]=\g\,{\rm Tr}\left(\frac{z}{1+z}\right)\,.
\een
In general, we cannot assume that $z$ and $(1+z)^{-1}$ commute, but since only these two factors appear under the trace (and in the same
power $n$), then they effectively do, so that there is no problem with the ordering and the meaning of terms such as this.

For the second derivative, we have
\ben
f^{(2)}(z)=\frac{\rmd}{\rmd z}\left(\frac{\g}{1+z}\right)=-\g\frac{1}{(1+z)^2}\,,
\een
so that
\ben
{\rm Tr}\left[f^{(2)}(z)z^{2}\right]=-\g\,{\rm Tr}\frac{z^{2}}{(1+z)^{2}}\,.
\een
For higher derivatives, we find similarly that
\be
{\rm Tr}\left[f^{(n)}(z)z^{n}\right]=\g(-1)^{n+1}(n-1)!\,{\rm Tr}\frac{z^{n}}{(1+z)^{n}}\,,
\ee
hence
\be
{\rm Tr}\left[f^{(n)}(0)z^{n}\right]=\g(-1)^{n+1}(n-1)!\,{\rm Tr}z^{n}.
\ee
This implies that
\be
{\rm Tr}\,\ln\left[(1+z)^{\g}\right]=\sum_{n=1}^{+\infty}\g(-1)^{n+1}\frac{(n-1)!}{n!}{\rm Tr}\,z^{n}=\g\,{\rm Tr}\sum_{n=1}^{+\infty}\frac{(-1)^{n+1}}{n}z^{n}=\g\,{\rm Tr}\ln(1+z)
\ee
because of the linearity of the trace, for any real exponent $\g$ and for any object $1+z$. The latter could be a differential operator, a matrix, and so on, of which we do not have to assume that $\left[z,(1+z)^{-1}\right]=0$. This implies, in particular, that
\be
{\rm Tr}\ln\B^{\g}=\g\,{\rm Tr}\ln\B
\ee
for the identification $1+z=\B$. Since we know from the functional traces tabulated in \cite{Barvinsky:1985an} that the result of ${\rm Tr}\ln\B$ is proportional to the background gravitational (or internal space) curvatures and it does not contain any monomials of the background scalar field $\phi$, nor of its derivatives, then we can neglect this contribution as far as the two types of UV divergences we are looking for ($\phi\B\phi$ and $\phi^4$) are concerned. In our computation, in fact, the curvature is zero since we consider the theory around flat Minkowski spacetime.


\subsection{Traces of fractional operators: second term in \Eqq{hesred}}\label{appB3}

As for the term $(\B^{\g-1}\phi^{2}) \B^{-\g}$, one notices that it is a total derivative for $\g>1$, hence for non-integer $\g$ we neglect it in the UV-divergent part of the effective action. The functional traces of $\B^{-\g}$ are non-zero only for integer $\g=0,1,2$. However, when $\g=2$ the background quantity $\B^{\g-1}\phi^{2} = \B\phi^{2}$ is, again, a total derivative. When $\g=0,1$, these terms are expressed via powers of spacetime (or internal) curvature, that we set to zero at the end of our computation. 
 
When we take the square of this term, we get four powers of the background field $\phi$, so that the divergence is only of the form $\phi^4$ with no derivatives at all. Higher powers of the term $(\B^{\g-1}\phi^{2}) \B^{-\g}$ produce higher powers in $\phi$, which are convergent. Focusing on the term $(\B^{\g-1}\phi^{2})\B^{-\g}[(\B^{\g-1}\phi^{2})\B^{-\g}]$ and using the Leibniz expansion \Eq{BgAB} of the d'Alembertian operator, we find that only the first term of the infinite expansion yields a background result with no derivatives acting on the field $\phi$, provided $\g=1$:
\be
(\B^{\g-1}\phi^{2})\B^{-\g}[(\B^{\g-1}\phi^{2})\B^{-\g}]\Big|_{\g=1}=
\phi^{2}\B^{-1}[(\phi^{2})\B^{-1}]=\phi^{4}\B^{-2}+O\left(\phi^2\N\phi^2\right),
\ee
where we omit the last term containing at least one derivative or one power of the d'Alem\-bert\-ian operator acting on background fields, when the total power of the fields $\phi$ is four. Therefore, from this term, we get a contribution only in the special case of $\g=1$ and it reads $\phi^4{\rm Tr}\B^{-2} \propto\phi^4$, where we used
\be\label{trB2}
{\rm Tr}\B^{-2}=\frac{1}{16\pi^2}\,.
\ee


\subsection{Discontinuity at \texorpdfstring{$\g=1$}{gamma=1}}\label{appB4}

Let us recall the standard computation in $\la\phi^{4}$ theory with the BV technique. From the Lagrangian
\be
\cL=\frac{1}{2}\phi\B\phi-\frac{\la_0}{4!}\phi^{4}=\cL_{\rm kin}+\cL_{\rm int}\,,
\ee
we have
\ban
\delta^{2}\cL_{\rm int}&=&-\frac{\la_0}{4!}12\phi^{2}\delta\phi\delta\phi=-\frac{\la_0}{2}\phi^{2}\delta\phi\delta\phi\,,\\
\Hes&=&\B-\frac{\la_0}{2}\phi^{2}=\B\left(1-\frac{\la_0}{2}\phi^{2}\B^{-1}\right),\\
\ln\Hes &=&-\frac{\la_0}{2}\phi^{2}\B^{-1}-\frac{1}{2}\left(-\frac{\la_0}{2}\phi^{2}\B^{-1}\right)^{2}+\dots\\
				&=&-\frac{\la_0}{2}\phi^{2}\B^{-1}-\frac{\la_0^{2}}{8}\phi^{4}\B^{-2}+\dots\,,
\ean				
hence				
\ben
{\rm Tr}\ln\Hes \stackrel{\text{\tiny\Eq{trB2}}}{=} -\frac{\la_0}{2}\phi^{2}{\rm Tr}\B^{-1}-\frac{1}{16\pi^{2}}\frac{\la_0^{2}}{8}\phi^{4}\,,
\een
which yields the beta functions
\ba
\beta_1 &=& 0\,,\\
\beta\left(-\frac{\la_0}{4!}\right) &=&-\frac{1}{16\pi^{2}}\frac{\la_0^{2}}{8}\qquad\Longrightarrow\qquad
\beta_{\la_0}=\frac{1}{16\pi^{2}}24\frac{\la_0^{2}}{8}=\frac{3\la_0^{2}}{16\pi^{2}}\,.
\ea

As pointed out in the main text, the case $\g=1$ is special among all the others with non-integer $\g$. Here we highlight the origin of this discontinuity in the limiting situation of $\g\to1$. In the general case $\g\neq1$, the only term of the reduced Hessian $\Hes_{\rm red}$ (\Eqq{Hred}) with no derivatives between background fields $\phi$ is $8\phi^2\B^{-1}$ (giving rise \emph{a posteriori} to divergent terms in the effective action quartic in the background $\phi$ and with no derivatives). In contrast, in the exceptional case $\g=1$ we need to include also the first term in \Eqq{Hred}, which reads $4\phi^2\B^{-1}$ and is not a total derivative. Overall, this gives the contribution $12\phi^2\B^{-1}$, which is $3/2$ times larger than in the $\g\neq 1$ case and produces a beta function $\beta_{\la_{0}}$ $(3/2)^2=9/4$ times larger than for other $\g>1$. This jump of value cannot be obtained continuously from the general $\g$-independent expression for any $\g\neq1$. This amazing and unexpected feature, probably persisting at higher loop levels, shows once again the uniqueness of the standard two-derivative theory with $\g=1$, which, as it seems, cannot be obtained as the continuous limit of the quantum theory with general $\g\neq1$. 

Still in the special case $\g=1$, one notices that the other parts of the reduced Hessian operator cancel in this situation, resulting only in the already discussed part with no derivatives between background fields:
\ben
\Hes_{\rm red,\,\g=1}= -12\la_0\phi^2\B^{-1}\,.
\een
This shows that there is no scalar wave-function renormalization when $\g=1$. With hindsight, we can say that here this absence is caused by the fact that the part of Hessian in \Eqq{Hred} containing derivatives between background fields is proportional to the factor $(\g-1)$, so that in this case the limit $\g\to1$ is continuous and vanishing. Moreover, this implies that the general value of this divergent contribution for the case of $\g>1$ is proportional to the factor $(\g-1)$ and this is verified by an explicit computation, where the full coefficient amounts to $g(\g-1)$ for the renormalization of the scalar wave-function. 


\section{Scalar propagator}\label{appSCA}

This appendix collects some results on the free-level propagator of a scalar field, both in fractional non-local and in higher-derivative local theories.


\subsection{Källén--Lehmann representation \Eq{kalegen2}}\label{appKL}

Starting from Cauchy's integral formula \Eq{opt} for the scalar-field propagator, we deform the contour $\tilde\G$ to circle around poles and pass around the cuts and split it into its main components: integrations $\int_{\rm poles}$ along mini-circles of radius $\ve>0$ around the poles, integrations $\int_{\rm b.p.}$ along mini-arcs of radius $\ve>0$ around the branch points, integrations $\int_{\rm cuts}$ along the branch cuts and an integration $\int_{\G_R}$ along a counter-clockwise arc $\G_R$ of radius $R$ closing the contour. Sending $R\to +\infty$, the contribution of the outer arc is zero if $\tilde G(z)$ falls off at $z\to\infty$, which will always be the case in our models (but we will check it explicitly later). Then, sending $\ve\to0^+$ one obtains
\be\label{ointsplit}
\oint_{\tilde\G}=\lim_{\ve\to0^+}\left(\int_{\rm poles}+\int_{\rm b.p.}+\int_{\rm cuts}\right)+\lim_{R\to+\infty}\int_{\G_R}.
\ee
Note that the integral on the cuts also depends on $\ve$ because one evaluates the Green's function $\tilde G(s\pm\rmi\ve)$ on lines parallel and infinitesimally close to the real axis. 

The integral on $\G_R$ can be parametrized by a phase. Calling $s=m^2$ the pole or branch point and $z=m^2+R\,\exp(\rmi\t)$, we have
\be\label{gR}
\frac{1}{2\pi\rmi}\int_{\G_R}\rmd z\,\frac{\tilde G(z)}{z+k^2} = \frac{1}{2\pi}\int_{\t_1}^{\t_2}\rmd\t\,\tilde G(m^2+R\,\rme^{\rmi\t})+\dots\,,
\ee
where $\t_1\leq\t\leq\t_2$ is a certain range for $\t$ and we omitted higher-order terms in $1/R$. In the limit $R\to +\infty$, this contribution should vanish.

Similarly, the integrals around the poles or the branch points can be parametrized as $z=m^2+\ve\,\exp(\rmi\t)$ and these contributions acquire the form
\ba
\frac{1}{2\pi\rmi}\int_{\rm pole/b.p.}\rmd z\,\frac{\tilde G(z)}{z+k^2} &=& \frac{\ve}{2\pi(k^2+m^2)}\int_{\t_1}^{\t_2}\rmd\t\,\rme^{\rmi\t}\tilde G(m^2+\ve\,\rme^{\rmi\t})\sum_{l=0}^{+\infty}(-1)^l\left(\frac{\ve\,\rme^{\rmi\t}}{k^2+m^2}\right)^l\nn
&=&\frac{1}{2\pi}\sum_{l=0}^{+\infty}(-1)^l\frac{\ve^{1+l}}{(k^2+m^2)^{1+l}}\int_{\t_1}^{\t_2}\rmd\t\,\rme^{\rmi\t(1+l)}\tilde G(m^2+\ve\,\rme^{\rmi\t})\label{ceps}\,,
\ea
where we used the generalized binomial series. In the limit $\ve\to 0$, when around a pole this integral gives a finite contribution (the residue). However, when around a branch point, it can either vanish or diverge depending on the form of $\tilde G$ (in our case, on the value of the fractional exponent $\g$). 

Then, in a well-defined theory the only non-vanishing contribution in \Eq{ointsplit} comes from the poles and the branch cuts and gives rise to the Källén--Lehmann representation
\be\label{kalegen}
\tilde G(-k^2)=\frac{1}{2\pi\rmi}\int_{\rm poles}\rmd s\,\frac{\tilde G(s)}{s+k^2}+\frac{1}{2\pi\rmi}\int_{\rm cuts}\rmd s\,\frac{\tilde G(s)}{s+k^2}\,.
\ee
In particular, for a massive theory with one branch cut, we obtain \Eq{kalegen2}.


\subsection{Feynman prescription of the fractional propagator}\label{appFEY}

In this section, we calculate the propagator \Eq{optF} with the causal (Feynman) prescription. From \Eqqs{gR} and \Eq{Gz}, we have
\ba
\frac{1}{2\pi\rmi}\int_{\G_R}\rmd z\,\frac{\tilde G(z)}{z+k^2} &=& \frac{1}{2\pi}\int_{\t_1}^{\t_2}\rmd\t\,(-R\,\rme^{\rmi\t})^{-\g}+\dots\nn
&=& \frac{1}{2\pi R^{\g}}\,\rme^{-\rmi\pi\g}\int_{\t_1}^{\t_2}\rmd\t\,\rme^{-\rmi\t\g}+O\left(\frac{1}{R^{1+\g}}\right)\nn
&=& \frac{1}{\g R^{\g}}\,\rme^{-\rmi\pi\g}\frac{\rme^{-\rmi\t_1\g}-\rme^{-\rmi\t_2\g}}{2\pi\rmi}+O\left(\frac{1}{R^{1+\g}}\right),\label{gR2}
\ea
where the sign in the phase $(-1)^{-\g}=\exp(-\rmi\pi\g)$ is determined by the choice of branch for the integrand; with $0\leq\t\leq 2\pi$, $-1=\exp(\rmi\pi)$. Therefore, for $\g>0$
the integral on the arc at infinity vanishes in the limit $R\to+\infty$, irrespectively of the values of $\t_1$ and $\t_2$. The condition \Eq{g0} coming from the arc remained unnoticed in \cite{mf1} but, anyway, it is the same as the condition of positivity of the spectral dimension.

The other contributions in \Eqq{ointsplit} are
\be\label{ointsplitF}
\oint_{\G_{\rm o}}=\lim_{\ve\to0^+}\left(\int_{C_\ve}+\int_{\rm cut^+}+\int_{\rm cut^-}\right).
\ee
The radius $\ve$ of the clockwise mini-arc $C_\ve$ is different from the $\e$ appearing in \Eqq{optF} and it can be taken to be $\ve<\e$ so that $-k^2$ never lies on the contour. Using \Eqq{Gz}, the general formula \Eq{ceps} can be specialized to a branch point. Ignoring the $\rmi\e$ prescription for $-k^2$, for $\g\neq n$ we need to consider only the $l=0$ term:
\ba
\frac{1}{2\pi\rmi}\int_{\rm b.p.}\rmd z\,\frac{\tilde G(z)}{z+k^2} &=& \frac{\ve}{m^2+k^2}\int_{\t_1}^{\t_2}\frac{\rmd\t}{2\pi}\,\rme^{\rmi\t}(-\ve\,\rme^{\rmi\t})^{-\g}+O(\ve^{2-\g})\nn
&=& \frac{\ve^{1-\g}}{m^2+k^2}\,\rme^{-\rmi\pi\g}\int_{\t_1}^{\t_2}\frac{\rmd\t}{2\pi}\,\rme^{\rmi\t(1-\g)}+O(\ve^{2-\g})\nn
&=& \frac{\ve^{1-\g}}{(1-\g)(m^2+k^2)}\,\rme^{-\rmi\pi\g}\frac{\rme^{\rmi\t_2(1-\g)}-\rme^{\rmi\t_1(1-\g)}}{2\pi\rmi}+O(\ve^{2-\g})\label{ceps2}\,.
\ea
In the case of $C_\ve$, $\t$ runs from $\t_1=2\pi$ to $\t_2=0$, so that
\ba
\tilde G_{\rm b.p.}(-k^2) &=& \lim_{\ve\to 0^+} \frac{1}{2\pi\rmi}\int_{C_\ve}\rmd z\,\frac{\tilde G(z)}{k^2+z-\rmi\e}\nn
&=& \lim_{\ve\to 0^+}\frac{\ve^{1-\g}}{(1-\g)(m^2+k^2-\rmi\e)}\,\rme^{-\rmi\pi\g}\left.\frac{\rme^{\rmi\t(1-\g)}}{2\pi\rmi}\right|_{\t_1=2\pi}^{\t_2=0}+O(\ve^{2-\g})\nn
&=& \lim_{\ve\to 0^+}\frac{\ve^{1-\g}}{(1-\g)(m^2+k^2-\rmi\e)}\,\rme^{-\rmi\pi\g}\frac{1-\rme^{-2\rmi\pi\g}}{2\pi\rmi}\nn
&=& \lim_{\ve\to 0^+} \frac{\ve^{1-\g}}{(1-\g)(m^2+k^2-\rmi\e)}\,\rme^{-2\rmi\pi\g}\frac{\sin(\pi\g)}{\pi}\label{ceps3}\,,
\ea
which vanishes only if $\g<1$. Obviously, the result does not change if we evaluate \Eq{ceps3} on a different branch, for instance the family of ``$+$ branches'' $2(1+j)\pi\leq\t\leq 2(2+j)\pi$ or the family of ``$-$ branches'' $-2(1+j)\pi\leq\t\leq -2j\pi$, where $j=0,1,2,\dots$. In the first case, $-1=\exp[(3+2j)\rmi\pi]$ and the phase in the last line of \Eq{ceps3} becomes $\exp[-2(3+2j)\rmi\pi\g]$; the principal branch corresponds to $j=-1$. In the second case, $-1=\exp[-(1+2j)\rmi\pi]$ and the phase in the last line of \Eq{ceps3} is replaced by $-\exp[2(1+2j)\rmi\pi\g]$.

The parameter $\ve$ also appears in the contribution of the branch cut through the integrand
\ba
\tilde G(s\pm\rmi\ve) &\stackrel{\text{\tiny\Eq{Gz}}}{=}& (m^2-s\mp\rmi\ve)^{-\g}\nn
&=& \exp\left[-\g\,{\rm Ln}(m^2-s\mp\rmi\ve)\right]\nn
&=& \exp\left[-\g\,\ln\sqrt{(s-m^2)^2+\ve^2}-\rmi\g\,{\rm Arg}(m^2-s\mp\rmi\ve)\right]\nn
&=& \frac{1}{[(s-m^2)^2+\ve^2]^{\frac{\g}{2}}}\exp\left[-\rmi\g\,{\rm Arg}(m^2-s\mp\rmi\ve)\right],\label{Gplusmin}
\ea
where Ln and Arg are, respectively, the principal value of the complex logarithm and of the argument (phase). Given a complex number $x+\rmi y$, for $x\neq 0$ the principal value of the argument can be expressed as
\be\label{argxy}
{\rm Arg}(x + \rmi y) =
\left\{\begin{matrix}
  \arctan\left(\frac{y}{x}\right)\hphantom{+\pi}\qquad \text{if }\, x > 0\,,\hphantom{\, y \geq 0\,,} \\
  \arctan\left(\frac{y}{x}\right) + \pi\qquad \text{if }\, x < 0 \,,\, y \geq 0\,, \\
  \arctan\left(\frac{y}{x}\right) - \pi\qquad \text{if }\, x < 0 \,,\, y < 0\,.
\end{matrix}\right.
\ee
In the case of \Eqq{Gplusmin}, $x=m^2-s<0$ and $y=\mp\ve\lessgtr 0$.\footnote{Although $\tilde G=(m^2-z)^{-\g}$ has a branch cut at $\Re\,z\geq m^2$ in the $(\Re\,z,\Im\,z)$ plane, when written as a function $\tilde G=(x+\rmi y)^{-\g}$ in the $(x=m^2-s,y=\mp\ve)$ plane it has a branch cut at $x<0$. Hence $-\pi<{\rm Arg}(x+\rmi y)\leq\pi$ and Arg is correctly parametrized by \Eqq{argxy}. When $y\to 0^-$ and $x<0$ (approach to the negative real axis from below), ${\rm Arg}\to-\pi$; when $x\to 0^\pm$ and $y<0$ (approach to the negative imaginary axis from either side), ${\rm Arg}\to-\pi/2$; when $x>0$ and $y\to 0^\pm$ (approach to the positive real axis from either side), ${\rm Arg}\to 0$; when $x\to 0^\pm$ and $y>0$ (approach to the positive imaginary axis from either side), ${\rm Arg}\to\pi/2$; when $y\to 0^+$ and $x<0$ (approach to the negative real axis from above), ${\rm Arg}\to\pi$.} Therefore,
\ba
\tilde G(s\pm\rmi\ve)&=& \frac{1}{[(s-m^2)^2+\ve^2]^{\frac{\g}{2}}}\exp\left[-\rmi\g\,\arctan\left(\frac{\mp\ve}{s-m^2}\right)\pm\rmi\pi\g\right]\nn
&=& \frac{1}{[(s-m^2)^2+\ve^2]^{\frac{\g}{2}}}\,\rme^{\pm\rmi\pi\g}\left(1\pm\frac{\rmi\g\ve}{s-m^2}\right)+O(\ve^2)\,.\label{Gplusmin3ve}
\ea
For non-integer $\g$, 
\be\label{Gplusmin3}
\lim_{\ve\to 0^+} \tilde G(s\pm\rmi\ve) =\frac{1}{|s-m^2|^\g}\,\rme^{\pm\rmi\pi\g}.
\ee
Then, the last two integrals in \Eq{ointsplitF} are
\ba
\tilde G_{\rm cut^\pm}(-k^2)&=&\frac{1}{2\pi\rmi}\int_{\rm cut^\pm}\rmd s\,\frac{\tilde G(s)}{s+k^2-\rmi\e}\nn
&=& \pm\lim_{\ve\to 0^+}\frac{1}{2\pi\rmi}\int_{m^2}^{+\infty}\rmd s\,\frac{\tilde G(s\pm\rmi\ve)}{s+k^2-\rmi\e}\nn
&\stackrel{\text{\tiny\Eq{Gplusmin3}}}{=}& \pm\frac{1}{2\pi\rmi}\int_{m^2}^{+\infty}\rmd s\,\frac{\rme^{\pm\rmi\pi\g}}{|s-m^2|^\g}\frac{1}{s+k^2-\rmi\e}\nn
&=&\int_{m^2}^{+\infty}\rmd s\,\frac{\rho_\pm(s)}{s+k^2-\rmi\e},\label{Gplus}
\ea
where we defined the complex spectral densities
\be\label{rhopm}
\rho_\pm(s)\coloneqq \pm\frac{1}{2\pi\rmi}\frac{\rme^{\pm\rmi\pi\g}}{|s-m^2|^\g}\,.
\ee
Overall,
\ba
\tilde G_{\rm F}(-k^2) &=&\tilde G_{\rm cut}(-k^2)=\tilde G_{\rm cut^+}(-k^2)+\tilde G_{\rm cut^-}(-k^2)\nn
&=& \int_{m^2}^{+\infty}\rmd s\,\frac{\rho_{\rm F}(s)}{s+k^2-\rmi\e},\label{GFeyapp}
\ea
where
\ba
\rho_{\rm F}(s) &=& \rho_+(s)+\rho_-(s)\nn
&=& \frac{1}{|s-m^2|^\g}\frac{\rme^{\rmi\pi\g}-\rme^{-\rmi\pi\g}}{2\pi\rmi}\nn
&=& \frac{1}{|s-m^2|^\g}\frac{\sin(\pi\g)}{\pi}\,.\label{rhoFeyapp}
\ea


\subsection{Propagator in local models (integer \texorpdfstring{$\g$}{gamma})}\label{appLOC}

In this section, we recall some properties of the scalar-field free-level propagator with the Feynman, Dyson and Anselmi--Piva prescriptions in higher-derivative models with $\g=n$, $1\leq n\in\mathbb{N}$, which include standard QFTs with $\g=1$.

\subsubsection{Feynman and Dyson prescriptions}

In these cases, the tree-level propagator is given exclusively by the contribution of the pole, since $\tilde G$ has no discontinuity on the real line and, therefore, the spectral function \Eq{rhoFey} vanishes: $\rho_{\rm F}=0$.

Around the pole at $z=m^2$, we use the full series \Eq{ceps2} and integrate $\tilde G(z)=(m^2-z)^{-n}$ full circle, isolating the $O(\ve^0)$ term ($l=n-1$):
\ba
\frac{1}{2\pi\rmi}\int_{\rm pole}\rmd z\,\frac{\tilde G(z)}{z+k^2} &=&\frac{1}{2\pi}\sum_{l=0}^{+\infty}(-1)^l\frac{\ve^{1+l}}{(k^2+m^2)^{1+l}}\int_{2\pi}^{0}\rmd\t\,\rme^{\rmi\t(1+l)}(-\ve\,\rme^{\rmi\t})^{-n}\nn
&=&\frac{1}{2\pi}\sum_{l=0}^{+\infty}(-1)^{l-n}\frac{\ve^{1+l-n}}{(k^2+m^2)^{1+l}}\int_{2\pi}^{0}\rmd\t\,\rme^{\rmi\t(1+l-n)}\nn
&=&\sum_{l=0}^{n-2}(-1)^{l-n}\frac{\ve^{1+l-n}}{(k^2+m^2)^{1+l}}\frac{1-\rme^{2\rmi\pi(1+l-n)}}{2\pi\rmi(1+l-n)}+\frac{1}{(k^2+m^2)^n}+O(\ve)\nn
&=&\frac{1}{(k^2+m^2)^n}+O(\ve)\label{cepsn}\,.
\ea
Therefore, with the Feynman or Dyson prescription
\be
\tilde G_{\rm F,D}(-k^2) = \lim_{\ve\to 0^+}\frac{1}{2\pi\rmi}\int_{C_\ve}\rmd z\,\frac{\tilde G(z)}{k^2+z\mp\rmi\e}=\frac{1}{(k^2+m^2\mp\rmi\e)^n}\,.
\ee

\subsubsection{Anselmi--Piva prescription}

Consider the scalar-field Lagrangian
\be\label{Lquad}
\cL=\frac12\,\phi\left(\B+\ell_*^2\B^2\right)\phi\,,
\ee
where $\ell_*$ is a length scale. Upon the decomposition $\phi=\phi_1+\phi_2$, with the local field redefinitions
\be
\phi_1=\left(1+\ell_*^2\B\right)\phi\,,\qquad \phi_2=-\ell_*^2\B\phi\,,\label{Lneven0}
\ee
\Eqq{Lquad} is recast as the two-field Lagrangian
\be
\cL=\frac12\,\phi_1\B\phi_1-\frac12\,\phi_2\left(\ell_*^{-2}+\B\right)\phi_2\,.
\ee
There are two modes, a standard particle field $\phi_1$ and a massive ghost $\phi_2$. Applying the Anselmi--Piva procedure exclusively on $\phi_2$ ($A=B=-1$ and $m^2=\ell_*^{-2}$ in \Eqq{plemsok}), one can make the theory unitary.

The general case with higher-order derivatives is more complicated but one can still diagonalize the kinetic term and single out the unstable modes. According to the Ostrogradsky theorem for a scalar field theory, a kinetic term with $2n$ derivatives corresponds to $n$ degrees of freedom, some of which are stable while the others are ghosts \cite{Asorey:1996hz,Ost50,PU,EW1,Woo15}. Here, however, we take a different perspective adapted to the physical problem at hand and content ourselves to separate stable modes dominating in the IR from modes dominating in the UV. Generally, in a field theory with fields of arbitrary rank, UV modes can be either stable or unstable. For unitarity of the theory, we must ensure that the unstable modes are absent from the beginning or that a special procedure be employed to remove all of them.

In the particular example of a scalar, generalizing the  above, the higher-derivative system
\be\label{Lquadn}
\cL=\frac12\,\phi\left(\B+\ell_*^{2n-2}\B^n\right)\phi
\ee
can be rewritten (up to total derivatives) for even $n\geq 2$ as
\ba
\cL &=& \frac12\,\phi_1\B\phi_1-\frac12\,\phi_2\left(\ell_*^{-2}+\ell_*^{2n-4}\B^{n-1}\right)\phi_2\,,\label{Lneven1}\\
\phi_1 &=& \left(1+\ell_*^{2n-2}\B^{n-1}\right)\phi\,,\qquad \phi_2=-\ell_*^{n}\B^{\frac{n}{2}}\phi\,,\label{Lneven2}
\ea
while for odd $n\geq 3$ such that $\lceil n\rceil$ is divisible by 4 as
\ba
\cL &=& \frac12\,\phi_1\B\phi_1-\frac12\,\phi_2\left(\ell_*^{-2}+\ell_*^{n-3}\B^{\frac{n-1}{2}}\right)\phi_2\,,\\
\phi_1 &=& \left[1+(1+\sqrt{2})\ell_*^{n-1}\B^{\frac{n-1}{2}}\right]\phi\,,\qquad \phi_2=-\sqrt{2+2\sqrt{2}}\,\ell_*^{\frac{n+1}{2}}\B^{\frac{n+1}{4}}\phi\,.\label{Lneven3}
\ea
Actually, for the definitions of $\phi_2$ in \Eq{Lneven2} and \Eq{Lneven3} we have a possibility of using a different overall sign. Above, we used two different formul\ae\ for $n$ even and odd since, in local higher-derivative field theory, one wants to avoid non-integer powers of the box operator. Moreover, contrary to the $n=2$ case \Eq{Lneven0}, for $n>2$ the definitions of $\phi_1$ and $\phi_2$ in \Eqqs{Lneven2} and \Eq{Lneven3} do not correspond to any simple decomposition of the original field $\phi$.

The spectrum is obtained treating both $\phi_1$ and $\phi_2$ as fundamental fields, regardless of what and how many degrees of freedom they encode in terms of the original field $\phi$. The field $\phi_1$ is canonical with a standard kinetic term. Even if the kinetic term of $\phi_2$ is not in the canonical form, one can apply the Anselmi--Piva procedure to $\phi_2$ and making it a fakeon, thus removing all ghosts from the spectrum. 


\section{Properties of the Källén--Lehmann representation}\label{appC}

In this appendix, we report the calculation of \Eqqs{optNfinal}--\Eq{psi2} and the full catalogue of splittings of the K\"allén--Lehmann representation of the scalar free-level propagator.


\subsection{Calculation of \Eqq{optNfinal}}\label{appGN}

To find the propagator, we integrate \Eq{GN} on a contour $\tilde\G=\G_N$ decomposable into several pieces:
\be
\tilde G_N(-k^2) = \oint_{\G_N}\rmd z\,\frac{\tilde G_N(z)}{z+k^2}\,,\label{optN}
\ee
where
\bs\label{ointsplitN}\ba
\oint_{\G_N}&=&\lim_{\ve\to0^+}\left(\int_{\rm b.p.}+\int_{\rm cuts}+\lim_{R\to+\infty}\int_{\rm arcs}\right),\\
\int_{\rm b.p.} &=& \sum_{i=1}^N\int_{C_\ve^i}\,,\\
\int_{\rm cuts} &=& \sum_{i=1}^N\int_{{\rm cut}_i} = \sum_{i=1}^N\left(\int_{{\rm cut}_i^+}+\int_{{\rm cut}_i^-}\right),\\
\int_{\rm arcs} &=& \sum_{i=1}^N\int_{\G_R^i}\,.
\ea\es
In the following sub-sections, we will study each contribution in all generality, later examining the cases $N=2$ and $N=3$ more in detail.

\subsubsection{Arcs at infinity}\label{arcsinfty}

The integrals on the arcs $\G_R^i$ at infinity vanish if condition \Eq{g0} holds. From \Eqqs{gR} and \Eq{GN} with $z=R\,\exp(\rmi\t)$, each arc spans a certain angular range $\t_1^i\leq\t\leq\t_2^i$ and yields
\ba
\frac{1}{2\pi\rmi}\int_{\G_R^i}\rmd z\,\frac{\tilde G_N(z)}{z+k^2} &=& \frac{1}{2\pi}\int_{\t_1^i}^{\t_2^i}\rmd\t\,\tilde G_N(R\,\rme^{\rmi\t})+\dots\nn
&=& \frac{c}{2\pi}\prod_{j=1}^N\int_{\t_1^i}^{\t_2^i}\rmd\t\,(-R\,\rme^{\rmi\t})^{-\g_j}+O\left(\frac{1}{R^{1+\g}}\right)\nn
&=& \frac{c(-1)^{\sum_j\g_j}}{2\pi R^{\g}}\int_{\t_1^i}^{\t_2^i}\rmd\t\,\rme^{-\rmi\t\g}+O\left(\frac{1}{R^{1+\g}}\right)\nn
&=& \frac{c(-1)^{\sum_j\g_j}}{\g R^{\g}}\frac{\rme^{-\rmi\t_1^i\g}-\rme^{-\rmi\t_2^i\g}}{2\pi\rmi}+O\left(\frac{1}{R^{1+\g}}\right)\nn
&\stackrel{R\to\infty}{\rightarrow}& 0\qquad {\rm iff}\quad\g>0\,.\label{gR2g2}
\ea
From now on, we will assume $\g>0$ and ignore this type of contributions.

\subsubsection{Branch points}

We can easily appreciate the trick of distributing the burden of the original branch point into several pieces. The contribution of each branch point $z_i=m^2\exp(\rmi\vp_i)$ is given by the integral \Eq{ceps} over an arc $C_\ve^i$ of radius $\ve$ centered at $z=z_i$. From \Eqq{ceps} with $z=z_i+\ve\,\exp(\rmi\t)$ and $\t_1\leq\t\leq\t_2$, we get
\ba
\tilde G_{C_\ve^i}(-k^2) &=& \frac{1}{2\pi\rmi}\int_{C_\ve^i}\rmd z\,\frac{\tilde G_N(z)}{z+k^2}\nn
&=& \frac{c\,\ve}{z_i+k^2}\Bigg[\prod_{j\neq i}(m^2-m^2\rme^{\rmi\De_{ij}})^{-\g_j}\Bigg]\rme^{\rmi\g_i\vp_i}\int_{\t_1}^{\t_2}\frac{\rmd\t}{2\pi}\,\rme^{\rmi\t} (-\ve\,\rme^{\rmi\t})^{-\g_i}\nn
&&+O(\ve^{2-\g_i})\nn
&=& \frac{c\,\ve^{1-\g_i}}{z_i+k^2}\Bigg[\prod_{j\neq i}(m^2-m^2\rme^{\rmi\De_{ij}})^{-\g_j}\Bigg]\rme^{\rmi\g_i(\vp_i-\pi)}\int_{\t_1}^{\t_2}\frac{\rmd\t}{2\pi}\,\rme^{\rmi\t(1-\g_i)}\nn
&&+O(\ve^{2-\g_i})\nn
&=& \frac{c\,\ve^{1-\g_i}}{(1-\g_i)(z_i+k^2)}\Bigg[\prod_{j\neq i}(m^2-m^2\rme^{\rmi\De_{ij}})^{-\g_j}\Bigg]\rme^{\rmi\g_i(\vp_i-\pi)}\left.\frac{\rme^{\rmi\t(1-\g_i)}}{2\pi\rmi}\right|_{\t=\t_1}^{\t=\t_2}\nn
&&+O(\ve^{2-\g_i})\nn
&=& \ve^{1-\g_i}f(\vp_i,\vp_{j\neq i})+O(\ve^{2-\g_i})\,,\label{zizj}
\ea
where 
\be\label{Deij}
\De_{ij}\coloneqq \vp_i-\vp_j\,.
\ee
The function $f\neq 0$ is well-defined because, by our definition, the limit $\De_{ij}\to 0$ is taken \emph{after} the limit $\ve\to 0$. Then, using the second equality in \Eq{GN}, \Eqq{zizj} tends to zero for all $i$ if, and only if,
\be\label{gNapp}
\g_i<1\qquad\Longrightarrow\qquad \g=\sum_{i=1}^N\g_i<N\,,
\ee
to be compared with the more restrictive bound \Eq{g1}.

\subsubsection{Branch cuts}

\begin{figure}
\bc
\includegraphics[width=12cm]{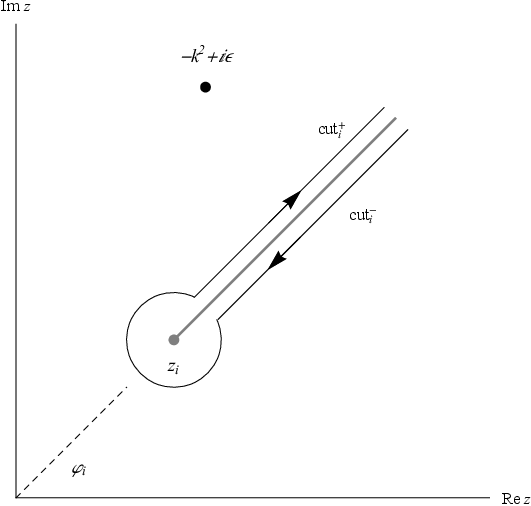}
\ec
\caption{\label{fig3} Generic branch point $z_i=\exp(\rmi\vp_i)\,m^2$ and associated branch cut at an angle $\vp_i$ in the complex plane.}
\end{figure}

We now turn to the calculation of the contribution of a generic radial branch cut oriented at an angle $0\leq\vp_i<2\pi$ starting from the branch point $z_i=m^2\exp(\rmi\vp_i)$ (Fig.~\ref{fig3}).

\begin{figure}
\bc
\includegraphics[width=5cm]{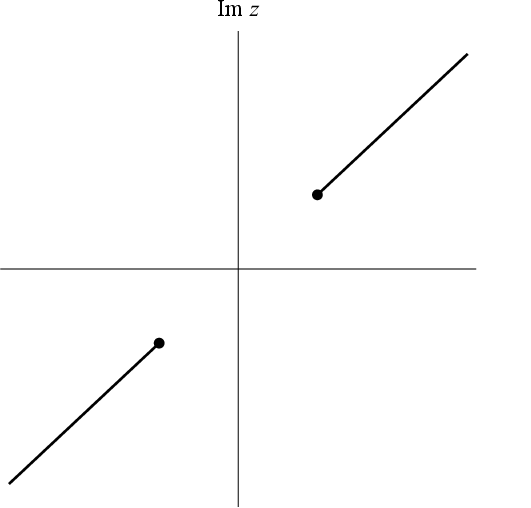}\includegraphics[width=5cm]{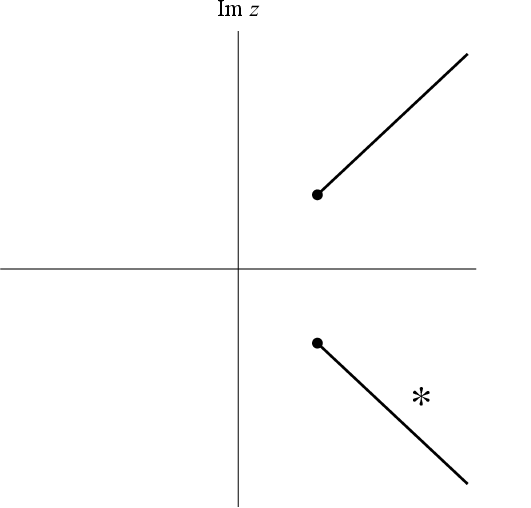}\includegraphics[width=5cm]{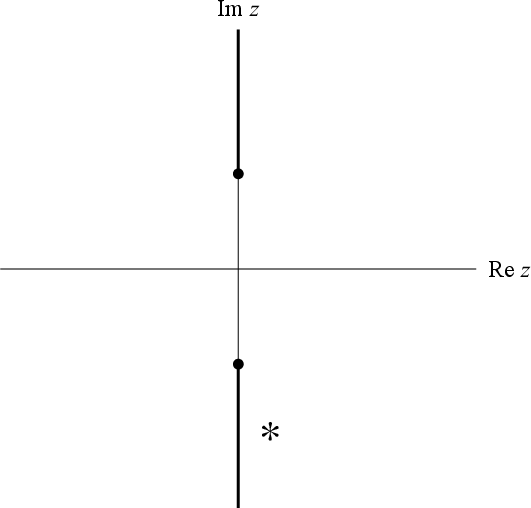}
\ec
\caption{\label{fig4} Left to right: pair of opposite, complex conjugate and both opposite and complex conjugate branch cuts in the $(\Re\,z,\Im\,z)$ plane. One of the cuts in the complex conjugate pair is marked with an asterisk $*$. The angle of the cuts in the first two panels is arbitrary.}
\end{figure}

Having already dispensed with the branch point itself, here we are interested only in the part of the contour running parallel to the branch cut away from (``${\rm cut}_i^+$'') and towards (``${\rm cut}_i^-$'') $z_i$; this orientation is consistent with a counter-clockwise orientation of the contour. Parametrizing the cut as
\be
z= s\,\rme^{\rmi\vp_i}\,,
\ee
we have to integrate along $s$ while keeping the angle $\vp_i$ fixed. Omitting the prescription for $-k^2$,
\ba
\tilde G_{{\rm cut}_i}(-k^2) &=& \frac{1}{2\pi\rmi}\int_{{\rm cut}_i}\rmd z\,\frac{\tilde G(z)}{z+k^2}
= \int_{m^2}^{+\infty}\rmd s\,\frac{\rho_i(s)}{s+\rme^{-\rmi\vp_i}k^2}\,,\label{Gcutgen}\\
\rho_i(s)&\coloneqq& \lim_{\ve\to 0^+}\frac{\tilde G[(s+\rmi\ve)\,\rme^{\rmi\vp_i}]-\tilde G[(s-\rmi\ve)\,\rme^{\rmi\vp_i}]}{2\pi\rmi}\,,\label{rhogen}
\ea
where we defined a generalized spectral function. This general formula, which we could not found anywhere in the literature, can be specialized to four typical cases:
\begin{enumerate}
\item $\vp_1=0$, $z_1=m^2$, $z=s$: real ordinary branch point, branch cut in the positive real axis. Equation \Eq{rhogen} becomes the well-known expression used in section \ref{sec3b}:
\be
\tilde G_{{\rm cut}_1}(-k^2) = \int_{m^2}^{+\infty}\rmd s\,\frac{\rho_1(s)}{s+k^2}\,,\qquad\rho_1(s) = \lim_{\ve\to 0^+}\frac{\tilde G(s+\rmi\ve)-\tilde G(s-\rmi\ve)}{2\pi\rmi}\,.\label{rhogen1}
\ee
\item $\vp_2=\pi$, $z_2=-m^2$, $z=-s$: real tachyonic branch point, branch cut in the negative real axis:
\be
\tilde G_{{\rm cut}_2}(-k^2) = \int_{m^2}^{+\infty}\rmd s\,\frac{\rho_2(s)}{s-k^2}\,,\qquad
\rho_2(s) = \lim_{\ve\to 0^+}\frac{\tilde G(-s-\rmi\ve)-\tilde G(-s+\rmi\ve)}{2\pi\rmi}\,.\label{rhogen2}
\ee
\item $\vp_3=\frac{\pi}{2}$, $z_3=\rmi\,m^2$, $z=\rmi s$: purely imaginary branch point, branch cut in the positive imaginary axis:
\be
\tilde G_{{\rm cut}_3}(-k^2) = \int_{m^2}^{+\infty}\rmd s\,\frac{\rho_3(s)}{s-\rmi k^2}\,,\qquad
\rho_3(s) = \lim_{\ve\to 0^+}\frac{\tilde G(\rmi s-\ve)-\tilde G(\rmi s+\ve)}{2\pi\rmi}\,.\label{rhogen3}
\ee
\item $\vp_4=-\frac{\pi}{2}$, $z_4=-\rmi\,m^2$, $z=-\rmi s$: purely imaginary branch point conjugate to $z_3$, branch cut in the negative imaginary axis:
\be
\tilde G_{{\rm cut}_4}(-k^2) = \int_{m^2}^{+\infty}\rmd s\,\frac{\rho_4(s)}{s+\rmi k^2}\,,\qquad\rho_4(s) = \lim_{\ve\to 0^+}\frac{\tilde G(-\rmi s+\ve)-\tilde G(-\rmi s-\ve)}{2\pi\rmi}\,.\label{rhogen4}
\ee
\end{enumerate}
It is useful to classify branch cuts into pairs of different type. We define a pair of \emph{opposite branch cuts} $i$ and $\bar\imath$ such that $\vp_{\bar\imath}=\vp_i\pm\pi$, and a pair of \emph{complex conjugate branch cuts} $i$ and $i^*$ such that $\vp_{i^*}=-\vp_i$ and $\g_{i^*}=\g_i$. 
 As shown in Fig.~\ref{fig4}, a pair of opposite cuts looks symmetric with respect to the origin but $\g_i$ and $\g_{\bar\imath}$ may differ, while a pair of complex conjugate cuts is symmetric under reflection across the real axis, with same fractional exponent. The only pair which is both opposite and complex conjugate is made of the two branch cuts on the imaginary axis, which we dubbed ``3'' and ``4'' in \Eqqs{rhogen3} and \Eqq{rhogen4}. The cuts ``1'' and ``2'' in \Eqqs{rhogen1} and \Eqq{rhogen2} are the only self-conjugate branch cuts.

\subsubsection{Spectral function}

For the theory to be unitary at the free level, the total integral in $s$ should be real-valued and such that the integrand be positive semi-definite in Euclidean signature ($k^2>0$). However, contrary to the single-cut case of section \ref{sec3}, here we have several contributions $\rho_i$ attached to different complex-valued denominators and we must combine them together into a real-valued quantity before imposing semi-positivity:
\be\label{rhotot}
\sum_i\frac{\rho_i(s)}{s+\rme^{-\rmi\vp_i}k^2}=\frac{\sum_i\rho_i(s)\prod_{j\neq i}\left(s+\rme^{-\rmi\vp_j}k^2\right)}{\prod_i\left(s+\rme^{-\rmi\vp_i}k^2\right)}\geq 0\,,
\ee
which we reduce to a strict inequality since we want a non-trivial propagator. Positivity is imposed on the numerator and the denominator separately, since the latter is nothing but the kinetic term in momentum space up to a global phase absorbed in the $\rho_i$. Therefore, we start with the condition
\be
\prod_{i=1}^N\left(s+\rme^{-\rmi\vp_i}k^2\right)>0\,.
\ee 
Clearly, the branch cut with $\vp_1=0$ is always allowed, while the opposite cut with $\vp_2=\pi$ is always excluded because it is associated with a factor with indefinite sign. In general, opposite pairs do not work. In fact, for each pair with $j\neq i$, we have
\be
p_{ij}\coloneqq\left(s+\rme^{-\rmi\vp_i}k^2\right)\left(s+\rme^{-\rmi\vp_j}k^2\right)=s^2+\rme^{-\rmi(\vp_i+\vp_j)}k^4+sk^2\left(\rme^{-\rmi\vp_i}+\rme^{-\rmi\vp_j}\right)\,.\label{pij}
\ee
If these cuts are opposite ($j=\bar\imath$), we get the expression
\be
p_{i\bar\imath}=s^2-k^4+2\rmi sk^2\sin\vp_i\,,
\ee
which is complex-valued unless $\vp_i=0,\pi$. However, in this case $p_{i\bar\imath}=s^2-k^4=(s+k^2)(s-k^2)$ has indefinite sign due to the factor $(s-k^2)$ coming from the cut at $\vp_2=\pi$. Therefore, we are forced to conclude that any other cut different from $\vp_1=0$ gives a well-defined contribution only if its paired into a complex conjugate couple. In particular, if $\vp_1=0$ is present, then $N$ must be odd, otherwise it must be even. For $j=i^*$, we have
\be\label{pii*}
p_{ii^*}=s^2+k^4+2sk^2\cos\vp_i>0\,,
\ee
for any $0<\vp_i<2\pi$.

The numerator on top of the denominator \Eq{pij} is found from
\be
\frac{\rho_i(s)}{s+\rme^{-\rmi\vp_i}k^2}+\frac{\rho_j(s)}{s+\rme^{-\rmi\vp_j}k^2} = \frac{q_{ij}}{p_{ij}}\,,
\ee
where
\ba
q_{ij}&\coloneqq& \rho_i(s)\left(s+\rme^{-\rmi\vp_j}k^2\right)+\rho_j(s)\left(s+\rme^{-\rmi\vp_i}k^2\right)\nn
&=& \left[(\Re\,\rho_i)(s+k^2\cos\vp_j)+(\Re\,\rho_j)(s+k^2\cos\vp_i)+(\Im\,\rho_i)k^2\sin\vp_j+(\Im\,\rho_j)k^2\sin\vp_i\right]\nn
&&+\rmi\left[(\Im\,\rho_i)(s+k^2\cos\vp_j)+(\Im\,\rho_j)(s+k^2\cos\vp_i)\right.\nn
&&\qquad\left.-(\Re\,\rho_i)k^2\sin\vp_j-(\Re\,\rho_j)k^2\sin\vp_i\right].
\ea
In particular,
\ba
q_{ii^*}&=& \left[(\Re\,\rho_i+\Re\,\rho_{i^*})(s+k^2\cos\vp_i)-(\Im\,\rho_i-\Im\,\rho_{i^*})k^2\sin\vp_i\right]\nn
&&+\rmi\left[(\Im\,\rho_i+\Im\,\rho_{i^*})(s+k^2\cos\vp_i)+(\Re\,\rho_i-\Re\,\rho_{i^*})k^2\sin\vp_i\right].\label{qij}
\ea
If we write each spectral function as the absolute value times a phase,
\be
\rho_i=|\rho_i|\,\rme^{-\rmi\psi_i}\,,
\ee
then reality of the left-hand side of \Eq{rhotot} is guaranteed if the imaginary part of \Eqq{qij} vanishes for any $k$, which happens when $\Im\,\rho_i=-\Im\,\rho_{i^*}$ and $\Re\,\rho_i=\Re\,\rho_{i^*}$:
\be\label{rhocon}
|\rho_i|=|\rho_{i^*}|\,,\qquad \psi_i=-\psi_{i^*}\quad \text{(mod $2\pi$)}\qquad\Longrightarrow\qquad \rho_{i^*}=\rho_i^*\,.
\ee
Positivity of the left-hand side of \Eq{rhotot} then stems directly from the fact that, if the denominator is positive, then we only need to ensure that $q_{ii^*}>0$. From the first line of \Eqq{qij},
\ba
q_{ii^*}&=& 2\Re\,\rho_i(s+k^2\cos\vp_i)-2\Im\,\rho_ik^2\sin\vp_i\nn
        &=& 2|\rho_i|\cos\psi_i(s+k^2\cos\vp_i)+2|\rho_i|\sin\psi_i k^2\sin\vp_i\nn
				&=& 2|\rho_i|\left[s\cos\psi_i+k^2\cos(\psi_i-\vp_i)\right].\label{qii*}
\ea
Therefore, since $s>0$ and $k^2>0$ in Euclidean signature, we get $q_{ii^*}>0$ if two inequalities are simultaneously satisfied:
\be\label{rhoiplus}
\cos\psi_i\geq 0\,,\qquad \cos(\psi_i-\vp_i)\geq 0\,,
\ee
where equality can hold only for one of the two. In particular, if $\psi_i=2\pi l$ with $l\in\mathbb{Z}$, then we only need $\cos\vp_i\geq 0$ for all $i$, that is,
\be
\rme^{-\rmi\psi_i}=+1\,,\qquad \rho_i=\rho_{i^*}=|\rho_i|>0\,,\qquad  \Big\{0\leq \vp_i\leq\frac{\pi}{2}\Big\}\cup\Big\{\frac{3\pi}{2}\leq\vp_i<2\pi\Big\}\,.\label{rhoiplusa}
\ee
Therefore, cuts in the first and fourth quadrant respect the stricter positivity condition \Eq{rhoiplusa}.

Overall, from \Eqqs{pii*} and \Eq{qii*} we have
\ban
\sum_i\frac{\rho_i(s)}{s+\rme^{-\rmi\vp_i}k^2} &=& \frac{\rho_1(s)}{s+k^2}+\sum_{(i,i^*)}\frac{q_{ii^*}}{p_{ii^*}}\\
&=& \frac{\rho_1(s)}{s+k^2}+2\sum_{(i,i^*)} \frac{|\rho_i|\left[s\cos\psi_i+k^2\cos(\psi_i-\vp_i)\right]}{s^2+k^4+2sk^2\cos\vp_i}\,,
\ean
where we isolated the $\vp_1=0$ cut if present and $(i,i^*)$ denotes the sum over pairs of complex conjugate cuts (one term per pair in the sum). Therefore, the final expression of the propagator \Eq{optN} is \Eqq{optNfinal} and the condition \Eq{rhotot} becomes \Eq{rhotot2}.

The next task is to calculate $\rho_i$ and find for which set of angles $\vp_i$ \Eqqs{rhocon} and \Eq{rhoiplus} (or the stricter condition \Eq{rhoiplusa}) are satisfied. We can follow the same steps used for the $N=1$ case \Eq{Gplusmin}, starting from the Green's function \Eq{GN} and the expression
\bs\ba
\tilde G_N[(s\pm\rmi\ve)\,\rme^{\rmi\vp_i}]&=& c\prod_{j=1}^N \left[m^2-(s\pm\rmi\ve)\,\rme^{\rmi\De_{ij}}\right]^{-\g_j}= c\prod_{j=1}^N (x_j^\pm+\rmi\,y_j^\pm)^{-\g_j}\,,\\
x_j^\pm&\coloneqq& -s\cos\De_{ij}\pm\ve\sin\De_{ij}+m^2\,,\\
y_j^\pm&\coloneqq& -s\sin\De_{ij}\mp\ve\cos\De_{ij}\,,
\ea\es
so that
\be
\tilde G_N(x_j^\pm+\rmi\,y_j^\pm) = c\prod_j\big|(x_j^\pm)^2+(y_j^\pm)^2\big|^{-\frac{\g_j}{2}}\rme^{-\rmi\g_j\,{\rm Arg}(x_j^\pm+\rmi\,y_j^\pm)}.\label{Gxy}
\ee
We apply \Eqq{argxy} for each branch cut separately and then send $\ve\to 0^+$. To understand this step better, we identify different categories for 
\ba
{\rm Arg}_{ji}^\pm&\coloneqq& \lim_{\ve\to 0^+}{\rm Arg}(x_j^\pm+\rmi\,y_j^\pm)\nn
&=&\lim_{\ve\to 0^+}{\rm Arg}[(m^2-s\cos\De_{ij}\pm\ve\sin\De_{ij})+\rmi(-s\sin\De_{ij}\mp\ve\cos\De_{ij})]\,,\label{arggen}
\ea
the principal value of the argument of the $j$th term in the product \Eq{Gxy} evaluated at, respectively, the outgoing ($+$) and incoming ($-$) side of the $i$th cut. ${\rm Arg}_{ji}^\pm$ can be of three types:
\begin{itemize}
\item $j=i$ ($\vp_j=\vp_i$), the contribution to the $i$th branch cut of the $i$th term in the product. In this case, since $\De_{ii}=0$, we have $x_j^\pm=m^2-s<0$ and $y_j^\pm=\mp\ve$, so that ${\rm Arg}_{ii}^\pm=\lim_{\ve\to 0^+}\arctan[\mp\ve/(m^2-s)]\pm \pi$ and
\be\label{argeq1}
{\rm Arg}_{ii}^\pm=\mp\pi\,.
\ee
The argument of the principal value has a discontinuity at each cut.
\item $j=\bar\imath$ ($\vp_j=\vp_i\pm\pi$), the contribution of the term opposite to the $i$th cut. Although we excluded a pairing of branch cuts into opposites, among the $N$ cuts of the theory there may still be two branch cuts which are accidentally opposite, but each one of them is always matched with a complex conjugate cut. Therefore, the pairing is always made between conjugate cuts. For $\De_{ij}=\pm\pi$, \Eqq{arggen} becomes
\be\label{argeq2a}
{\rm Arg}_{\bar\imath i}^\pm=\lim_{\ve\to 0^+}{\rm Arg}(m^2+s+\pm\rmi\ve)=\lim_{\ve\to 0^+}{\rm arctan}\left(\frac{\pm\ve}{m^2+s}\right)=0\,.
\ee
\item $j\neq i$ ($\vp_j\neq \vp_i,\vp_i\pm\pi$): the branch cuts $i$ and $j$ are oriented at a generic non-trivial, non-opposite position. The contribution of the $j$th argument at the $i$th branch cut is \Eq{arggen}. Since $\vp_j$ and $\vp_i$ are, in this case, always different and such that $\vp_j\neq\vp_i\pm\pi$, then in particular $\De_{ij}\neq 0,\pi$, so that $\sin\De_{ij}$ never vanishes and we can ignore the $\ve$ term in the imaginary part in \Eqq{arggen}. A similar fate is reserved for the $\ve$ term in the real part, since $\De_{ij}$ is $s$-independent. Therefore, we can take the limit $\ve\to 0$ directly inside the Arg function and obtain
\be\label{argeq2}
{\rm Arg}_{ji}^\pm={\rm Arg}_{ji}\coloneqq {\rm Arg}[(m^2-s\cos\De_{ij})-\rmi s\sin\De_{ij}]\,.
\ee
Note that this expression is valid also the cases $j=\bar\imath$ (opposite pairs, in agreement with \Eqq{argeq2a}) and $j=i^*$ ($\vp_j=-\vp_i\neq 0,\pi$), the contribution of the term conjugate to the $i$th cut. Then, one immediately gets $\De_{ii^*}=2\vp_i=-\De_{i^*i}$ and
\ba
{\rm Arg}_{i^*i}+{\rm Arg}_{ii^*} &=& {\rm Arg}[m^2-s\cos(2\vp_i)-\rmi s\sin(2\vp_i)]\nn
&&+{\rm Arg}[m^2-s\cos(2\vp_i)+\rmi s\sin(2\vp_i)]\nn
&=& {\rm Arg}\{[m^2-s\cos(2\vp_i)]^2+[s\sin(2\vp_i)]^2\}\qquad\textrm{(mod $2\pi$)}\nn
&=&0\,.\label{argeq3}
\ea
\end{itemize}
Another useful relation coming from \Eq{argeq2} is between the argument of the contribution of the $\vp_1=0$ branch cut to the spectral function of the cut $i$ and of its conjugate $i^*$:
\be\label{arg1i}
\hspace{-.2cm} {\rm Arg}_{1i}= {\rm Arg}[(m^2-s\cos\vp_i)-\rmi s\sin\vp_i]\,,\qquad {\rm Arg}_{1i^*}= {\rm Arg}[(m^2-s\cos\vp_i)+\rmi s\sin\vp_i]\,,
\ee
which implies
\be
{\rm Arg}_{1i}+{\rm Arg}_{1i^*}=0\,.\label{arg1j}
\ee

The contribution \Eq{rhogen} of the $i$th branch cut to the spectral function is obtained from \Eqqs{Gxy}, \Eq{argeq1} and \Eq{argeq3}:
\ba
\rho_i(r) &=& \lim_{\ve\to 0^+} \frac{\tilde G_N(x_j^++\rmi\,y_j^+)-\tilde G_N(x_j^-+\rmi\,y_j^-)}{2\pi\rmi}\nn
&=& \frac{c}{\prod_j\big|s^2+m^4-2m^2s\cos\De_{ij}\big|^{\frac{\g_j}{2}}}\frac{\rme^{-\rmi\sum_j\g_j\,{\rm Arg}_{ji}^+}-\rme^{-\rmi\sum_j\g_j\,{\rm Arg}_{ji}^-}}{2\pi\rmi}\nn
&=& \frac{c\,\rme^{-\rmi\sum_{j\neq i}\g_j\,{\rm Arg}_{ji}}}{|s-m^2|^{\g_i}\prod_{j\neq i}\big|s^2+m^4-2m^2s\cos\De_{ij}\big|^{\frac{\g_j}{2}}}\frac{\rme^{-\rmi\g_i\,{\rm Arg}_{ii}^+}-\rme^{-\rmi\g_i\,{\rm Arg}_{ii}^-}}{2\pi\rmi}\nn
&\stackbin{\text{\tiny\Eq{argeq1}}}{=}& \frac{c\,\rme^{-\rmi\sum_{j\neq i}\g_j\,{\rm Arg}_{ji}}}{|s-m^2|^{\g_i}\prod_{j\neq i}\big|s^2+m^4-2m^2s\cos\De_{ij}\big|^{\frac{\g_j}{2}}}\frac{\rme^{\rmi\pi\g_i}-\rme^{-\rmi\pi\g_i}}{2\pi\rmi}\nn
&=& \frac{c\,\rme^{-\rmi\sum_{j\neq i}\g_j\,{\rm Arg}_{ji}}}{|s-m^2|^{\g_i}\prod_{j\neq i}\big|s^2+m^4-2m^2s\cos\De_{ij}\big|^{\frac{\g_j}{2}}}\frac{\sin(\pi\g_i)}{\pi}\nn
&\stackbin{\text{\tiny\Eq{cphase}}}{=}&\frac{\rme^{-\rmi\sum_j\g_j\vp_j-\rmi\sum_{j\neq i}\g_j\,{\rm Arg}_{ji}}}{|s-m^2|^{\g_i}\prod_{j\neq i}\big|s^2+m^4-2m^2s\cos\De_{ij}\big|^{\frac{\g_j}{2}}}\frac{\sin(\pi\g_i)}{\pi}\,.\label{masterapp}
\ea
Since all branch cuts except $\vp_1=0$ (if present) must be paired into complex conjugates and since $\g_{i^*}=\g_i$, we can rewrite the phase
\be
\psi_i = \g_i\vp_i+\sum_{j\neq i}\g_j(\vp_j+{\rm Arg}_{ji})\label{psi1}
\ee
as
\ba
\psi_1 &=& 0\,,\qquad \text{if $\vp_1=0$ present}\,,\\
\psi_{i\neq 1} &=& \g_i(\vp_i+\vp_{i^*})+\g_1\,{\rm Arg}_{1i}+\g_i\,{\rm Arg}_{i^*i}+\sum_{j\neq 1,i,i^*}\g_j(\vp_j+{\rm Arg}_{ji})\nn
&=& \g_1\,{\rm Arg}_{1i}+\g_i\,{\rm Arg}_{i^*i}+\sum_{j\neq 1,i,i^*}\g_j(\vp_j+{\rm Arg}_{ji})\nn
&=& \g_1\,{\rm Arg}_{1i}+\g_i\,{\rm Arg}_{i^*i}+\sum_{(j,j^*)\neq (i,i^*)}\g_j({\rm Arg}_{ji}+{\rm Arg}_{j^*i})\nn
&=& \g_1\,{\rm Arg}_{1i}+\g_i\,{\rm Arg}_{i^*i}+\sum_{(j,j^*)\neq (i,i^*)}\g_j\cA_{ji}\,,\label{psi2app}
\ea
where in the last two steps we used the fact that the sum on $j\neq 1$ must have an even number of elements, paired into complex conjugates such that there is always a $j^*$ for which $\vp_{j^*}=-\vp_j$. 

\subsubsection{Reality}

The reality condition \Eq{rhocon} holds because
\ba
\psi_{i^*} &=& \g_1\,{\rm Arg}_{1i^*}+\g_i\,{\rm Arg}_{ii^*}+\sum_{(j,j^*)\neq (i,i^*)}\g_j\cA_{ji^*}\nn
			     &\stackbin{\text{\tiny\Eq{arg1j}}}{=}& -\g_1\,{\rm Arg}_{1i}+\g_i\,{\rm Arg}_{ii^*}+\sum_{(j,j^*)\neq (i,i^*)}\g_j\cA_{ji^*}\nn		
					 &\stackbin{\text{\tiny\Eq{argeq3}}}{=}& -\g_1\,{\rm Arg}_{1i}-\g_i\,{\rm Arg}_{i^*i}+\sum_{(j,j^*)\neq (i,i^*)}\g_j\cA_{ji^*}\nn
					 &=& -\left[\g_1\,{\rm Arg}_{1i}+\g_i\,{\rm Arg}_{i^*i}+\sum_{(j,j^*)\neq (i,i^*)}\g_j\cA_{ji}\right]\nn
					 &\stackbin{\text{\tiny\Eq{psi2}}}{=}& -\psi_i\,,
\ea
where we used the fact that $\cA_{ji^*}=-\cA_{ji}$.

\subsubsection{Positivity}

From the bounds \Eq{g0} and \Eq{gN}, we have that $0<\g_i<1$ implies $\sin(\pi\g_i)>0$, so that $|\rho_i|>0$ consistently with our assumptions.

If the minimal positivity condition \Eq{rhoiplus} holds, then the allowed values of $\vp_i$ lie in certain regions of the parameter space we will explore the next sub-section. In the case of the stricter positivity condition \Eq{rhoiplusa} and in the presence of the $\vp_1=0$ cut, we have
\be\label{posicon1}
\psi_{i\neq 1}=\g_1\,{\rm Arg}\left(m^2-s\,\rme^{\rmi\vp_i}\right)+\g_i\,{\rm Arg}\left(m^2-s\,\rme^{2\rmi\vp_i}\right)+\sum_{(j,j^*)\neq (i,i^*)}\g_j\cA_{ji}=0\qquad \text{(mod $2\pi$)}\,,
\ee
while in its absence the first term disappears.


\subsection{Splittings of the propagator}\label{appC2}

\begin{figure}
\includegraphics[width=4.9cm]{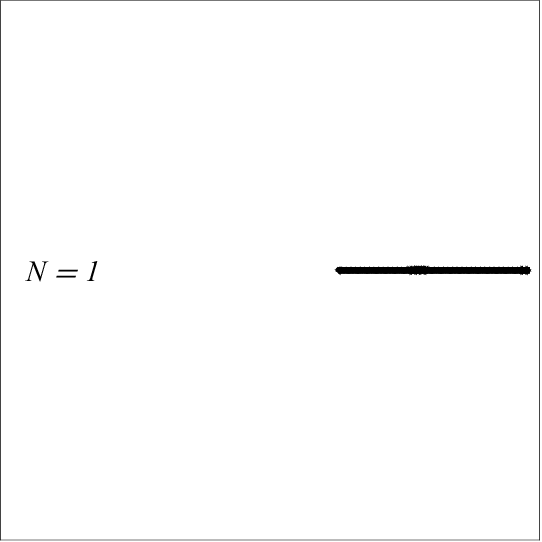}\,\,\includegraphics[width=4.9cm]{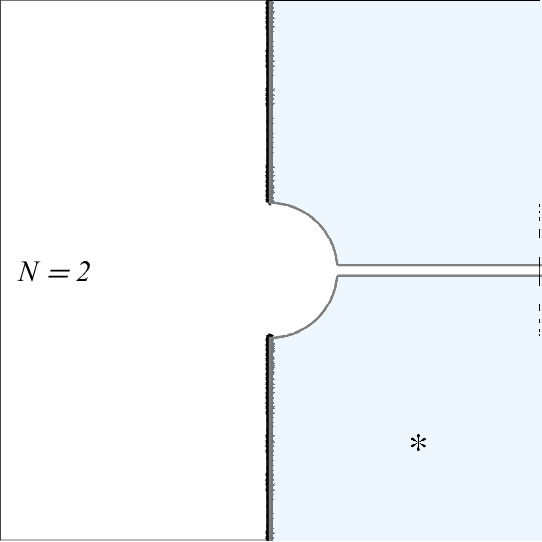}\,\,\includegraphics[width=4.9cm]{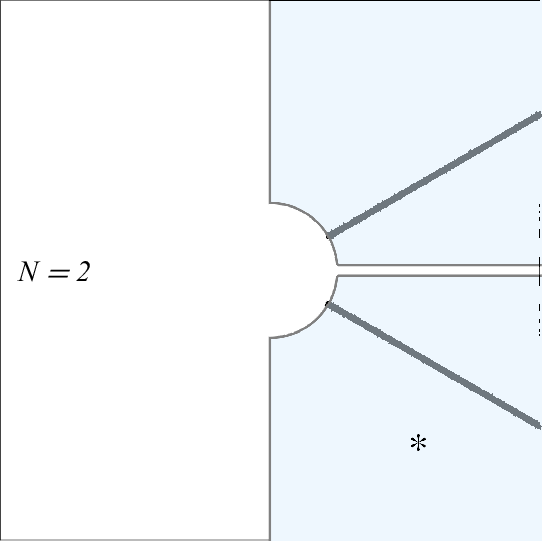}\\
\includegraphics[width=4.9cm]{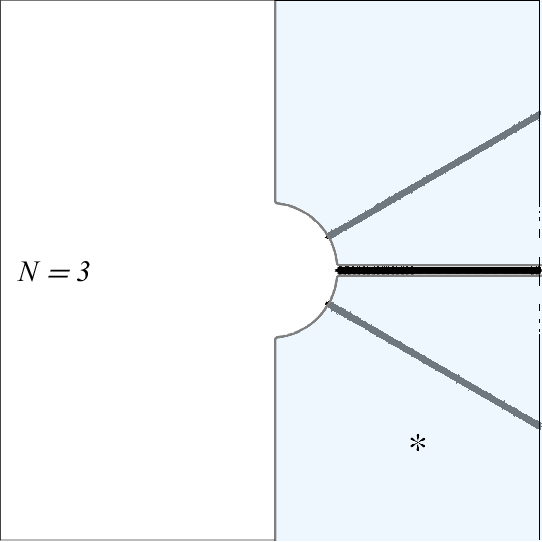}\,\,\includegraphics[width=4.9cm]{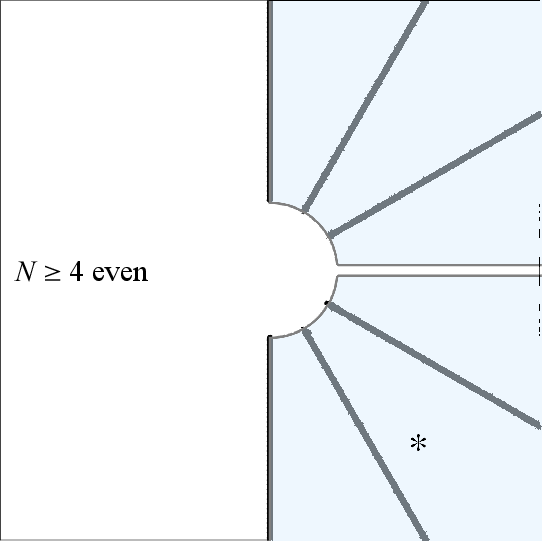}\,\,\includegraphics[width=4.9cm]{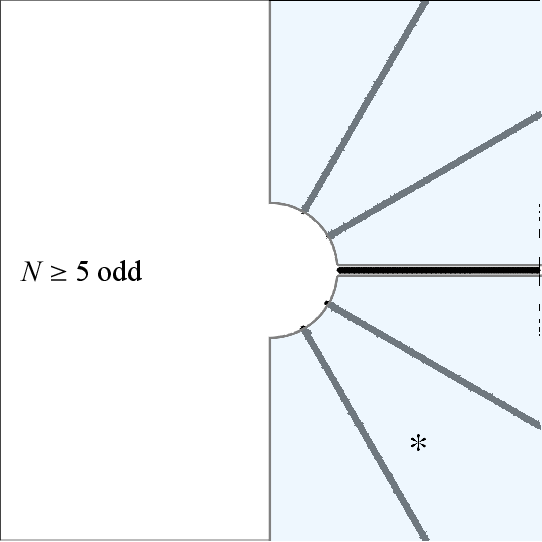}
\caption{\label{fig5} Examples of branch-cut configurations in the $(\Re\,z,\Im\,z)$ plane giving rise to unitary theories when $\g_i=\g/N$ for all $i$: $N=1$ (non-renormalizable), $N=2$ (at most higher-loop super-renormalizable), $N=3$ (non-renormalizable), $N\geq 4$ even or odd (non-renormalizable). In the $N=1$ case, only the cut at $\vp_1=0$ is allowed. In the $N=2$ plots, the pair of cuts can be chosen at any angle inside the shaded regions including at $\vp_i=\pi/2,3\pi/2$. An asterisk $*$ marks the fourth quadrant, the area of the cuts conjugate to those in the first quadrant. The cuts on the imaginary axis are excluded in odd-$N$ configurations.}
\end{figure}

Let us study the properties of the propagator \Eq{optNfinal} depending on the order $N$ of the splitting. Figure \ref{fig5} summarizes the cases for which the theory has a positive-definite Källén--Lehmann representation and is therefore unitary.

\subsubsection{\texorpdfstring{$N=1$}{N=1} cut}

Here there is no splitting. From \Eqq{psi1}, we have $\psi_1=\g\vp_1$, so that the only viable branch cut is the one on the positive real axis, $\vp_1=0$, and $\rho_1>0$ (Fig.~\ref{fig5}). All the other cuts carry a non-trivial phase and $\rho_1\in\mathbb{C}$, except the cut at $\vp_1=\pi$ for which $\mathbb{R}\ni\rho_1<0$ and the left-hand side of \Eqq{rhotot2} has indefinite sign.

As we saw in section \ref{sec3}, the resulting theory is unitary but non-renormalizable, since $0<\g<1$.

\subsubsection{\texorpdfstring{$N=2$}{N=2} cuts}

This case is studied in the main text (section \ref{secun}) since it is the only one leading to a renormalizable theory.

\subsubsection{\texorpdfstring{$N=3$}{N=3} cuts}

For $N=3$, the only possibility is to include the $\vp_1=0$ cut plus a complex conjugate pair. Then,
\be
\psi_i= \g_1\,{\rm Arg}\left(m^2-s\,\rme^{\rmi \vp_i}\right)+\g_i\,{\rm Arg}\left(m^2-s\,\rme^{2\rmi \vp_i}\right).\label{posicon3}
\ee
Excluding $\vp_i=0,\pi$, to solve the stricter condition \Eq{posicon1} we choose for simplicity
\be\label{g1gi}
\g_1=\g_i\,,
\ee
so that, taking the large-$s$ limit of \Eqq{posicon3},
\be\label{psiN3b}
\psi_i=\g_i[(\vp_i-n\pi)+(2\vp_i-n\pi)]=\g_i(3\vp_i-2n\pi)\,,\qquad n=1,2\,,
\ee
which vanishes for the angles
\be\label{2pi3}
\vp_2=\frac{2\pi}{3}\,,\qquad \vp_3=\frac{4\pi}{3}\,,
\ee
which are the only $(s,m)$-independent roots in the interval $0<\vp_i<2\pi$.
This pair of complex conjugate branch cuts is in the second and third quadrant, i.e., they are not in the range specified by \Eq{posicon1}. Therefore, the condition \Eq{posicon1} is too restrictive.

\begin{figure}
\bc
\includegraphics[width=7cm]{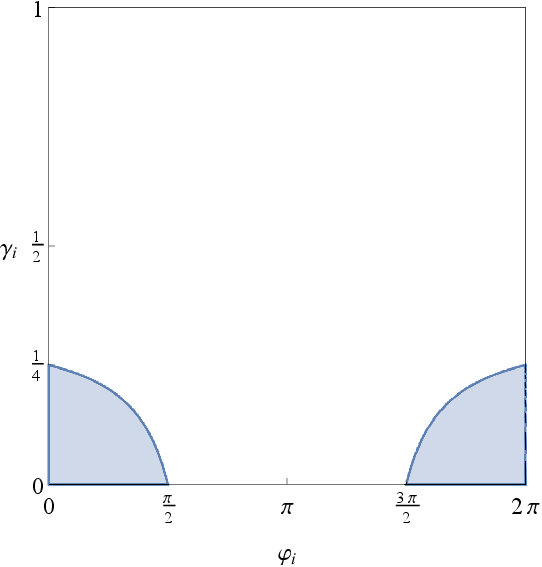}
\ec
\caption{\label{fig7} Angle $\vp_i$ and exponent $\g_1=\g_i$ satisfying the positivity conditions \Eq{rhoiplus} and \Eqq{posicon3} for $N=3$. Shaded areas are the regions in the parameter space where free-level unitarity is guaranteed.}
\end{figure}

Under the weaker constraints \Eq{rhoiplus} with the condition \Eq{g1gi} and $\psi_i$ given by \Eqq{posicon3}, the allowed regions in the $z$ and the $(\vp_i,\g_i)$ planes are shown in Figs.~\ref{fig5} and \ref{fig7}, respectively:
\be
\hspace{-1cm}\text{non-renormalizable:\hspace{.5cm} $0<\g_i\leq \frac14$}\,,\qquad \left\{0<\vp_i\leq \frac{\pi}{2}\right\}\cup \left\{\frac{3\pi}{2}\leq\vp_i<2\pi\right\}.\label{N3cond}
\ee
The theory is non-renormalizable because $\g<3/4$. 

With all $\g_{i}=\g/3$, the kinetic term and the propagator are, respectively,
\ba
\cK_{3}(\B)
&=&-\left(m^2-\B\right)^{\frac{\g}{3}}\left(m^4-2m^2\cos\vp\B+\B^2\right)^{\frac{\g}{3}},
\ea
and
\be
\tilde{G}_{3}(-k^2)=\frac{1}{\left(m^2+k^2\right)^{\frac{\g}{3}}\left(m^4+2m^2k^2\cos\vp +k^4\right)^{\frac{\g}{3}}}=\int_{m^2}^{+\infty}\! \rmd s\,\cG_3(s,k^2)\,,
\ee
where
\ba
\cG_3(s,k^2)&=& \frac{\sin\left(\frac{\pi\g}{3}\right)}{\pi}\!\left\{\frac{1}{\left(s+k^2\right)\left(s-m^2\right)^{\frac{\g}{3}}\left(s^2+m^4-2m^2s\cos\vp\right)^{\frac{\g}{3}}}+\frac{2}{s^2+k^4+2sk^2\cos\vp}\right.\nn
&&\left.\times\frac{s\cos\psi(s,\vp)+k^2\cos[\psi(s,\vp)-\vp]}{\left(s-m^2\right)^{\frac{\g}{3}}\left(s^2+m^4-2m^2s\cos\vp\right)^{\frac{\g}{6}}\left[s^2+m^4-2m^2s\cos(2\vp)\right]^{\frac{\g}{6}}}\right\},\\
\psi(s,\vp)&=&\frac{\g}{3}\left[{\rm Arg}\left(m^2-s\,\rme^{\rmi\vp}\right)+{\rm Arg}\left(m^2-s\,\rme^{2\rmi\vp}\right)\right].
\ea

\begin{figure}
\bc
\includegraphics[width=7cm]{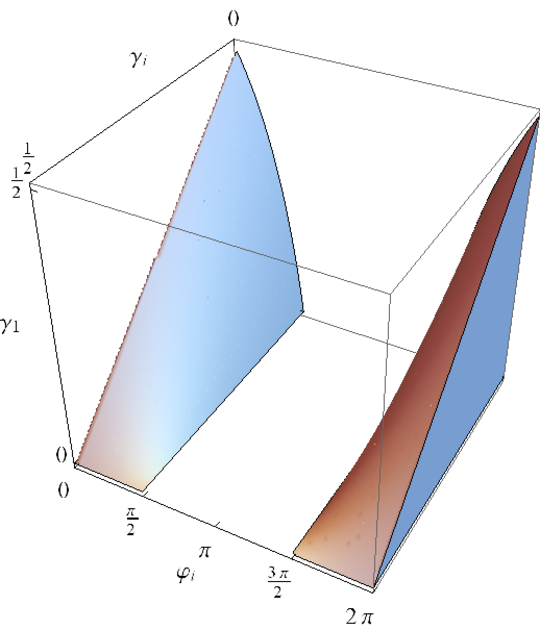}
\ec
\caption{\label{fig8} Angle $\vp_i$ and exponents $\g_1$ and $\g_i\neq\g_1$ satisfying the positivity conditions \Eq{rhoiplus} and \Eqq{posicon3} for $N=3$.}
\end{figure}

If we forgo the assumption \Eq{g1gi} and allow a different fractional exponent for the pair of conjugate cuts, we do not improve the situation. As one can see in Fig.~\ref{fig8}, $\g_i + \g_1 <1/2$, so that $\g=2\g_i + \g_1<1$.

\subsubsection{\texorpdfstring{$N\geq 4$}{N>=4} cuts}

\begin{figure}
\bc
\includegraphics[width=7cm]{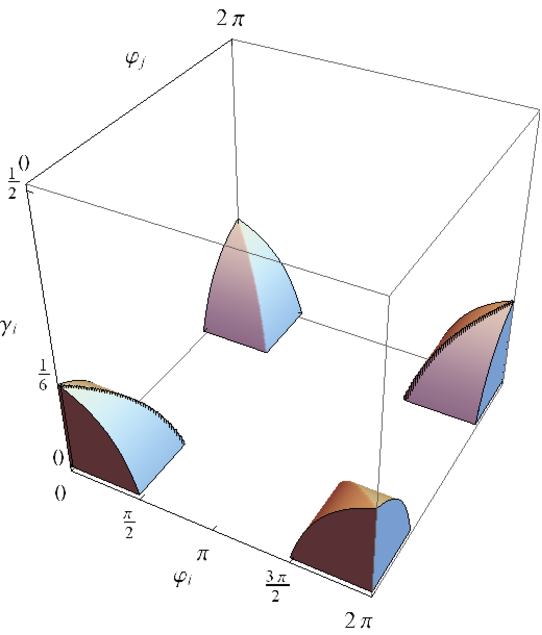}
\ec
\caption{\label{fig9} Angles $\vp_i$ and $\vp_j$ and exponent $\g_i$ satisfying the positivity conditions \Eq{rhoiplus} and \Eqq{posicon3} (with $\g_1=0$) for $N=4$. We assume $\vp_i\neq\vp_j$. Positivity conditions are applied to both angles switching the index $i$ with $j$.}
\end{figure}

When $N\geq 4$ is even, it is sufficient to select the $N/2$ pairs of conjugate cuts in the allowed regions in Fig.~\ref{fig5}. The allowed region in the $(\vp_i,\g_i)$ plane changes non-trivially with respect to Fig.~\ref{fig6}, due to the $\cA_{ij}$ terms in \Eqq{psi2}. When $N=4$, the allowed angles $\vp_i$ are still contained within the strips $\vp_i\in(0,\pi/2]\cup[3\pi/2,1)$ but the $\g_i$ are smaller than $1/6$ (Fig.~\ref{fig9}).

Therefore, the theory is non-renormalizable because $\g<2/3$. For general even or odd $N$, if we assume
\be\label{gigN}
\g_i=\frac{\g}{N}\qquad\forall\,i\,,
\ee
then all the angles are limited to the strips $\vp_i\in(0,\pi/2]\cup[3\pi/2,1)$ and the upper bound on $\g_i$ is (see below for the proof)
\be\label{empibou}
\g=N\g_i<\frac{N}{2(N-1)}\,,
\ee
which worsens as $N$ increases. Relaxing the symmetric configuration \Eq{gigN} does not help, as the $N=3$ example of Fig.~\ref{fig8} shows. In the case $N=2$, there is a discontinuity at the angles \Eq{pi2} (Fig.~\ref{fig6}) where we can cross the bound \Eq{empibou}. 

We now prove the upper bound \Eq{empibou}. Here all the $\g_i$ are assumed to be equal, $\g_i=\g/N$. When they are not, the upper bound \Eq{empibou} is no longer equidistributed among the fractional exponents and it is further decreased for some $\g_i$, as shown in Fig.~\ref{fig8}.

Let $N$ be even. From \Eqqs{psi2} and \Eq{Aji},
\be\label{eqE1}
\psi_{i}=\frac{\g}{N}\left[{\rm Arg}_{i^{*}i}+\sum_{j\neq i}^\textrm{pairs}\left({\rm Arg}_{ji}+{\rm Arg}_{j^{*}i}\right)\right],
\ee
where the sum on pairs is such that $(j,j^{*})\neq(i,i^{*})$ and
\ben
{\rm Arg}_{ki}={\rm Arg}\left[m^2-s\,\rme^{\rmi\left(\vp_i-\vp_k\right)}\right].
\een
Assume now that all angles are small ($\vp_i\ll1$) and pairwise different ($\vp_i\neq\vp_j$ for $i\neq j$). Then, $\vp_i-\vp_k\ll1$ and
\ba
{\rm Arg}_{ki}&=&{\rm Arg}\left\{m^2-s-\rmi\,s(\vp_i-\vp_k)+O\left[(\vp_i-\vp_k)^2\right]\right\}\nn
&\stackbin{\text{\tiny\Eq{argxy}}}{=}&\pi\,{\rm sgn}(\vp_k-\vp_i)-s(\vp_i-\vp_k)+O\left[\left(\vp_i-\vp_k\right)^{2}\right],\label{eqE2}
\ea
where we used the fact that $s>m^2$. Then, since $\vp_i>0$ in our convention,
\ba
\psi_i&\simeq&\frac{\g}{N}\Bigg\{\pi\,{\rm sgn}(-2\vp_i)+\pi\sum_{j\neq i}^\textrm{pairs}\left[{\rm sgn}(\vp_j-\vp_i)+{\rm sgn}(-\vp_j-\vp_i)\right]\Bigg\}\nn
&=&\frac{\pi\g}{N}\Bigg[-{\rm sgn}\,\vp_i+\sum_{j\neq i}^\textrm{pairs}{\rm sgn}(\vp_j-\vp_i)-\sum_{j\neq i}^\textrm{pairs}{\rm sgn}(\vp_j+\vp_i)\Bigg]\nn
&=&\frac{\pi\g}{N}\Bigg[-1+\sum_{j\neq i}^\textrm{pairs}{\rm sgn}(\vp_j-\vp_i)-\sum_{j\neq i}^\textrm{pairs}1\Bigg]\nn
&=&\frac{\pi\g}{N}\Bigg[-1+\sum_{j\neq i}^\textrm{pairs}{\rm sgn}(\vp_j-\vp_i)-\frac{N-2}{2}\Bigg]\nn
&=&\frac{\pi\g}{N}\Bigg[\sum_{j\neq i}^\textrm{pairs}{\rm sgn}(\vp_j-\vp_i)-\frac{N}{2}\Bigg].\label{eqE3}
\ea
One of the positivity conditions \Eq{rhoiplus} is $\cos\psi_i\geq0$, so that $-{\pi}/{2}\leq\psi_i\leq{\pi}/{2}$ in general. However, it turns out that $\psi_i<0$. In fact, the maximum of the expression $\sum_{j\neq i}^\textrm{pairs}{\rm sgn}(\vp_j-\vp_i)$ is $(N-2)/2$ when $\vp_i$ is the smallest angle in the set, while the minimum is $-(N-2)/2$ when $\vp_i$ is the largest angle. Thus, the minimum of $\psi_i$ for any $i$ is
\be\label{eqEpsi}
\psi_i\geq\frac{\pi\g}{N}\left(-\frac{N-2}{2}-\frac{N}{2}\right)=\frac{\pi\g}{N}(1-N)\,.
\ee
Requiring that
\ben
-\frac{\pi}{2}\leq \frac{\pi\g}{N}(1-N)\,,
\een
we get the bound \Eq{empibou}, since $1-N<0$:
\be\label{appEbou}
\frac{\g}{N}\leq\frac{N}{2(N-1)}\,.
\ee

So far, we assumed $N$ to be even. When it is odd, we must add the term $(\g/N){\rm Arg}_{1i}$ to \Eqq{eqE1}, where $\vp_1=0$. Then, in \Eqq{eqE3} we have an extra term proportional to ${\rm Arg}_{1i}=\pi\,{\rm sgn}(-\vp_i)+O(\vp_i)=-\pi+O(\vp_i)$, while now $\sum_{j\neq i}^\textrm{pairs}1=(N-3)/2$:
\ba
\psi_i&\simeq&\frac{\pi\g}{N}\Bigg[-1-1+\sum_{j\neq i}^\textrm{pairs}{\rm sgn}(\vp_j-\vp_i)-\sum_{j\neq i}^\textrm{pairs}1\Bigg]\nn
&=&\frac{\pi\g}{N}\Bigg[-2+\sum_{j\neq i}^\textrm{pairs}{\rm sgn}(\vp_j-\vp_i)-\frac{N-3}{2}\Bigg]\nn
&=&\frac{\pi\g}{N}\Bigg[\sum_{j\neq i}^\textrm{pairs}{\rm sgn}(\vp_j-\vp_i)-\frac{N+1}{2}\Bigg].\label{eqE5}
\ea
The minimum of the remaining sum is $-(N-3)/2$, so that we get again the right-hand side of \Eq{eqEpsi},
\be
\psi_i\geq\frac{\pi\g}{N}\left(-\frac{N-3}{2}-\frac{N+1}{2}\right)=\frac{\pi\g}{N}(1-N)\,,
\ee
and everything else follows through to the bound \Eq{appEbou}.

Having shown that the bound \Eq{empibou} (\Eqq{appEbou}) holds for cuts all close to one another near the positive semi-axis (small angles $\vp_i\ll 1$), we prove that the region with all $\vp_i\to 0$ is the global extremum (actually maximum) of the hyper-surface $\cos\psi_i=0$ for all $\psi_i$, so that
for any $i$ and $j$
\be\label{maxi}
\left.\frac{\p\psi_i}{\p\vp_j}\right|_{\vp_k=0}=0\,.
\ee
Using \Eqq{eqE2} and calling ``signs'' all terms made explicit above and given by the sign of differences of angles, for even $N$ we have
\ba
\psi_i&=&\frac{\g}{N}\left[{\rm signs}-2s\vp_i-s\sum_{j\neq i}^\textrm{pairs}(\vp_i-\vp_j)-s\sum_{j\neq i}^\textrm{pairs}(\vp_i+\vp_j)\right]\nn
&=&\frac{\g}{N}\left[{\rm signs}-2s\vp_i-2s\sum_{j\neq i}^\textrm{pairs}\vp_i\right]\nn
&=&\frac{\g}{N}\left[{\rm signs}-2s\vp_i-2s\vp_i\frac{N-2}{2}\right]\nn
&=&\frac{\g}{N}\left({\rm signs}-Ns\vp_i\right),
\ea
and we get the same result for odd $N$ upon replacing
\ben
-2s\vp_i-2s\sum_{j\neq i}^\textrm{pairs}\vp_i\to -3s\vp_i-2s\sum_{j\neq i}^\textrm{pairs}\vp_i=-3s\vp_i-2s\vp_i\frac{N-3}{2}=-Ns\vp_i\,.
\een
Therefore, for $i\neq j$, when we do not allow for any coincident angles, we get \Eqq{maxi} for any $N\geq 2$, since the signs are taken as constants.

\subsection{No-go theorem on the Källén--Lehmann representation}\label{appD}

In this appendix, we prove that, if the propagator admits the representation \Eq{kalegen2},
\be\label{kalegen2app}
\tilde G(-k^2)=\int_{m^2}^{+\infty}\rmd s\,\frac{\rho(s)}{s+k^2}\,,
\ee
then it cannot scale faster than $k^{-2}$ in the UV if $\rho(s)\geq 0$. In particular, if the UV propagator scales as $k^{2\g}$ for $\g>1$, then the spectral function is not positive definite. We assume that the operations of taking the limit $k^2\to\infty$ and integrating in $s$ commute.

Consider the general mathematical problem of the limit
\ben
\lim_{k^2\to\infty}\left[k^{2\g}\int_{s_{-}}^{s_{+}}\rmd s\,\frac{\rho(s)}{s+k^2}\right],
\een
where $\g>1$, $s_{-}< s\leqslant s_{+}$ (with the possibility that $s_{+}=+\infty$) and $\rho(s)> 0$. Using the fact that $\rho$, $s_{-}$ and $s_{+}$ are, by assumption, independent of $k$, we have
\be
\hspace{-.2cm}\lim_{k^2\to\infty}\left[k^{2\g}\int_{s_{-}}^{s_{+}}\rmd s\,\frac{\rho(s)}{s+k^2}\right]=\!\int_{s_{-}}^{s_{+}}\!\rmd s\,\rho(s)\lim_{k^2\to\infty}\left[\frac{k^{2\g}}{s+k^2}\right]= \!\int_{s_{-}}^{s_{+}}\!\rmd s\,\rho(s)\,(+\infty)=+\infty,\label{appDeq1}
\ee
where in the last step we used positivity of $\rho(s)$, which implies that $\int_{s_{-}}^{s_{+}}\rmd s\,\rho(s)>0$. If $\rho(s)=0$ for all $s_{-}\leqslant s\leqslant s_{+}$, then the function is zero. If instead $\int_{s_{-}}^{s_{+}}\rmd s\,\rho(s)=0$, then we get the indefinite expression 
\ben
\lim_{k^2\to\infty}\left[k^{2\g}\int_{s_{-}}^{s_{+}}\rmd s\,\frac{\rho(s)}{s+k^2}\right]= \int_{s_{-}}^{s_{+}}\rmd s\,\rho(s)\,(+\infty)=0\times(+\infty)\,,
\een
which may also give finite values of the limit. If this happens, then $\tilde G(-k^2)\sim k^{-2\g}$ in the UV for $\g>1$. However, if $\rho$ is positive semi-definite and non-trivial, then its integral cannot be identically zero.

Therefore, the function \Eq{kalegen2app} cannot have a large-$k$ asymptotics $\sim k^{-2\g}$ for
$\g>1$, since his UV behaviour is only $\sim k^{-2}$:
\be\label{appDeq2}
\lim_{k^2\to\infty}\left[k^{2}\int_{s_{-}}^{s_{+}}\rmd s\,\frac{\rho(s)}{s+k^2}\right]=A\,,
\ee
where $0<A<+\infty$. If $\int_{s_{-}}^{s_{+}}\rmd s\,\rho(s)<0$ and finite, then the conclusion is the same, up to a flip of the sign of \Eqqs{appDeq1} and of $A$.

The above theorem rests upon forgetting about the technical origin of \Eqq{kalegen2app} and taking it at face value as an integral on the real line. However, the assumption that the operations of taking the limit $k^2\to\infty$ and integrating in $s$ commute is a delicate point. This was already recognized in \cite[section 10.7]{Wei95} in the case of the Källén--Lehmann representation of the quantum propagator of a standard two-derivative QFT but, if anything, in our case caveats can only pile up. Equation \Eq{kalegen2app} is the contribution of a branch cut in a complex contour, where all the other terms (branch point and arc at infinity) have already been shown to give a zero result. However, we have seen in the main text that vanishing of the contribution of the branch point requires $\g<1$, in contradiction with the hypothesis of the theorem. Moreover, sending $k^2$ to infinity may be inconsistent with having thrown away the arc at infinity. In other words, the limit $k^2\to\infty$ cannot be taken directly on the complex representation \Eq{opt}. Hence, for fractional theories the theorem does not hold.


\section{Example of a doubly split propagator}\label{appF}

In this appendix, we consider the propagator \Eq{ww} with $N=3$ and $\tilde N=2$. The calculation is relatively simple because all branch cuts are straight and we can use the same technology as for the single splitting. 

\begin{figure}
\bc
\includegraphics[width=7cm]{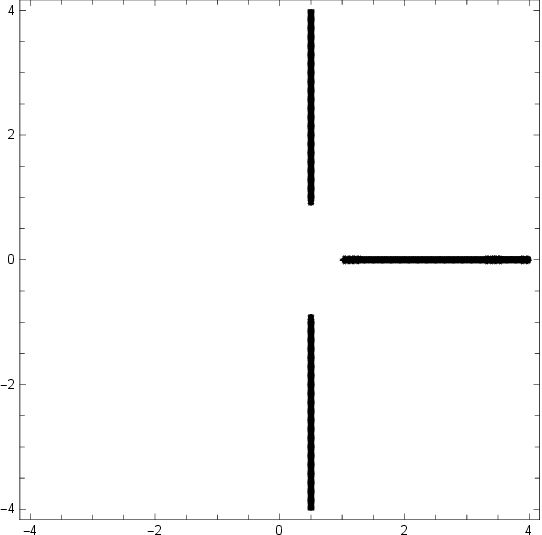}
\ec
\caption{\label{fig10} Branch cuts of the propagator \Eq{ww23}. The branch point on the positive real axis is at $z=m^2$, while the other two branch points are located at $(\Re\,z,\Im\,z)=(m^2\cos\vp,\pm m^2\sin\vp)$.}
\end{figure}

Using \Eqqs{w2} and \Eq{w3}, the $(\tilde N,N)=(2,3)$ UV propagator \Eq{ww} is
\be\label{ww23}
\tilde G(z) = \left[m^4+\left(m^2-z\right)^\frac23\left(m^4-2m^2z\cos\vp+z^2\right)^\frac23\right]^{-\frac{\g}{2}},\qquad 0<\vp<\frac{\pi}{2}\,,
\ee
possessing one branch cut in the positive real axis and two conjugate cuts parallel to the imaginary axis (Fig.~\ref{fig10}).

We dub the cut on the real axis ``1'' and parametrize it with $z=s$. The vertical cuts, dubbed ``3'' and ``4'' to use the same notation as in the main text, are parametrized by $z=m^2\cos\vp+\rmi s\sin\vp$ and $z=m^2\cos\vp-\rmi s\sin\vp$, respectively. Plugging these parametrizations into
\be
\tilde G_{{\rm cut}_i}(-k^2) = \frac{1}{2\pi\rmi}\int_{{\rm cut}_i}\rmd z\,\frac{\tilde G(z)}{z+k^2}\,,\qquad i=1,3,4\,,
\ee
we find the Källén--Lehmann representation of the free propagator \Eq{ww23}:
\be\label{propdou3}
\tilde G(-k^2) = \sum_{i=1,3,4}\tilde G_{{\rm cut}_i}(-k^2)=\int_{m^2}^{+\infty}\rmd s\left[\frac{\rho_1(s)}{s+k^2}+\frac{\rho_3(s)}{s-\rmi \tilde k^2}+\frac{\rho_4(s)}{s+\rmi \tilde k^2}\right],
\ee
where
\ba
\hspace{-.9cm}\rho_1(s) &\coloneqq& \lim_{\ve\to 0^+}\frac{\tilde G(s+\rmi\ve)-\tilde G(s-\rmi\ve)}{2\pi\rmi}\,,\\
\hspace{-.9cm}\rho_3(s) &\coloneqq& \lim_{\ve\to 0^+}\frac{\tilde G(m^2\cos\vp+\rmi s\,\sin\vp-\ve)-\tilde G(m^2\cos\vp+\rmi s\,\sin\vp+\ve)}{2\pi\rmi}\,,\label{rho3dou}\\
\hspace{-.9cm}\rho_4(s) &\coloneqq& \lim_{\ve\to 0^+}\frac{\tilde G(m^2\cos\vp-\rmi s\,\sin\vp+\ve)-\tilde G(m^2\cos\vp-\rmi s\,\sin\vp-\ve)}{2\pi\rmi}=\rho_3^*(s)\,,\\
\hspace{-.9cm}\tilde k^2 &\coloneqq& \frac{k^2+m^2\cos\vp}{\sin\vp}\,.
\ea
Then, \Eqq{propdou3} becomes
\be\label{propdou32}
\tilde G(-k^2) = \int_{m^2}^{+\infty}\rmd s\left[\frac{\rho_1(s)}{s+k^2}+2\frac{s\,\Re\,\rho_3(s)-\tilde k^2\Im\,\rho_3(s)}{s^2+\tilde k^4}\right],
\ee
which is manifestly real.

From now on, we assume without any loss of generality that
\be
0<\vp<\frac{\pi}{2}\qquad\Longrightarrow\qquad \cos\vp>0\,,\quad \sin\vp>0\,,\quad \tilde k^2>0\,,
\ee
where the last inequality holds in Euclidean signature. Therefore, if we do not impose the Anselmi--Piva procedure, unitarity is achieved if the integrand in \Eqq{propdou32} is positive semi-definite in Euclidean signature, which is achieved if, and only if,
\be\label{3condi}
\rho_1(s)\geq 0\,,\qquad \Re\,\rho_3(s)\geq 0\,,\qquad \Im\,\rho_3(s)\leq 0\,,\qquad \forall\,s\in[m^2,+\infty)\,.
\ee
A direct calculation gives
\be
\tilde G(s\pm\rmi\ve)=\cA(s)\,\exp\left[\pm\frac{\rmi}{2}\g\a(s)\right],
\ee
where
\ba
\cA(s)&\coloneqq&\left|m^8+A^2(s)-m^4A(s)\right|^{-\frac{\g}{4}},\\
A(s)  &\coloneqq& \left|(s-m^2)(s^2+m^4-2m^2s\cos\vp)\right|^\frac23,\\
\a(s) &\coloneqq& \left\{\begin{matrix}
 {\rm arctan}\!\left(\frac{A\,\sin\frac{2\pi}{3}}{m^4+A\,\cos\frac{2\pi}{3}}\right)\hphantom{+\pi}\qquad\hspace{.1cm} \text{if }\, s \leq \tilde s\,, \\
	\\
  {\rm arctan}\!\left(\frac{A\,\sin\frac{2\pi}{3}}{m^4+A\,\cos\frac{2\pi}{3}}\right)+\pi\qquad \text{if }\, s > \tilde s\,,
\end{matrix}\right.
\ea
and $\tilde s$ is the only real solution of
\be
x(\tilde s)=m^4+A(\tilde s)\,\cos\frac{2\pi}{3}=0\,.
\ee
Therefore,
\be
\rho_1(s)=\cA\,\frac{\sin\frac{\g\a}{2}}{\pi}\,,
\ee
so that the first positivity condition in \Eq{3condi} is respected provided $\sin(\a\g/2)\geq 0$ for all $s$. Since the angle $\a$ is positive and varies monotonically from 0 to $2\pi/3$, the positivity condition reads $4n\pi /\a(s)\leq\g\leq (4n+2)\pi/\a(s)$ for $n=0,1,2,\dots$. However, we can pick only $n=0$. To see this, we note that, when $s\to m^2$, one has $A\to 0$, $\cA\to m^{-2\g}$ and $\a\to 0$, while in the limit $s\to+\infty$ one has $A\to s^2$, $\cA\to s^{-\g}$ and $\a\to 2\pi/3$. On one hand, when $s\to m^2$ the $n=0$ interval is $0<\g<+\infty$, while all the other intervals become progressively displaced towards infinity. On the other hand, when $s\to+\infty$ we have $6n\leq\g\leq 3(2n+1)$, so that the only non-trivial intersection of the intervals at $s\to m^2$ and $s\to+\infty$ occurs for $n=0$:
\be\label{poscond1}
0<\g\leq 3\,.
\ee

To get $\rho_3$, we need
\be\label{GAt}
\tilde G_\pm(s)\coloneqq \lim_{\ve\to 0^+}\tilde G(m^2\cos\vp+\rmi s\,\sin\vp\pm\ve)=\cB_\pm(s)\,\exp\left[-\frac{\rmi}{2}\g\tau_\pm(s)\right],
\ee
where
\ba
\cB_\pm(s) &\coloneqq& \left|m^8+B^2(s)+2m^4B(s)\,\cos\left\{\frac23[\de(s)\mp\pi]\right\}\right|^{-\frac{\g}{4}},\\
B(s)       &\coloneqq& \left|(s^2-m^4)\sin^2\vp\right|^\frac{2}{3}\left|m^4(1-\cos\vp)^2+s^2\sin^2\vp\right|^\frac{1}{3},\\
\de(s)     &\coloneqq& {\rm arctan}\frac{s\,\sin\vp}{m^2(1-\cos\vp)}\,,\\
\tau_+(s)  &\coloneqq& {\rm arctan}\frac{y_+}{x_+} = -{\rm arctan}\left\{\frac{B\,\sin\left[\frac23(\de-\pi)\right]}{m^4+B\,\cos\left[\frac23(\de-\pi)\right]}\right\},\\
\tau_-(s)  &\coloneqq& \left\{\begin{matrix}
  \hspace{-.1cm}-{\rm arctan}\left\{\frac{B\,\sin\left[\frac23(\de+\pi)\right]}{m^4+B\,\cos\left[\frac23(\de+\pi)\right]}\right\}\hphantom{+\pi}\qquad\hspace{.1cm} \text{if }\, s \leq \bar s\,, \\
	\\
  -{\rm arctan}\left\{\frac{B\,\sin\left[\frac23(\de+\pi)\right]}{m^4+B\,\cos\left[\frac23(\de+\pi)\right]}\right\}-\pi\qquad \text{if }\, s > \bar s\,.
\end{matrix}\right.
\ea
Here the critical value $\bar s\in\mathbb{R}$ can be found numerically for any $\vp$ from the transcendental equation
\be
x_-(\bar s)=m^4+B(\bar s)\,\cos\left\{\frac23[\de(\bar s)+\pi]\right\}=0\,.
\ee

\begin{figure}
\bc
\includegraphics[width=7cm]{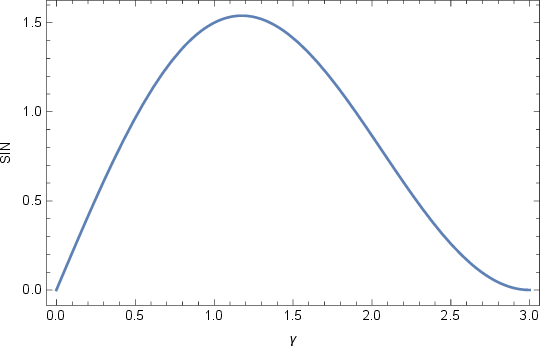}\qquad\includegraphics[width=7cm]{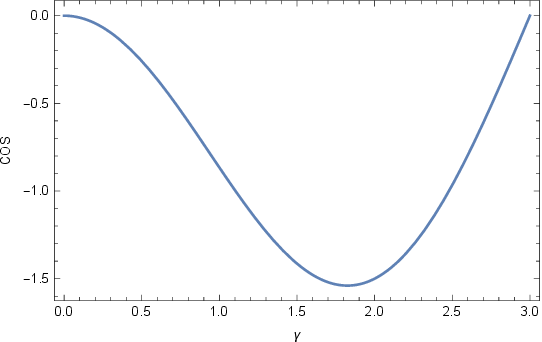}
\ec
\caption{\label{fig11} The functions \Eq{poscond2} in the range \Eq{poscond1}.}
\end{figure}

Plugging \Eqq{GAt} into \Eq{rho3dou}, we obtain
\be
\rho_3(s) = \frac{\tilde G_--\tilde G_+}{2\pi\rmi}=\frac{\cB_-\,\rme^{-\frac{\rmi}{2}\g\tau_-}-\cB_+\,\rme^{-\frac{\rmi}{2}\g\tau_+}}{2\pi\rmi}\,,
\ee
so that
\ba
2\pi\Re\,\rho_3(s) &=& \cB_+\sin\frac{\g\tau_+}{2}-\cB_-\sin\frac{\g\tau_-}{2}\,,\\
2\pi\Im\,\rho_3(s) &=& -\left(\cB_-\cos\frac{\g\tau_-}{2}-\cB_+\cos\frac{\g\tau_+}{2}\right),
\ea
and the last two conditions in \Eq{3condi} become
\be\label{3condi3}
\cB_+\sin\frac{\g\tau_+}{2}-\cB_-\sin\frac{\g\tau_-}{2}\geq 0\,,\qquad \cB_-\cos\frac{\g\tau_-}{2}-\cB_+\cos\frac{\g\tau_+}{2}\geq 0\,,\qquad \forall\,s\in[m^2,+\infty)\,.
\ee
We can verify these conditions easily by imposing them in the limiting cases $s\to m^2$ and $s\to+\infty$. We have
\ba
s\to m^2:\qquad&& B \to 0\,,\\
&& \de\to \de(m^2)\,,\nn
&& \cB_\pm\to m^{-2\g}\,,\nn
&& \tau_+\to 0\,,\qquad \tau_-\to 0\,,\nn
s\to +\infty:\quad && B \to s^2\sin^2\vp\,,\\
&& \de\to \frac{\pi}{2}\,,\nn
&& \cB_\pm\to B^{-\frac{\g}{2}}=(s\,\sin\vp)^{-\g}\,,\nn
&& \tau_+\to \frac{\pi}{3}\,,\qquad \tau_-\to -\pi\,,\non
\ea
so that in the limit $s\to m^2$ both conditions in \Eq{3condi3} are saturated to zero, while in the limit $s\to +\infty$ we get ${\rm SIN}\geq 0$ and ${\rm COS}\geq 0$, where
\be\label{poscond2}
{\rm SIN}\coloneqq \sin\frac{\g\pi}{6}+\sin\frac{\g\pi}{2}\,,\qquad {\rm COS}\coloneqq \cos\frac{\g\pi}{2}-\cos\frac{\g\pi}{6}\,.
\ee
Note that these conditions are independent of $\vp$, so that they must hold for any $0<\vp<\pi/2$. As anticipated in the main text, tree-level unitarity is never achieved because the second condition is never satisfied in the allowed range \Eq{poscond1}, as one can see in Fig.~\ref{fig11}. This is not a problem as long as we make the Anselmi--Piva prescription on the propagator. One of the goals of this appendix was to show that the spectral functions associated with the branch cuts of the model \Eq{ww23} are all well defined, i.e., real and finite, so that the Anselmi--Piva prescription can indeed be applied.


\end{document}